\documentclass[12pt]{iopart}
\usepackage{harvard}
\citationmode{abbr}
\usepackage{graphicx}
\usepackage{subcaption}
\pdfoutput=1
\usepackage{hyperref}
\usepackage[utf8]{inputenc}
\usepackage[export]{adjustbox}
\usepackage{wrapfig}
\graphicspath{{./}{figures/}}
\usepackage{lineno}
\nolinenumbers
\begin{document}

\title[Open clusters Haffner 22 and Melotte 71]{A deep investigation of two poorly studied open clusters Haffner 22 and Melotte 71
in Gaia era.}

\author{D. Bisht$^1$, Quingfeng Zhu$^1$, R. K. S. Yadav$^2$ , Geeta Rangwal$^3$, Devesh P. Sariya$^4$, Alok Durgapal$^5$ and Ing-Guey Jiang$^4$}\address{$^1$ Key Laboratory
for Researches in Galaxies and Cosmology, University of Science and
Technology of China, Chinese Academy of Sciences, Hefei, Anhui, 230026, China}\address{$^2$ Aryabhatta Research Institute of Observational Sciences,
           Manora Peak, Nainital 263129, India}\address{$^3$ Indian Institute of Astrophysics, Koramangala II BLock, Bangalore, 560034, India}
\address{$^4$ Department of Physics and Institute of Astronomy, National Tsing-Hua University, Hsin-Chu, Taiwan}
\address{$^5$ Center of Advanced Study, Department of physics, D.S.B. campus, Kumaun University, Nainital, 263002, India}\ead{dbisht@ustc.edu.cn}

\vspace{10pt}
\begin{indented}
\item[]March 2022
\end{indented}

\begin{abstract}
This paper presents a deep investigation of two open clusters, Haffner 22 and Melotte 71, using astrometric and photometric
data from Gaia~EDR3. We identified 382 and 597 most probable cluster members with membership probability higher than $50 \%$.
Mean proper motions in RA and DEC are estimated as ($-1.631\pm0.009$, $2.889\pm0.008$) and ($-2.398\pm0.004$,
$4.210\pm0.005$) ~mas~yr$^{-1}$
for Haffner 22 and Melotte 71, respectively. A comparison of observed CMDs with the theoretical isochrones leads to an age of $2.25\pm0.25$
and $1.27\pm0.14$ Gyr for these clusters. The distances $2.88\pm0.10$ and $2.28\pm0.15$ kpc based on the parallax are comparable with the
values derived by the isochrone fitting method.
Five and four blue straggler stars (BSS) are identified as cluster members in Haffner 22 and Melotte 71. Based on the
relative number of high-velocity (binary) and single stars, we inferred binary fractions for both clusters in the range of
$\sim$ $10\% \le f_{bin} \le 14\%$, for both core and oﬀ-core regions. We found binary content is larger in the core region. Mass
function slope is in good agreement with the Salpeter's value for Melotte 71 (x=$1.23\pm0.38$ within mass range 1-3.4 $M_{\odot}$) while
it is quite a flat slope for Haffner 22 (x=$0.63\pm0.30$ within mass range 1-2.3 $M_{\odot}$).
Evidence for the existence of mass-segregation effect is observed in both clusters. Using the Galactic potential model, Galactic orbits
are derived, indicating that both clusters follow a circular path around the Galactic center, evolving slowly.
\end{abstract}

%
%
%
%
%

\section{Introduction} \label{sec:intro}

\begin{figure*}
\begin{center}
\hbox{
\includegraphics[width=8.5cm, height=6.5cm]{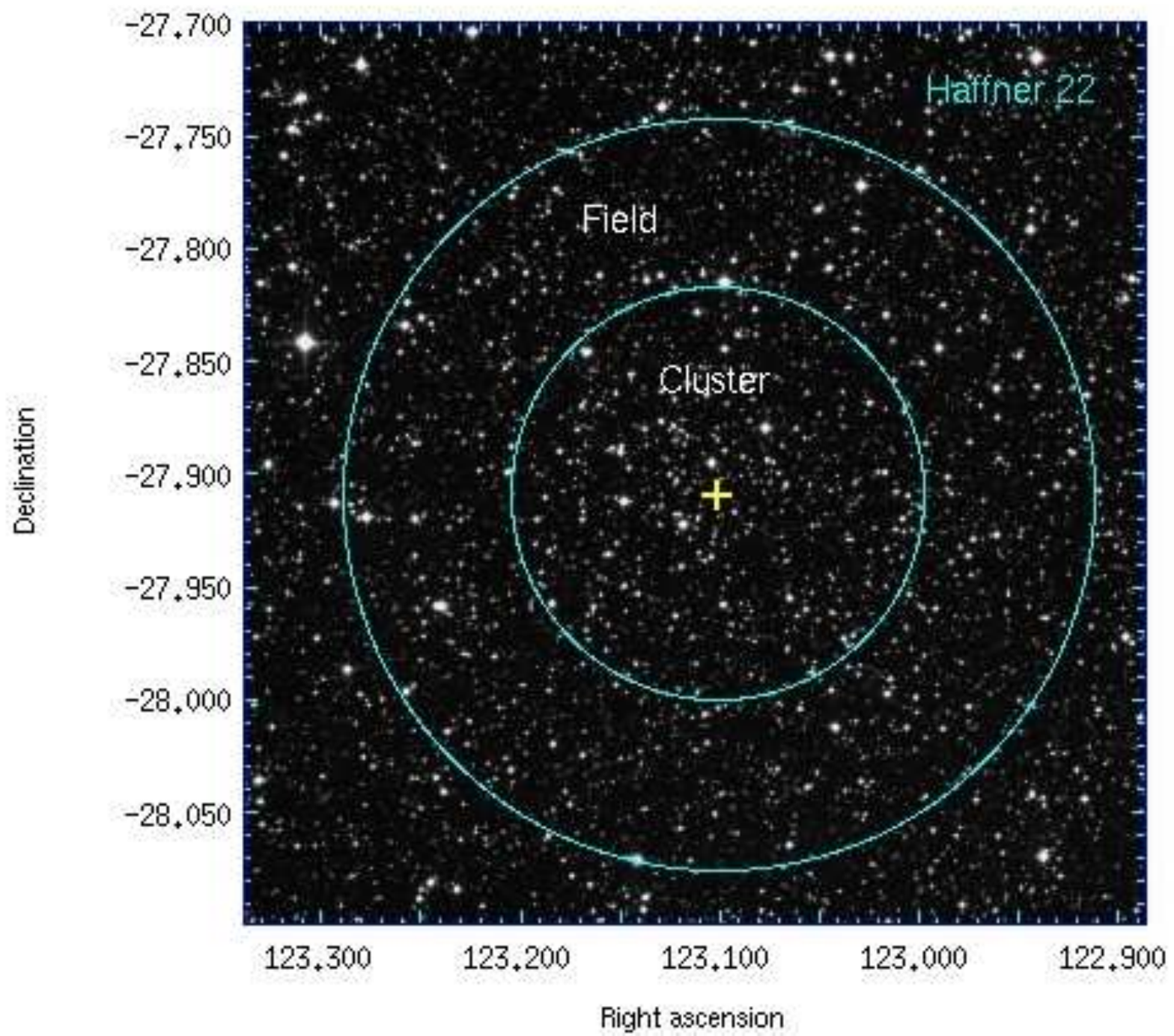}
\hspace{-1cm}\includegraphics[width=8.0cm, height=6.5cm]{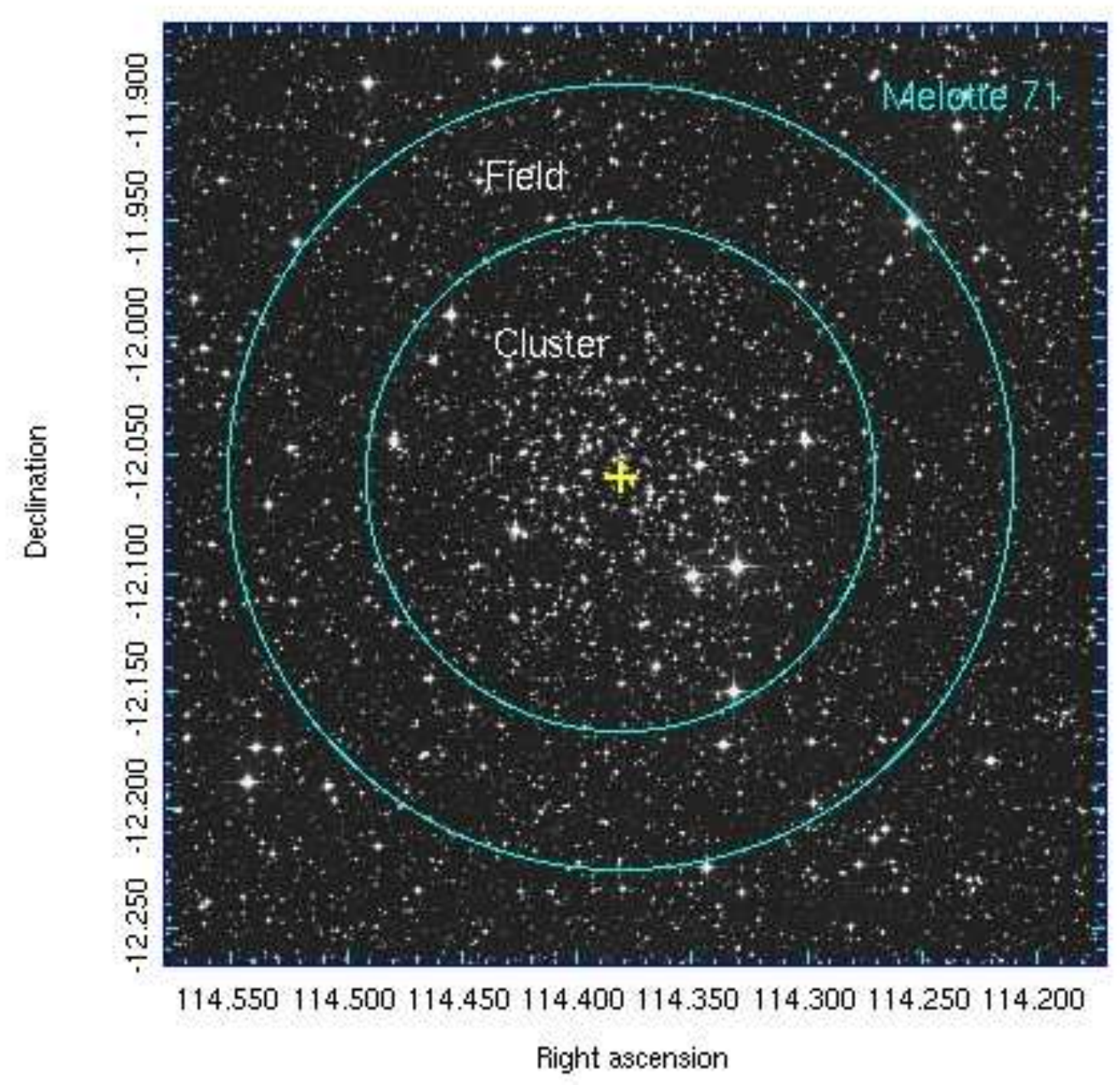}
}
\caption{Identification maps of clusters Haffner 22 and Melotte 71 taken from the Leicester Database and Archive Service. 
The inner-circle indicates the radius of clusters, while the outer circle indicates the radius of data extraction in this study.
Plus sign indicates the cluster center.}
\label{id}
\end{center}
\end{figure*}

Open clusters (OCs) in the Milky Way span a wide range in ages, distances, and chemical compositions ( Dias et al. 2002;
Kharchenko et al. 2013; Cantat-Gaudin et al. 2020). The (early) Third Gaia Data Release 3 (hereafter EDR3; Gaia Collaboration et al. 2020)
of Gaia mission was made public on 3$^{rd}$ December 2020. This data contains the central coordinates, proper motions in both the
right ascension and declination and parallax $(\alpha, \delta,
\mu_{\alpha}cos\delta, \mu_{\delta}, \pi)$ for more than 1.4 billion sources within the limit of 3 to 21 mag in $G$ band. We
can provide an estimation of cluster membership to enhance our understanding of the fundamental
parameters of OCs. Cantat-Gaudin et al. (2018) provided a membership probability catalog for 1229 OCs using Gaia DR2 for
stars up to 18 mag in the $G$ band. Recently they added few more clusters to get a catalog of 1481 OCs
(Cantat-Gaudin \& Anders 2020). Liu \& Pang (2019) used the Friend of Friend (FoF) method to find 2443 OCs and select their
probable members. The FoF method is based on the galaxy group finder algorithm proposed by Yang et al. (2005).
Sim et al. (2019) listed 655 cluster candidates (207 new candidates) based on the visual inspection of the stellar distributions in proper motion
space and spatial distributions in the Galactic coordinates space. Monteiro et al. (2020) investigated 45 OCs using the maximum
likelihood method to estimate membership in cluster regions. Ferreira et al. (2020) discovered 25 new OCs and identified
probable members using a decontamination procedure to the three-dimensional astrometric space.

Both clusters are very sparse and located in the third Galactic quadrant. They are situated very near to the Galactic
disk hence highly contaminated by field stars. Very few studies are available for these clusters. 
Haffner 22 and Melotte 71 fields of view contain a large number of field stars so
it is very necessary to separate those stars from the actual cluster stars
to identify accurate fundamental parameters.
Mass function and segregation are not well known for Haffner 22, while for Melotte 71, it has been done without selecting
cluster members. The Gaia EDR3 data enable us to distinguish cluster members and study the structural properties and
the dynamical status of clusters.
The cluster parameters and mass function (MF) derived using the cluster members would significantly enhance the
knowledge of these poorly studied open clusters.
These objects are old age OCs and have identical locations in the Galaxy. So, we
can compare their dynamical behaviour on the position in the Galaxy. We include an orbital
analysis of these
two candidates for the first time using Gaia EDR3 data. Orbits of OCs are essential to understand the influence of tides and the
formation and evolution processes of the clusters.
The primary aim of paper is to select the probable cluster members, obtain the fundamental parameters, binary
fraction, the mass function slope, and explain the Galactic orbits of Haffner 22 and Melotte 71 in the Gaia era.
The identification maps for both
clusters are shown in Fig.~\ref{id}, which is taken from the Leicester
Database and Archive Service.

The available information for target clusters is as follows:

{\bf Haffner 22}:  ($\alpha_{2000} = 8^{h}12^{m}27^{s}$, $\delta_{2000}=-27^{\circ} 54^{\prime} 00^{\prime\prime}$;
$l$=246.775 deg, $b$=3.377 deg) (Dias et al. 2002). Kharchenko et al. (2013) cataloged the proper motions, distance, reddening and log(age) value of Haffner 22
as ($-$4.52, 6.90) mas yr$^{-1}$, 2796 pc, 0.21 mag and 9.19, respectively using 2MASS and PPMXL data. Dias et al. (2014) derived
proper motion values of this cluster
as $-1.95\pm1.80$ and $2.67\pm1.57$ mas yr$^{-1}$ based on UCAC4 catalog. A catalog of cluster membership has been provided by
Sampedro et al. (2017) based on UCAC4 data. Cantat-Gaudin et al. (2018) has made a catalog for the cluster members and estimated
the physical parameters of Haffner 22 based on Gaia DR2 data. We have given comparision table for the fundamental parameters in Table \ref{comp_para}.

{\bf Melotte 71}: ($\alpha_{2000} = 7^{h}37^{m}30^{s}$, $\delta_{2000}=-12^{\circ} 4^{\prime} 00^{\prime\prime}$;
$l$=228.949 deg, $b$=4.498 deg) (Dias et al. 2002). Sampedro et al. (2017) cataloged the log(age), distance and reddening values
as 8.37, 3154 pc and 0.11 mag, respectively.
Kharchenko et al. (2013) cataloged the proper motions, distance, reddening and log(age) value of Melotte 71 as (-0.94, 4.72) mas yr$^{-1}$,
2473 pc, 0.10 mag and 8.97, respectively.  Brown et al. (1996) analyzed the  high-dispersion echelle spectra for two or three stars
in the old anticenter disk cluster Melotte 71 and obtained [Fe/H] value as $-0.3\pm0.2$ dex. Mermilliod et al. (1997) have used new
CORAVEL radial-velocity observations and $UBV$ photometry of 24 red giants in the field of Melotte 71 and obtained mean radial
velocity as $50.14\pm0.14$ km/sec. Twarog et al. (2006) have presented a CCD photometry on the intermediate-band $uvbyCaH\beta$
system for Melotte 71. Interstellar reddening, age and distance modulus have been estimated as $0.20\pm0.004$,
$0.9\pm0.1$ Gyr and $12.2\pm0.1$ mag.


Binary stars are a unique tool to gather valuable
information about the stellar properties. They play a vital role in the dynamical evolution of star clusters (Sollima et al., 2008).
The observational and theoretical works suggest that the stars in clusters may be born originally with large binary fractions, and
thus the majority of stars are found in binary systems
(Duquennoy \& Mayor 1991; Kroupa 1995a,b; Griffin \& Suchkov
2003; Goodwin \& Kroupa 2005; Kouwen-hoven et al. 2005; Rastegaev 2010). Many of the Milky Way's open and globular clusters
show a binary fraction with a rising binary
frequency towards the cluster core, which is interpreted to be the
result of mass segregation (e.g., Mathieu \& Latham 1986; Geller \&
Mathieu 2012; Milone et al. 2012).
Binary stars can affect the observational parameters of a star cluster, such as the velocity dispersion and the stellar mass function.
Therefore, the distribution of the binary fraction in OCs is
of crucial importance for several fields in astrophysics.

The outlook of the paper is as follows. Section 2 describes the used data. Section 3 deals with the study of proper motion, determination
of distance using parallax,
determining the membership probability of stars, and identifying the BSS. The structural properties, derivation of fundamental parameters
using the most probable cluster members, and the study of binary fraction in both clusters have been carried out in Section 4.
The dynamical analysis of the clusters is
discussed in Section 5. The cluster's orbit is studied in Section 6. Finally, the conclusions are presented in Section 7.

\section{Data used}

We have used a photometric and astrometric database of 10 arcmin radius circular field from the Gaia~EDR3
(Gaia Collaboration et al. 2020) catalog for the clusters
Haffner 22 and Melotte 71.
No quality cuts have been made in this selection. If we increase our selection radius from 10 arcmin then more members can be
found, but this won't change the main results discussed in this paper significantly.
The total number of stars within the applied radius were 6172 and 8043 for clusters Melotte 71 and Haffner 22.
The main quantities contained by the above catalog are:
positions $(\alpha, \delta)$,
parallaxes and proper motions ($\mu_{\alpha} cos\delta , \mu_{\delta}$) up to a limiting magnitude of $G=21$ mag.
The uncertainties in the parallax values are $\sim$ 0.02--0.03 milliarcsecond (mas) for sources at $G\le15$ mag and
$\sim$ 0.07 mas for sources with $G\sim17$ mag.
 The uncertainties in the corresponding proper motion
components are $\sim$ 0.01--0.02 mas yr$^{-1}$ (for $G\le15$ mag), $\sim$0.05 mas yr$^{-1}$ (for $G\sim17$ mag),
$\sim$0.4 mas yr$^{-1}$ (for $G\sim20$ mag) and $\sim$1.4 mas yr$^{-1}$ (for $G\sim21$ mag). In this paper we have
used stars upto 20$^{th}$ G mag. Using parallax and proper motion conditions,
we have removed 1950 and 1220 stars from Haffner 22 and Melotte 71 respectively.

\begin{figure*}
\begin{center}
\hbox{
\includegraphics[width=8.5cm, height=8.5cm]{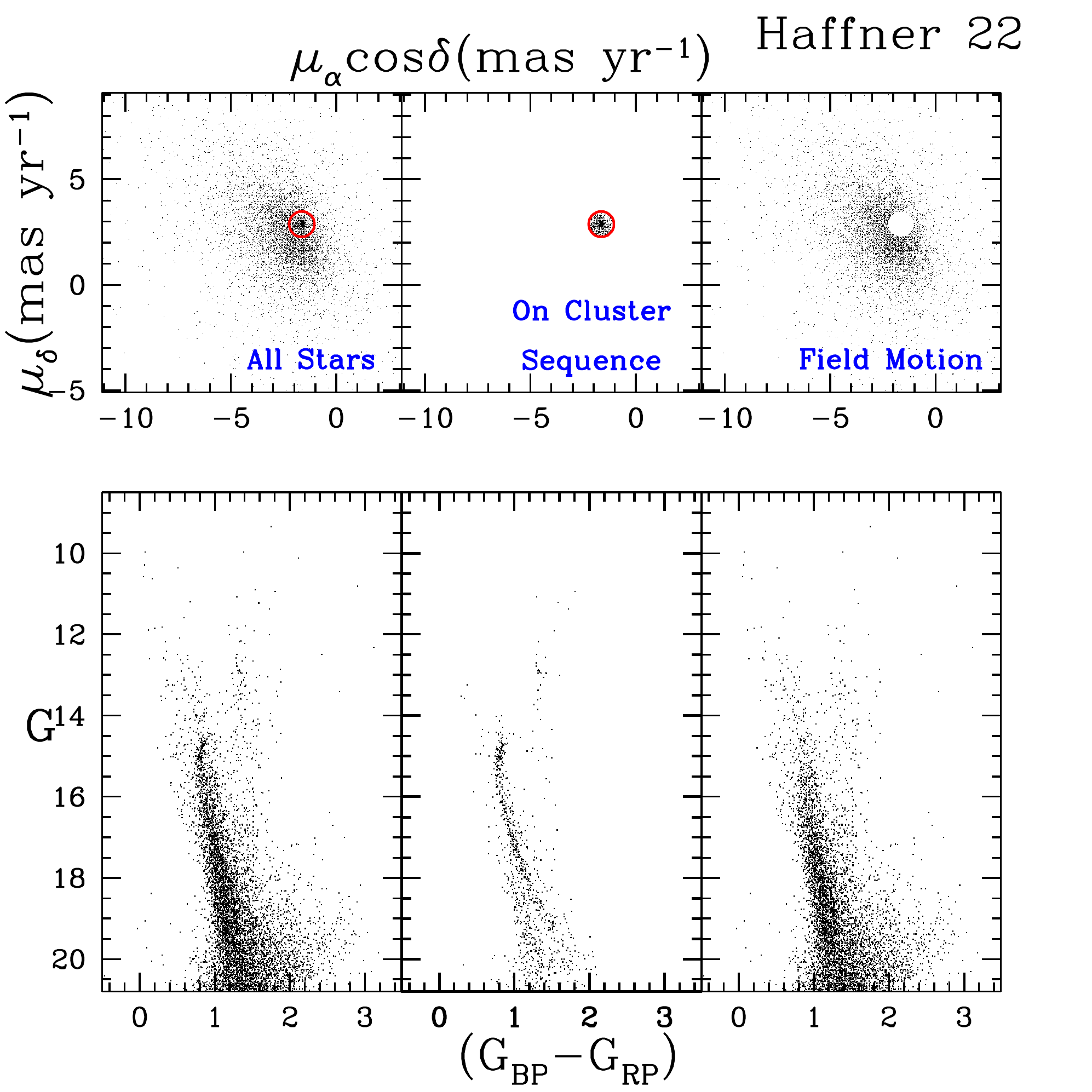}
\includegraphics[width=8.5cm, height=8.5cm]{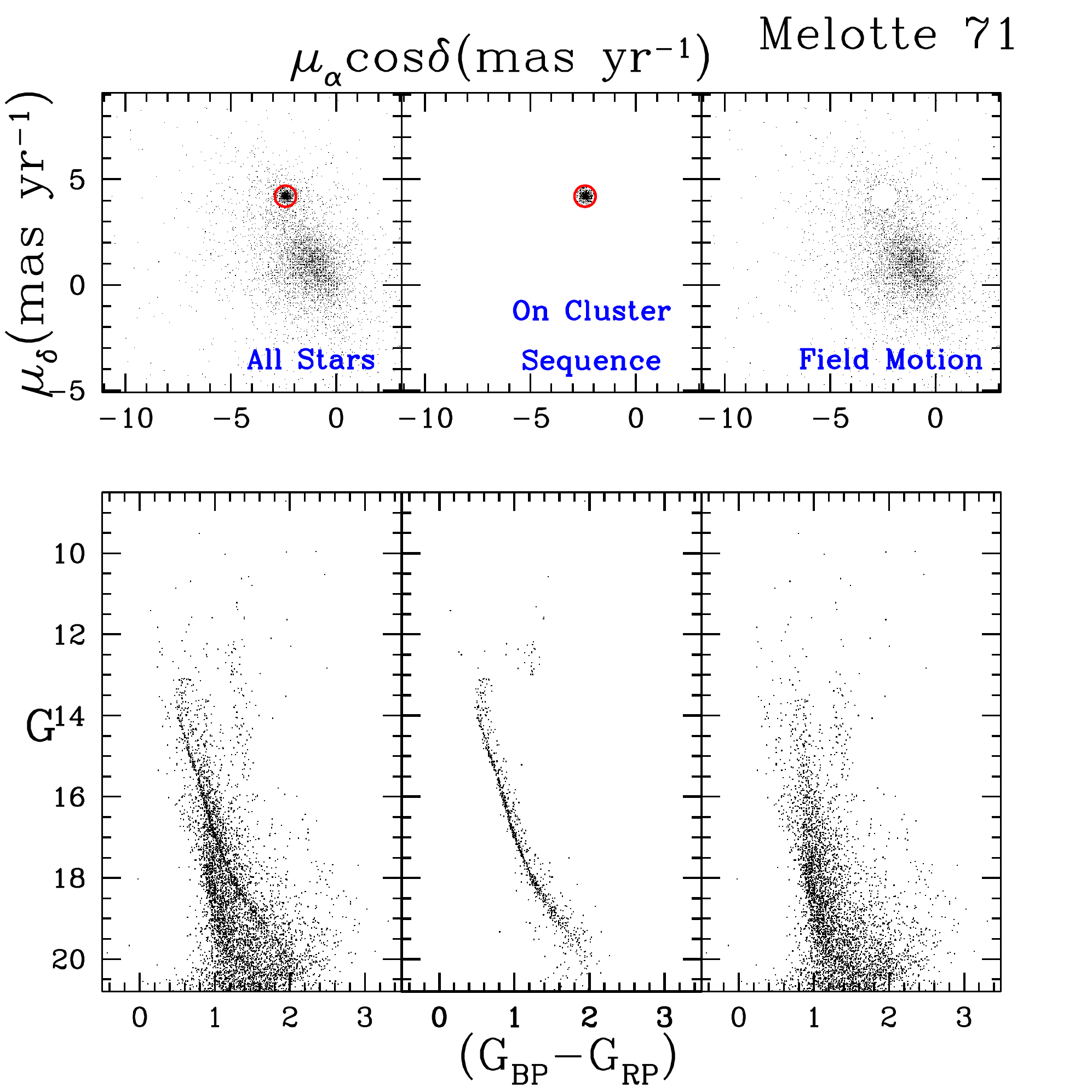}
}
\caption{(Top panels) Proper-motion vector point diagrams (VPDs) for Haffner 22 and Melotte 71. (Bottom panels)
$G$ versus $(G_{BP}-G_{RP})$ color magnitude diagrams. (Left panel) The entire sample. (Center) Stars within
the circle of 0.6 and 0.5 ~mas~yr$^{-1}$ radius for the clusters Haffner 22 and Melotte 71 centered around the
mean proper motion of the cluster. (Right) Probable background/foreground field stars in the direction of
this object.}
\label{vpd}
\end{center}
\end{figure*}

\section{Proper motions and Field star separation}

It is required to have precise information about proper motions to differentiate member stars from field stars. We used the kinematical
data from the Gaia EDR3 catalog to separate field stars from cluster stars. PMs, $\mu_{\alpha} cos{\delta}$ and $\mu_{\delta}$
are plotted as vector point diagrams (VPDs) in the top panels of Fig. \ref{vpd}. The bottom panels show $G$ versus $(G_{BP}-G_{RP})$
CMDs. The left panel in the CMDs shows all-stars present in the cluster's area, while the middle and right panels show the possible
cluster members and non-member stars, respectively. By visual inspection, we define the center and radius of the cluster members
in
VPD for a preliminary analysis. This selection was performed to minimize the field star contamination and keep the maximum possible
number of lower mass stars. A circle of 0.6  mas~ yr$^{-1}$ for Haffner 22 while 0.5 mas~ yr$^{-1}$ for Melotte 71 around the center
of the member stars distribution in the VPDs characterize our membership criteria. The picked radius is an agreement between losing
cluster members with poor PMs and the involvement of non-member stars. We have also used parallax for the reliable estimation of
cluster members. A star is considered a probable cluster member if it lies inside the circle in VPD and has a parallax value
within 3$\sigma$ from the mean cluster parallax. Finally, the main sequence of the cluster is separated.
These stars have a PM error of $\le$ 0.5 ~ mas~ yr$^{-1}$.

\begin{figure}
\begin{center}
\centering
\hbox{
\includegraphics[width=4.0cm, height=4.0cm]{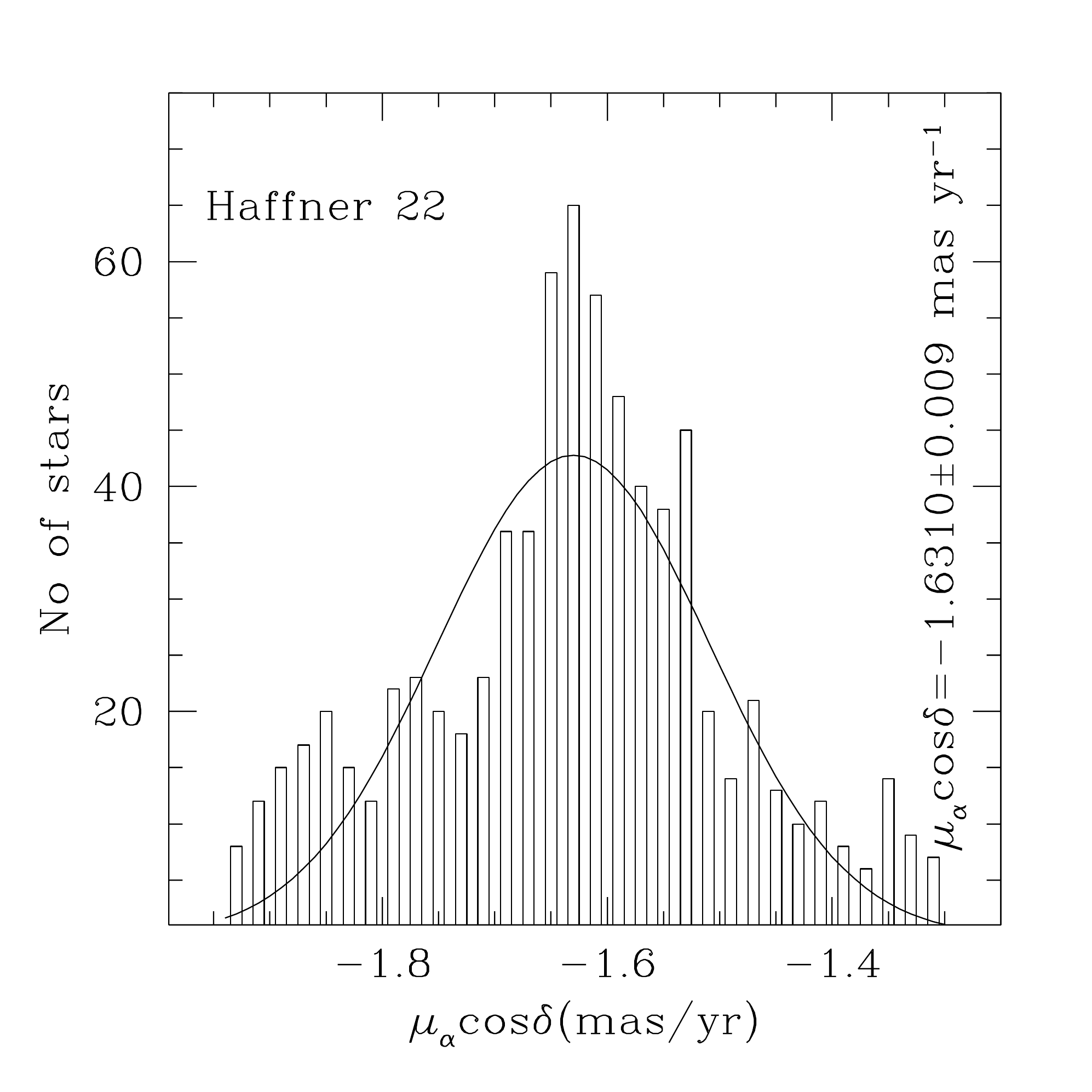}
\includegraphics[width=4.0cm, height=4.0cm]{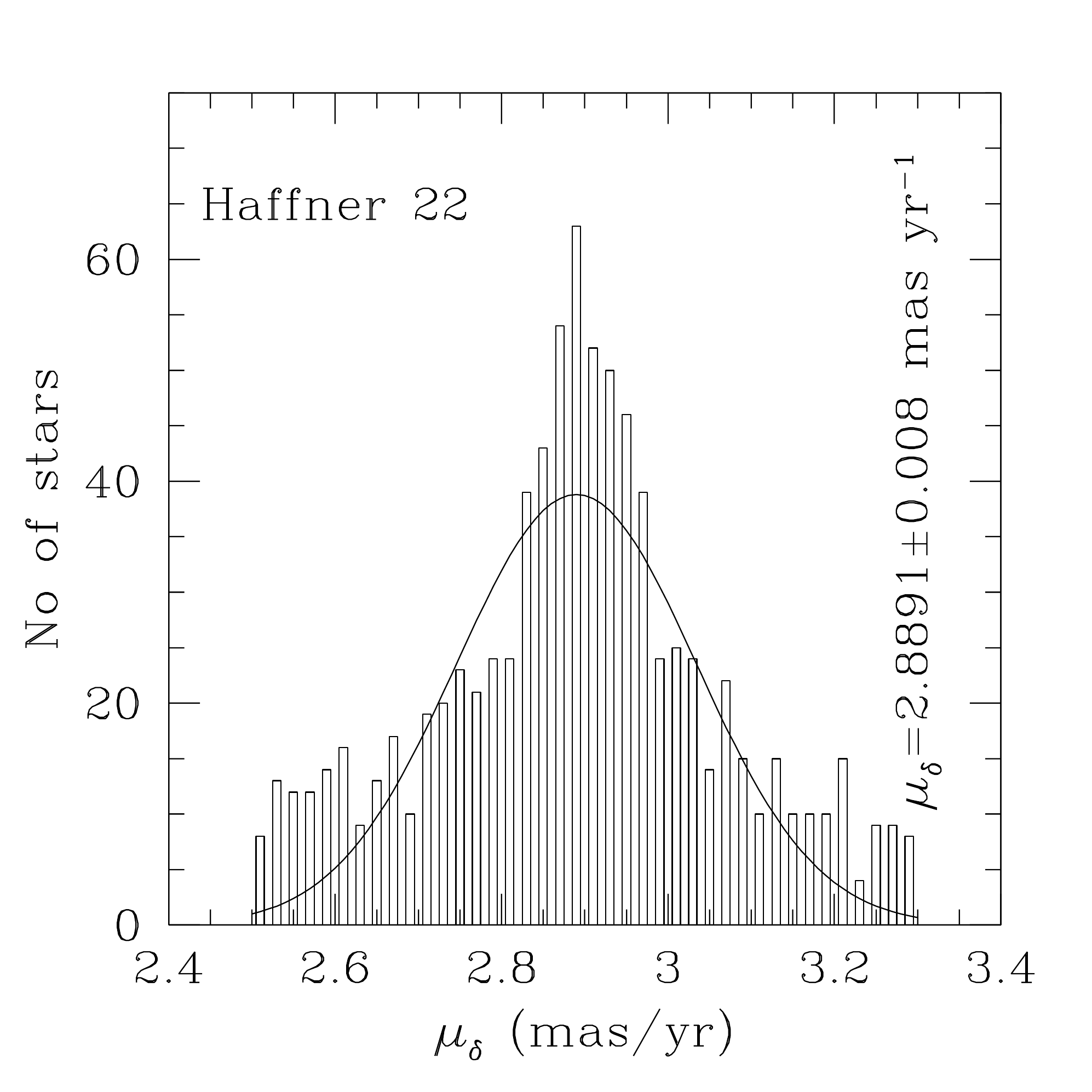}
}
\hbox{
\includegraphics[width=4.0cm, height=4.0cm]{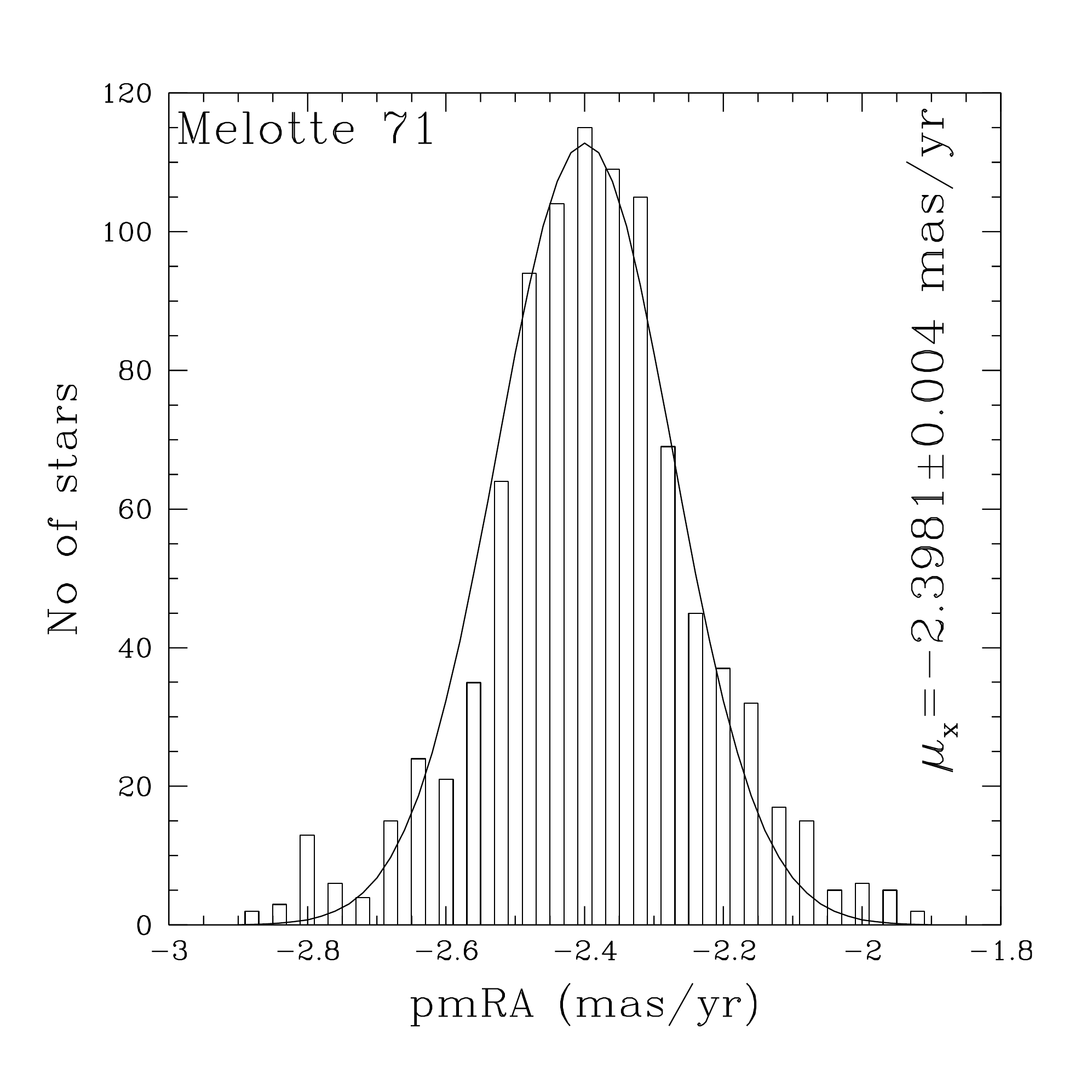}
\includegraphics[width=4.0cm, height=4.0cm]{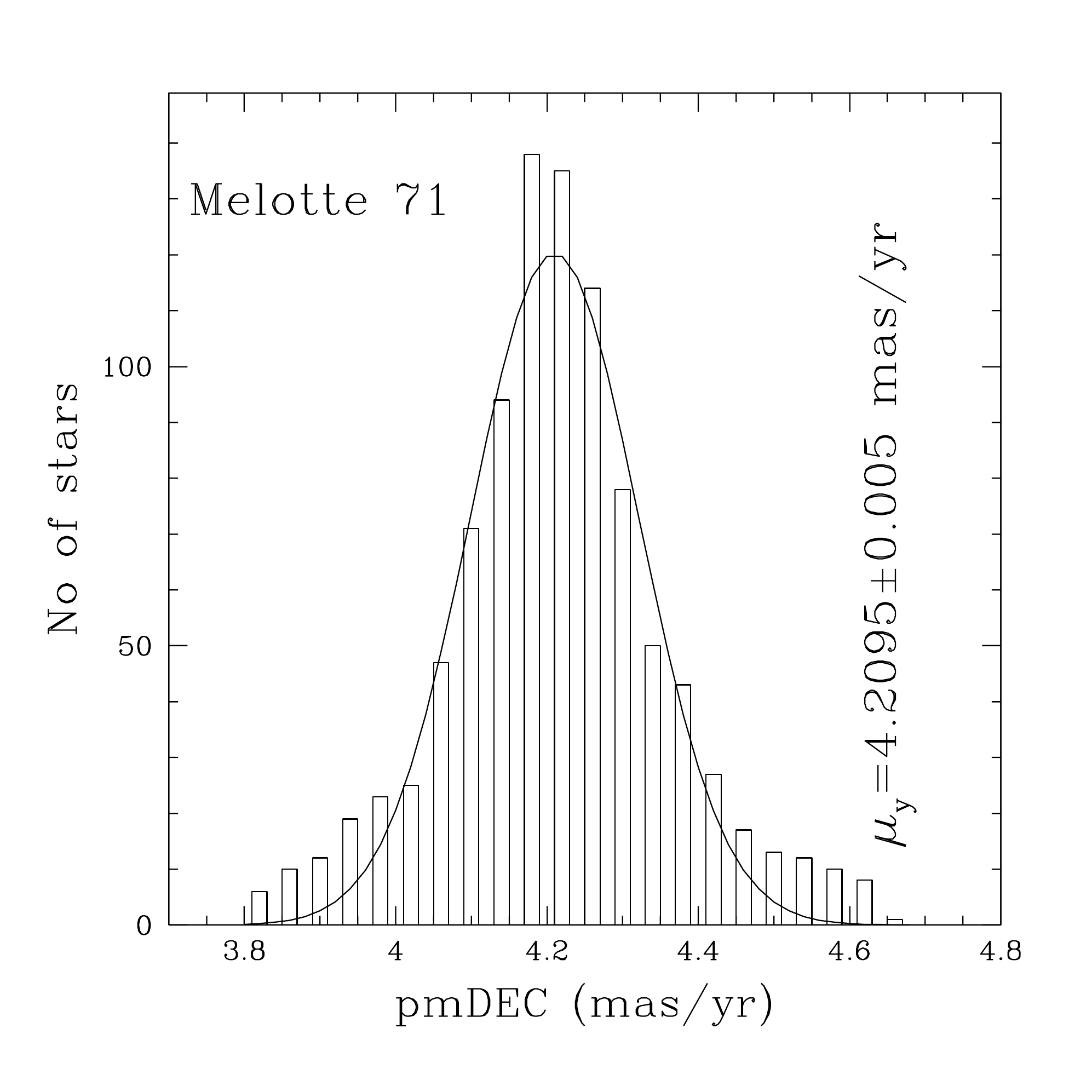}
}
\caption{Proper motion histograms of 0.1 mas~yr$^{-1}$ bins in right ascension and declination of the clusters Haffner 22
and Melotte 71. The Gaussian function fit to the central bins provides the mean values in both directions as shown
in each panel.
}
\label{pm_hist}
\end{center}
\end{figure}

For the precise estimation of mean proper motion, we deal only with the probable cluster members based on the clusters' VPDs and CMDs as shown
in Fig. \ref{pm_hist}. By fitting the Gaussian function on the constructed histograms, we determined the mean proper motion in the
directions of RA and DEC as ($-1.631\pm0.009$, $2.889\pm0.008$) and ($-2.398\pm0.004$, $4.210\pm0.005$) ~mas~yr$^{-1}$
for Haffner 22 and Melotte 71. From the peak of the Gaussian distribution, we found the mean proper motion in
RA and DEC directions for both the clusters and are listed in Table \ref{para}. The estimated values of mean proper motions for each cluster
are in fair agreement with the values given by Liu \& Pang (2019) and Cantat-Gaudin et al. (2018). Cantat-Gaudin et al. (2018) catalog
report the membership probabilities of stars up to $18^{th}$ magnitude in the $G$ band for both clusters. We derived membership probabilities
of the stars up to $20^{th}$ magnitude in the $G$ band and the method we used has been discussed in the next section.

\subsection{Distance of clusters using parallax}

\begin{figure}
\begin{center}
\hbox{
\includegraphics[width=4.0cm, height=4.0cm]{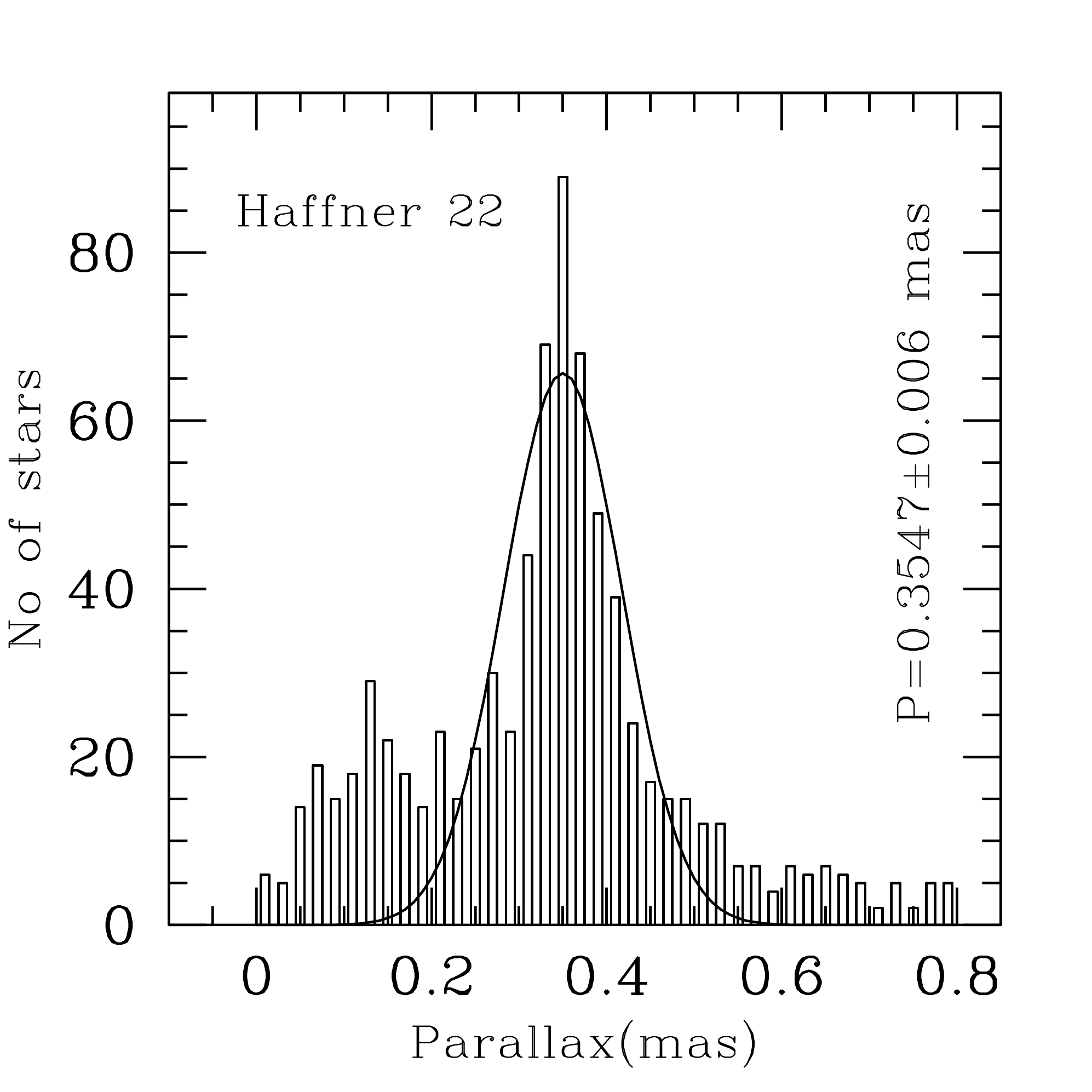}
\includegraphics[width=4.0cm, height=4.0cm]{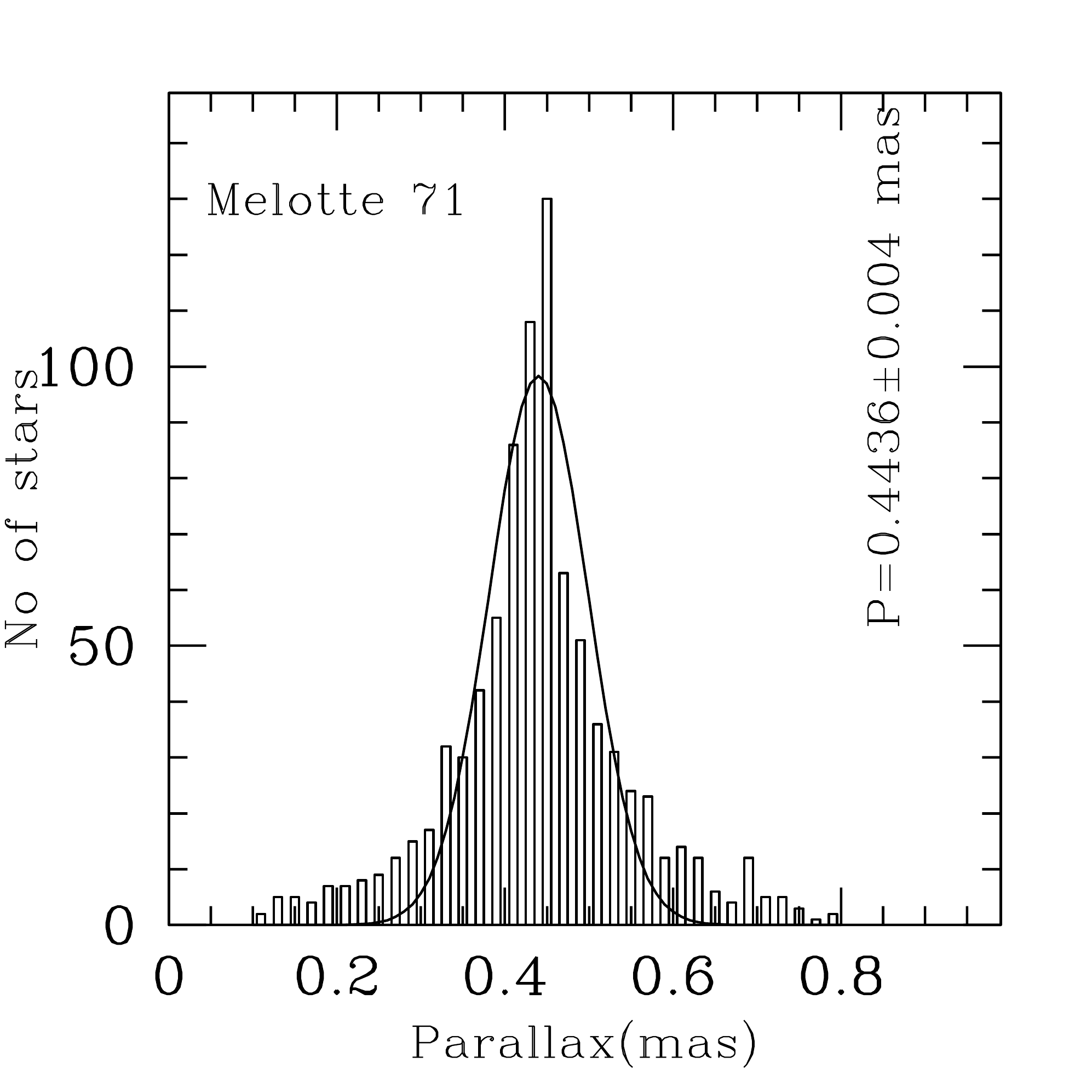}
}
\caption{Histogram of parallax for the clusters Haffner 22 and Melotte 71. The Gaussian function fitting to the central
bins provides mean value of parallax.}
\label{parallax}
\end{center}
\end{figure}

We have used the parallax of stars to obtain the distance of clusters Haffner 22 and Melotte 71. The Gaia~EDR3 parallax
has been corrected for these clusters after using zero-point offset ($-$0.017 mas) as given by Lindegren et al. (2020).
The histograms of parallax using probable members in both clusters with 0.15 mas bins are shown in Fig. \ref{parallax}.
The mean parallax is estimated as $0.3547\pm0.006$ mas and $0.4436\pm0.004$ mas for the clusters Haffner 22 and Melotte 71
and the corresponding distance values (reciprocal of cluster parallax) are $2.82\pm0.05$ kpc and $2.25\pm0.07$ kpc.
As listed in Table 4., our obtained
value of mean parallax for both objects is very close to the value given by Liu \& Pang (2019) and Cantat-Gaudin et al. (2018).
We have also used the method discussed by Bailer-Jones et al. (2018) for distance estimation from cluster parallax.
Finally, our obtained values are $2.88\pm0.10$ kpc and  $2.28\pm0.15$ kpc for the clusters Haffner 22 and Melotte 71, respectively.
These values of cluster distance are in fair agreement with the values obtained from the isochrone fitting method as
described above. The distances calculated using the trigonometric parallaxes are more
accurate as compared to the other techniques because this method is not dependent on the
intrinsic properties of the object. As discussed by Bailer-Jones (2015) the parallax
data from Gaia have corresponding error values. Which can affect the result if we calculate
distances by direct inverting the parallax values. So out of the three distances calculated by
us in this article, we prefer the distance obtained by using the method described by Bailer-Jones et al. (2018).

\subsection{Membership Probability}
\label{MP}

Open clusters are located within the densely populated Galactic plane and are usually contaminated by many foreground/background
stars. It is required to discriminate between cluster members and non-members to acquire more reliable cluster fundamental parameters.
The astrometric membership determination from the Gaia catalog has become more precise than using ground-based data (Dias et al., 2018).
We used the membership determination method from Balaguer-N\'{u}\~{n}ez et al. (1998) for the clusters Haffner 22 and Melotte 71.
Many authors have previously used this method (Bellini et al. 2009, Bisht et al. 2020a, 2020b, 2021a,2021b; Sariya et al. 2021a,2021b).

For the
cluster and field star distributions, two distribution functions ($\phi_c^{\nu}$) and ($\phi_f^{\nu}$) are constructed
for a particular i$^{th}$ star. The values of frequency distribution functions are given as follows:

\begin{center}
   $\phi_c^{\nu} =\frac{1}{2\pi\sqrt{{(\sigma_c^2 + \epsilon_{xi}^2 )} {(\sigma_c^2 + \epsilon_{yi}^2 )}}}$

$\times$ exp$\{{ -\frac{1}{2}[\frac{(\mu_{xi} - \mu_{xc})^2}{\sigma_c^2 + \epsilon_{xi}^2 } + \frac{(\mu_{yi} - \mu_{yc})^2}{\sigma_c^2 + \epsilon_{yi}^2}] }\}$ \\
\end{center}
\begin{center}
and\\
\end{center}
\begin{center}
$\phi_f^{\nu} =\frac{1}{2\pi\sqrt{(1-\gamma^2)}\sqrt{{ (\sigma_{xf}^2 + \epsilon_{xi}^2 )} {(\sigma_{yf}^2 + \epsilon_{yi}^2 )}}}$

$\times$ exp$\{{ -\frac{1}{2(1-\gamma^2)}[\frac{(\mu_{xi} - \mu_{xf})^2}{\sigma_{xf}^2 + \epsilon_{xi}^2}}
-\frac{2\gamma(\mu_{xi} - \mu_{xf})(\mu_{yi} - \mu_{yf})} {\sqrt{(\sigma_{xf}^2 + \epsilon_{xi}^2 ) (\sigma_{yf}^2 + \epsilon_{yi}^2 )}} + \frac{(\mu_{yi} - \mu_{yf})^2}{\sigma_{yf}^2 + \epsilon_{yi}^2}]\}$\\
\end{center}

where ($\mu_{xi}$, $\mu_{yi}$) are the PMs of $i^{th}$ star. The PM errors are represented by ($\epsilon_{xi}$, $\epsilon_{yi}$).
The cluster's PM center is given by ($\mu_{xc}$, $\mu_{yc}$) and ($\mu_{xf}$, $\mu_{yf}$) represent the center of field PM values.
The intrinsic PM dispersion for the cluster stars is denoted by $\sigma_c$, whereas $\sigma_{xf}$ and $\sigma_{yf}$ provide the
intrinsic PM dispersion's for the field populations. The correlation coefficient $\gamma$ is calculated as:\\

\begin{center}
$\gamma = \frac{(\mu_{xi} - \mu_{xf})(\mu_{yi} - \mu_{yf})}{\sigma_{xf}\sigma_{yf}}$.
\end{center}

Stars with PM errors $\le$0.5 mas~yr$^{-1}$ have been used to determine $\phi_c^{\nu}$ and $\phi_f^{\nu}$. A group of stars
is found at $\mu_{xc}$=$-$1.631 mas~yr$^{-1}$, $\mu_{yc}$=2.889 mas~yr$^{-1}$ for Haffner 22 and $\mu_{xc}$=$-$2.398
mas~yr$^{-1}$, $\mu_{yc}$=4.210 mas~yr$^{-1}$ for Melotte 71. Assuming a distance of 2.90 and 2.30 kpc for the clusters under
study and radial velocity dispersion of 1 km s$^{-1}$ for the open star clusters (Girard et al. 1989), the expected dispersion
($\sigma_c$) in PMs would be $\sim$0.08 and 0.10 mas~yr$^{-1}$ for the clusters Haffner 22 and Melotte 71. For field region
stars, we have estimated ($\mu_{xf}$, $\mu_{yf}$) = ($-$2.2, 1.3) mas yr$^{-1}$ for Haffner 22 and ($\mu_{xf}$,
$\mu_{yf}$) = ($-$1.5, 0.8) mas yr$^{-1}$ for Melotte 71 and ($\sigma_{xf}$, $\sigma_{yf}$) = (1.2, 1.9), (2.3, 1.9)
mas yr$^{-1}$ for both the objects.

\begin{figure*}
\begin{center}
\hbox{
\includegraphics[width=7.5cm, height=7.5cm]{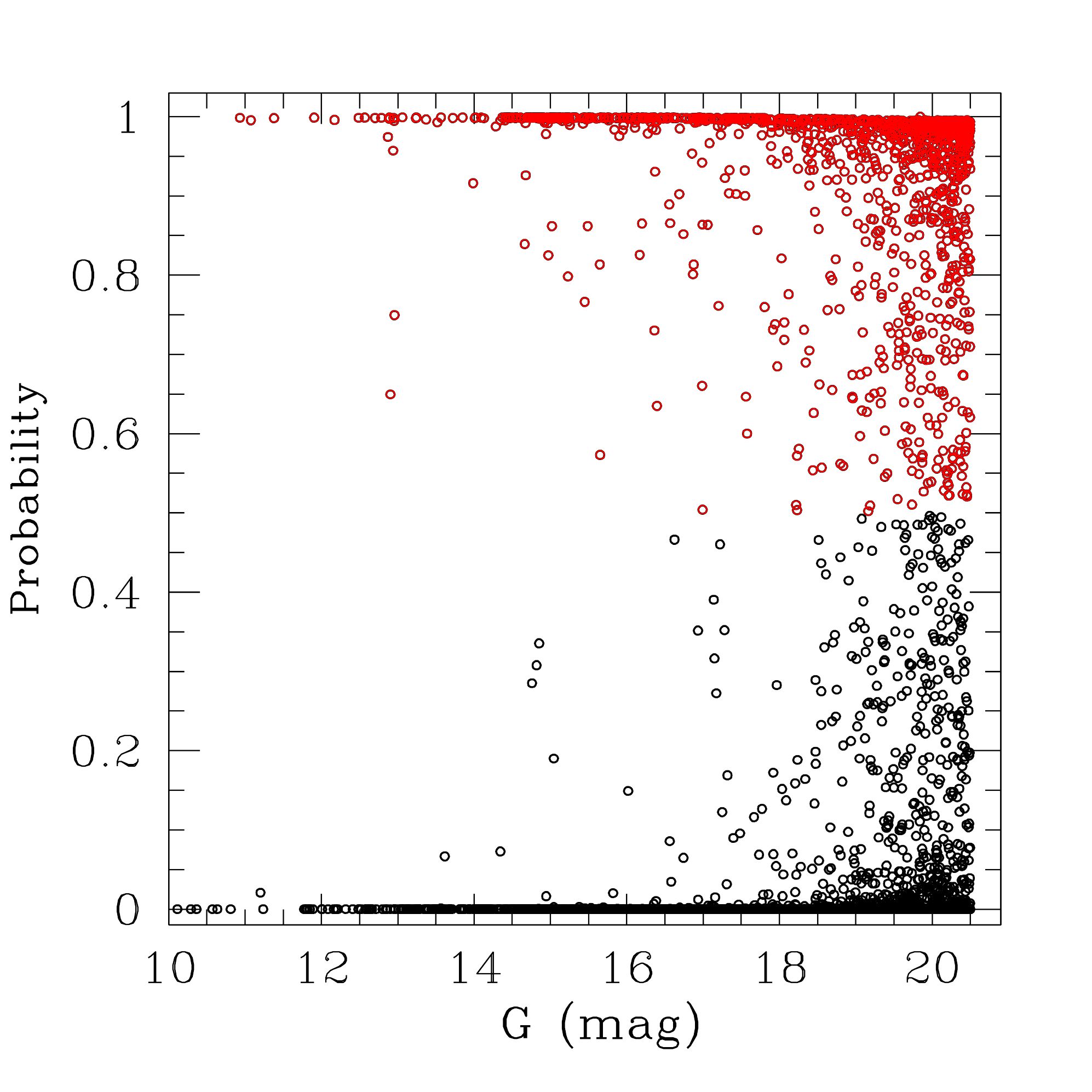}
\includegraphics[width=7.5cm, height=7.5cm]{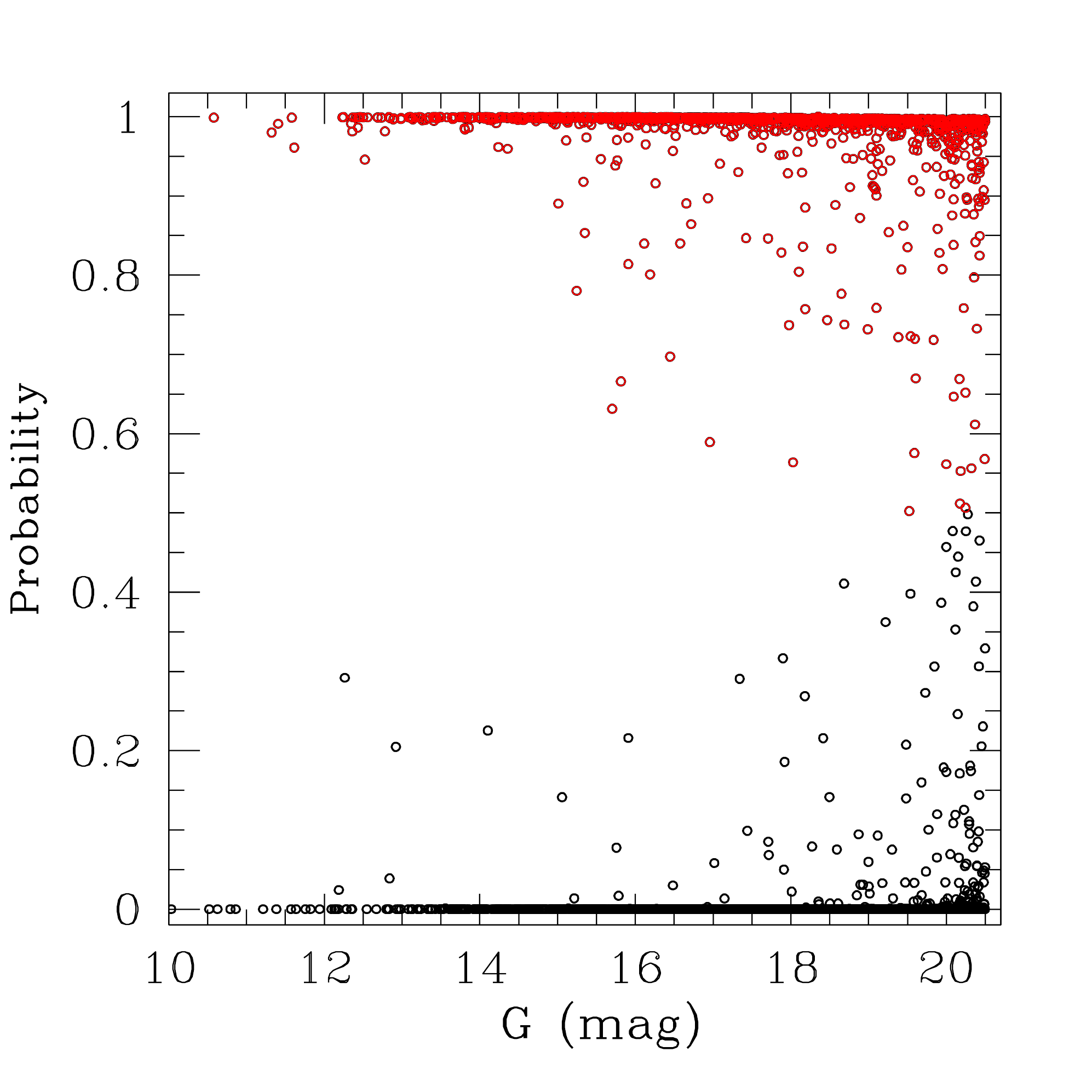}
}
\caption{Membership probability as a function of $G$ magnitude. The red circles show the cluster members
with membership probability higher than 50$\%$ in both the panels.}
\label{members}
\end{center}
\end{figure*}

\begin{figure*}
\begin{center}
\hbox{
\includegraphics[width=7.5cm, height=7.5cm]{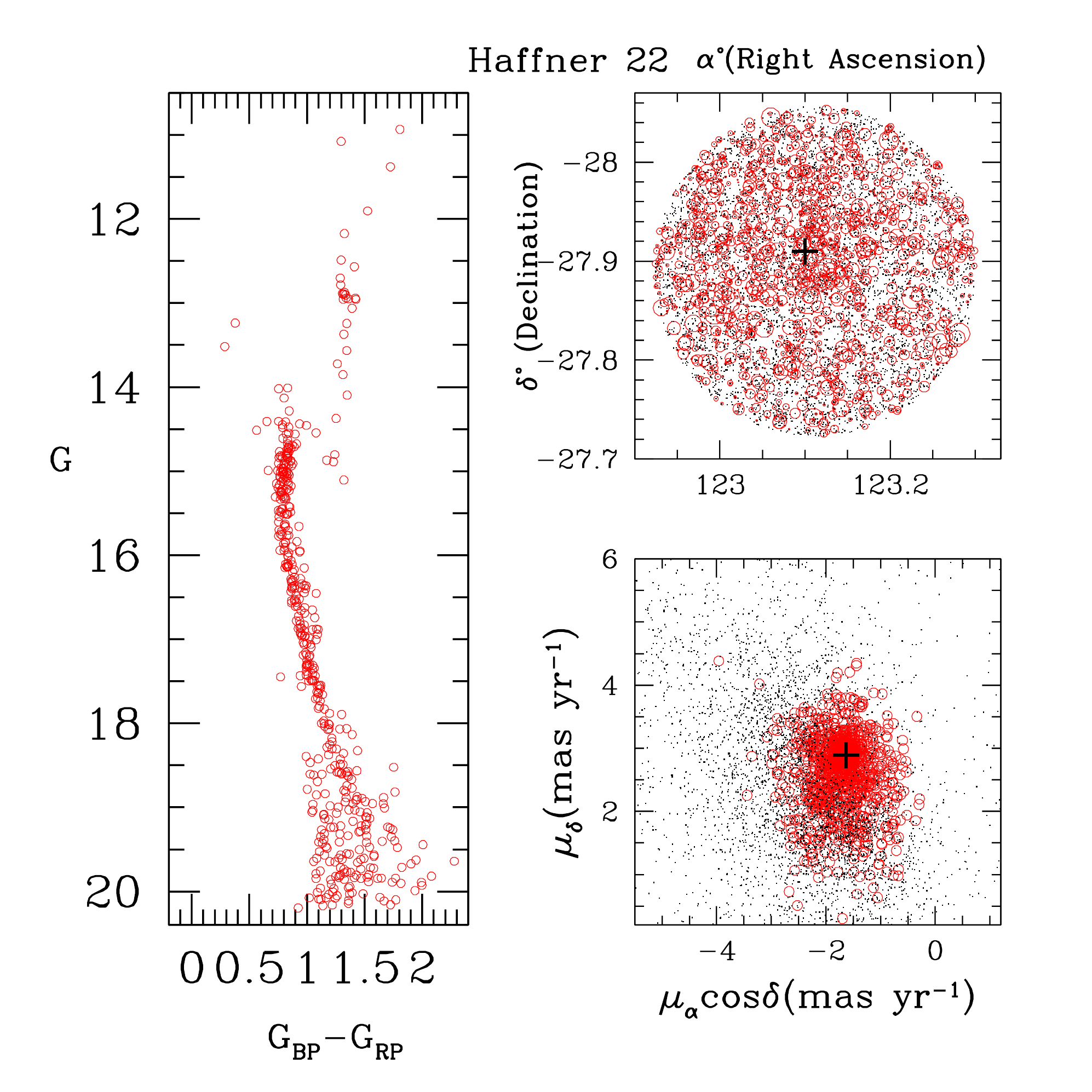}
\includegraphics[width=7.5cm, height=7.5cm]{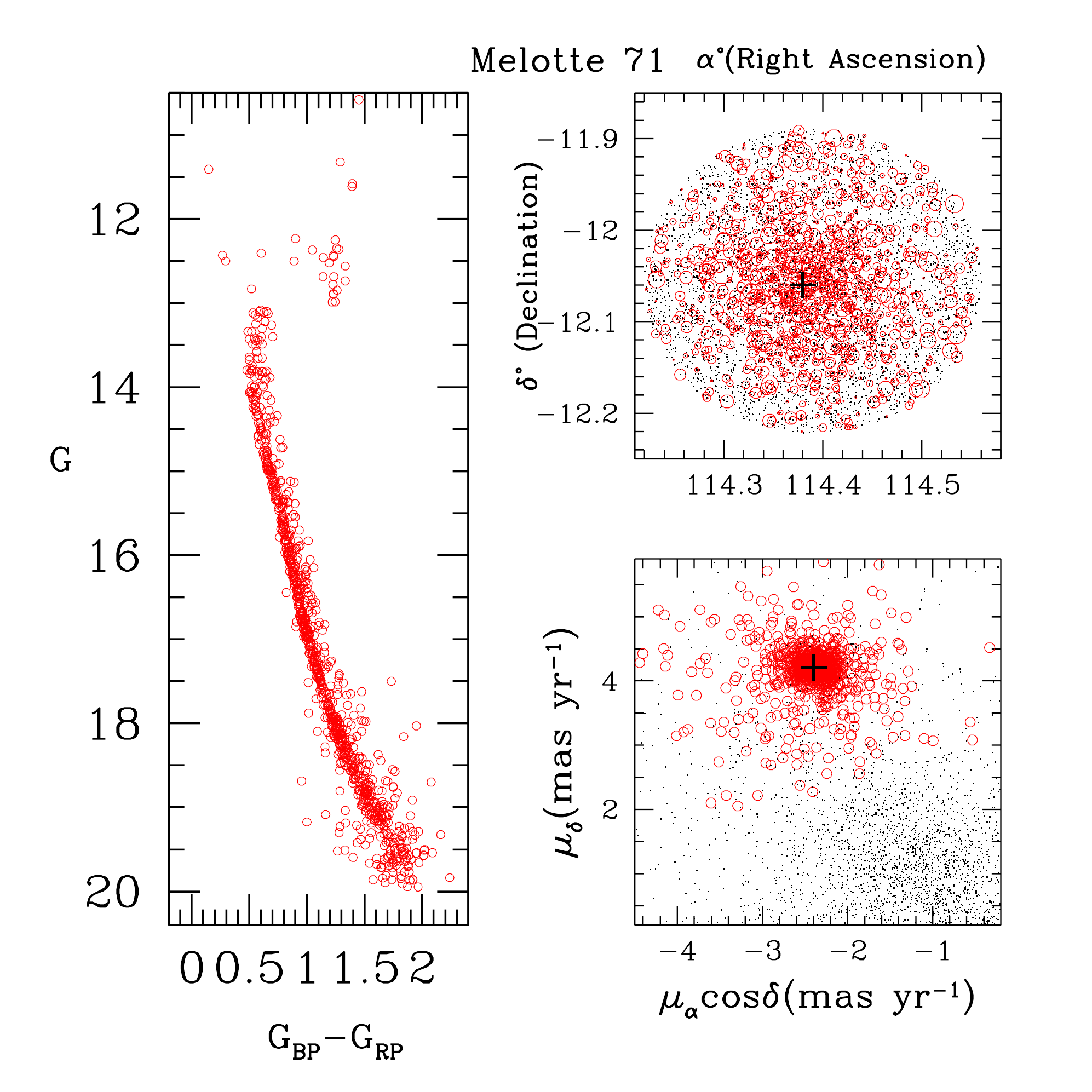}
}
\vspace{-0.1cm}\caption{($G, G_{BP}-G_{RP}$) CMD, identification chart and proper motion distribution of member stars
with membership probability higher than $50\%$. Black dots in proper motion distribution are field stars with membership probability
less than $50\%$. The plus sign indicates the cluster center in position and proper motions.}
\label{members1}
\end{center}
\end{figure*}

We identified 382 and 597 stars as cluster members for Haffner 22 and Melotte 71 with membership probability higher than
$50\%$ and $G\le20$ mag. In Fig. \ref{members}, we plotted membership probability versus $G$ magnitude for both the clusters.
In Fig. \ref{members1}, we have plotted $G$ versus ($G_{BP}-G_{RP}$) CMD, the identification chart and proper motion distribution
using stars with membership probability higher than $50\%$. In proper motion distribution of clusters, we have plotted field region
stars also as shown by black dots.

\begin{figure*}
\begin{center}
\hbox{
\includegraphics[width=7.5cm, height=7.5cm]{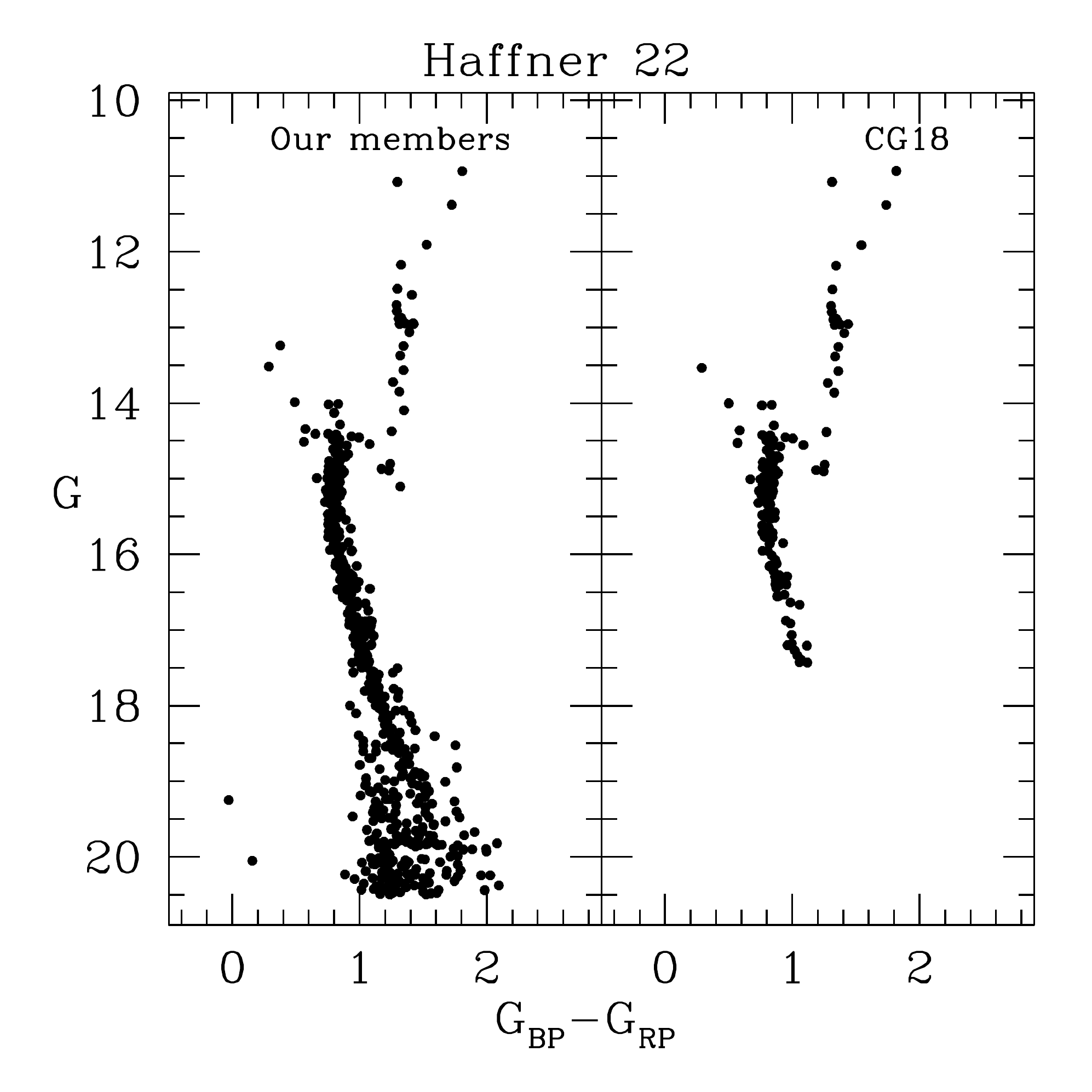}
\includegraphics[width=7.5cm, height=7.5cm]{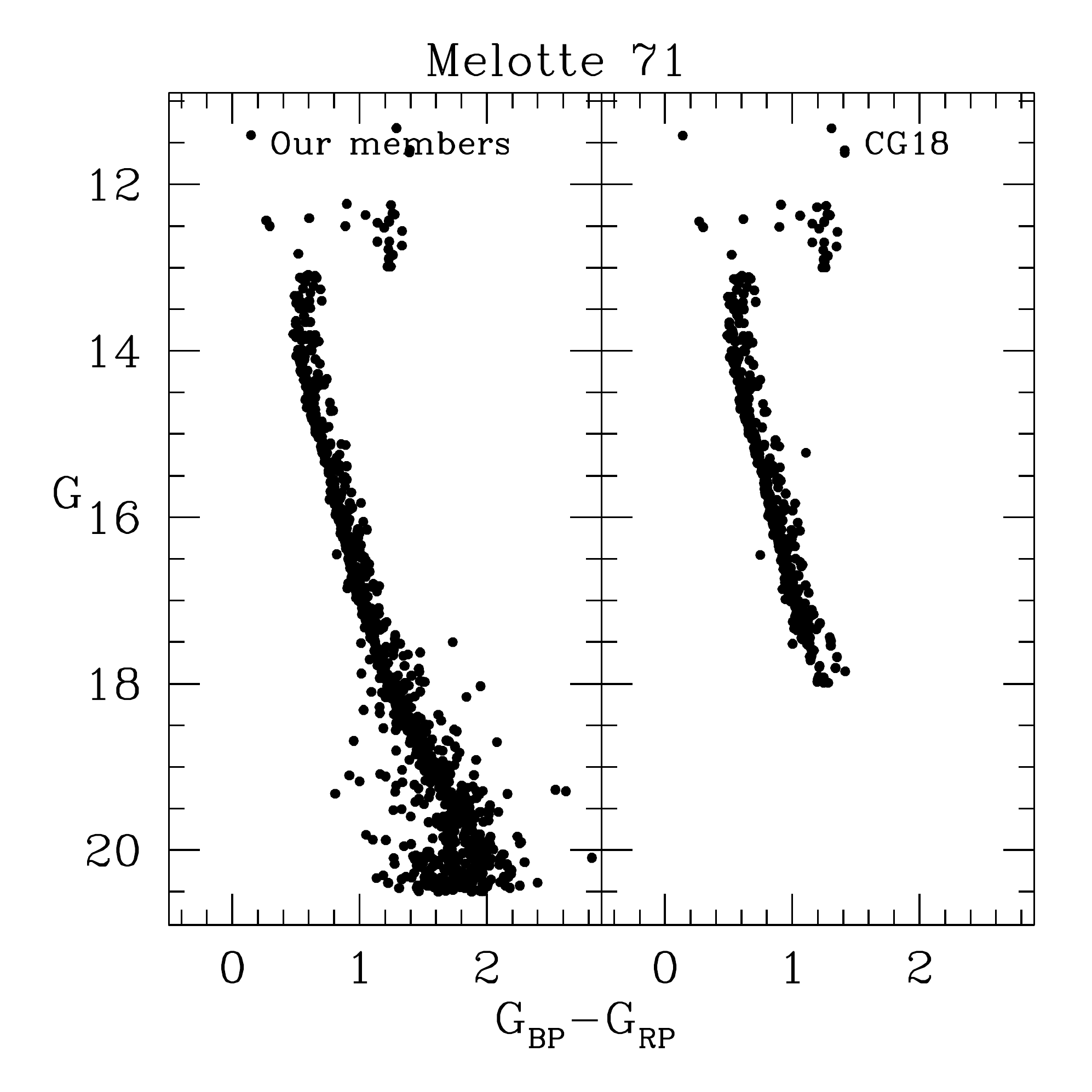}
}
\caption{The (G, $G_{BP}-G_{RP}$) CMDs of our identified members and members from
         Cantat-Gaudin et al. 2018 (CG18) catalog. All the stars plotted here have a membership
         probability higher than 50$\%$.}
\label{cmd_comp}
\end{center}
\end{figure*}

Membership probability has been determined for the clusters Haffner 22 and Melotte 71 by Cantat-Gaudin et al. (2018) up to 18.0 mag
using the Gaia-DR2 catalog. To make a comparison on membership probability, we have plotted $G, G_{BP}-G_{RP}$ CMDs using our
membership catalog and Cantat-Gaudin et al. (2018) catalog as shown in Fig \ref{cmd_comp}. We used only probable members with
membership probability higher than 50$\%$. Our membership determination is clearly atleast 2 magnitude deeper than Cantat-Gaudin et al. (2018).
This enhances our precision in determining various parameters of star clusters, varying from distance, extinction, and mass function.

\subsection{Determination of the effectiveness of probabilities}

We can not avoid the contamination of background/foreground stars
through the observational projection effect. We can review quantitatively
how adequate the results of our membership estimation were.

To ascertain the effectiveness of our membership determination, we used the expression given by Shao \& Zhao (1996):\\

$E=1-\frac{N\times\Sigma[P_{i}(1-P_{i})]}{\Sigma P_{i}\Sigma(1-P_{i})}$\\

where $N$ is the total number of stars with membership probability higher than 50$\%$ and $P_{i}$ indicates the probability
of $i^{th}$ star of the cluster. We obtained the effectiveness $(E)$ values as 0.48 and 0.51 for the clusters Haffner 22 and
Melotte 71. Shao \& Zhao (1996) calculated the effectiveness of membership determination for 43 OCs as ranges from
0.20 to 0.90 with a peak value of 0.55. Our estimated values of the effectiveness of membership determination are lying
within the above range and approaching a slightly higher side as well.

\begin{figure}
\begin{center}
\hbox{
\includegraphics[width=4.0cm, height=4.0cm]{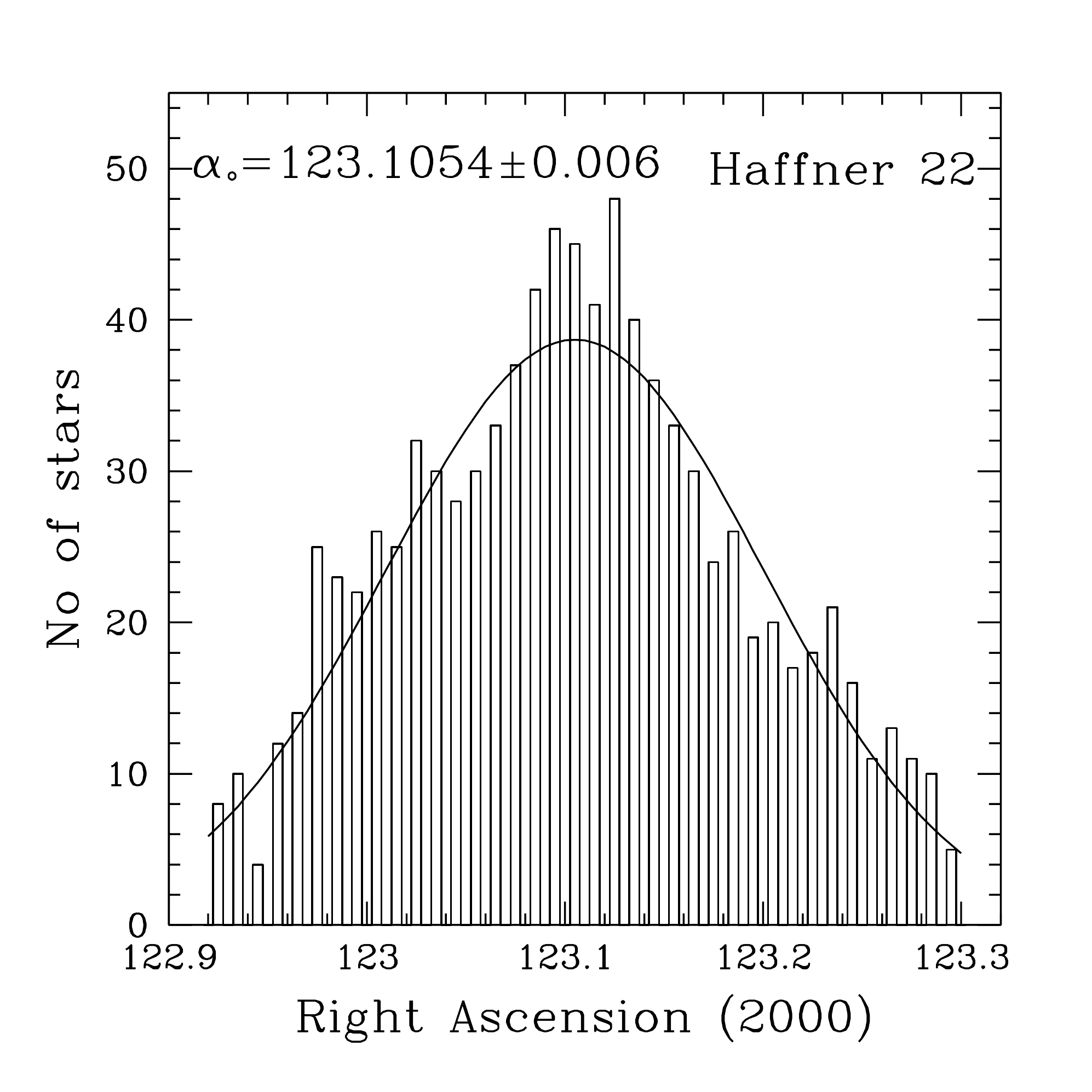}
\includegraphics[width=4.0cm, height=4.0cm]{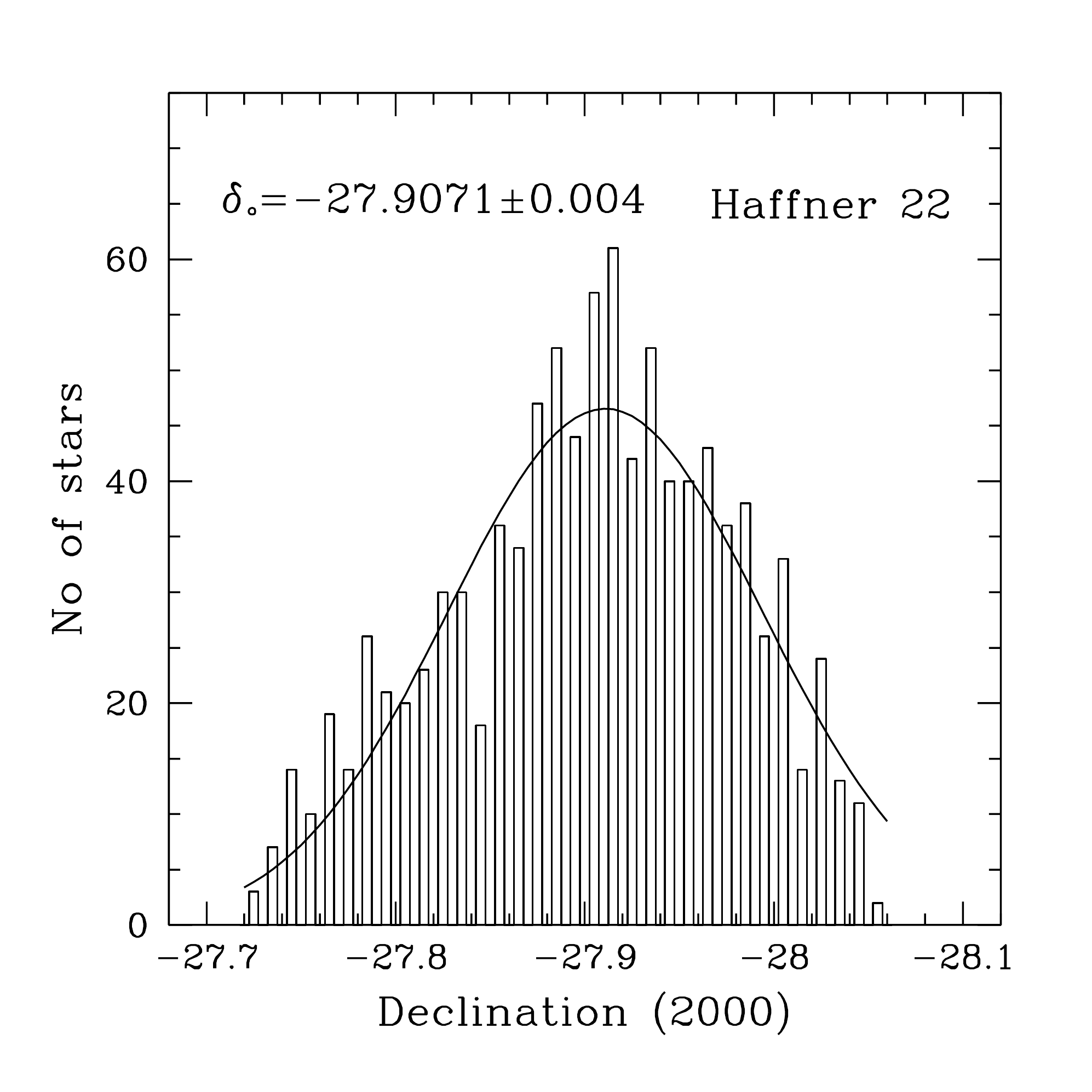}
}
\hbox{
\includegraphics[width=4.0cm, height=4.0cm]{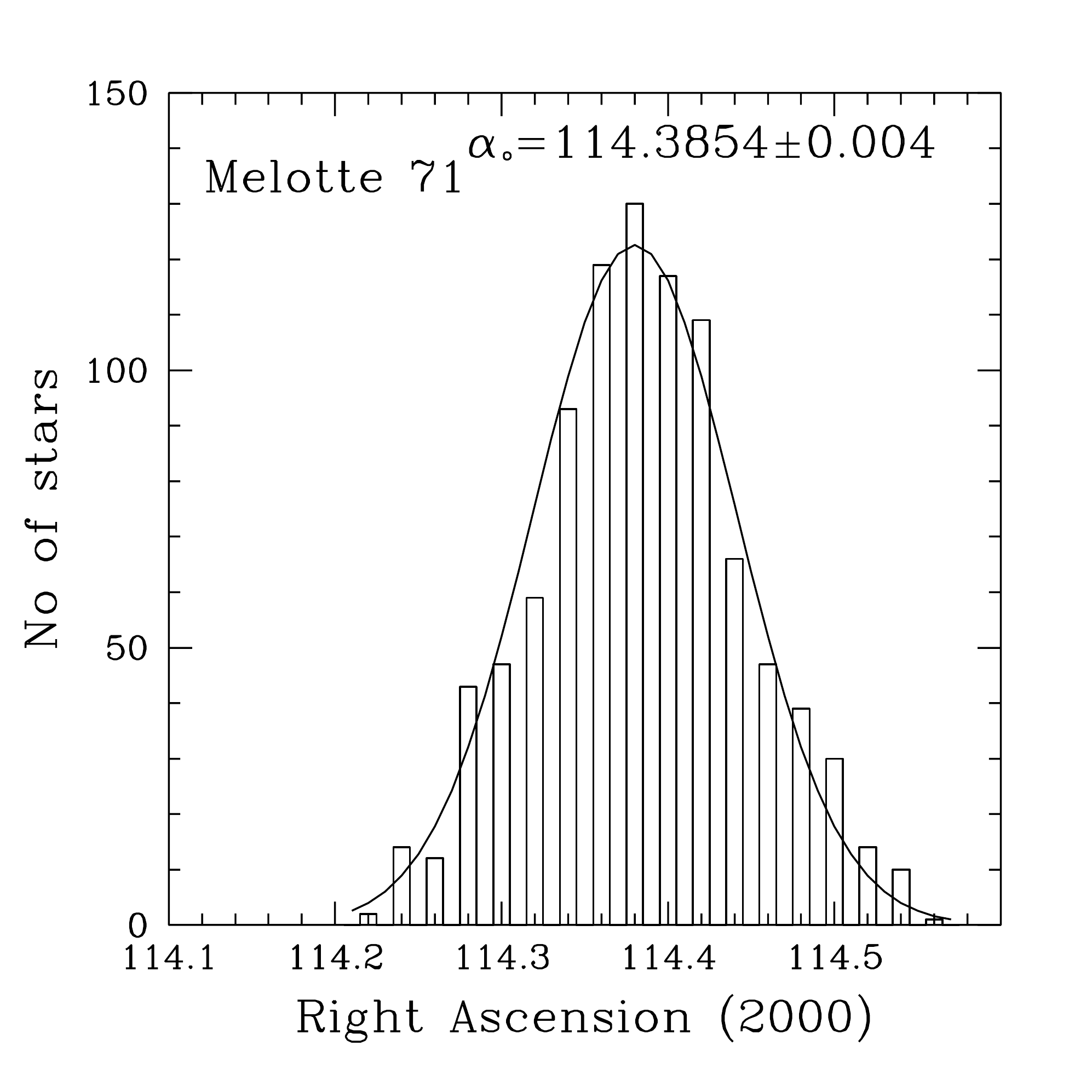}
\includegraphics[width=4.0cm, height=4.0cm]{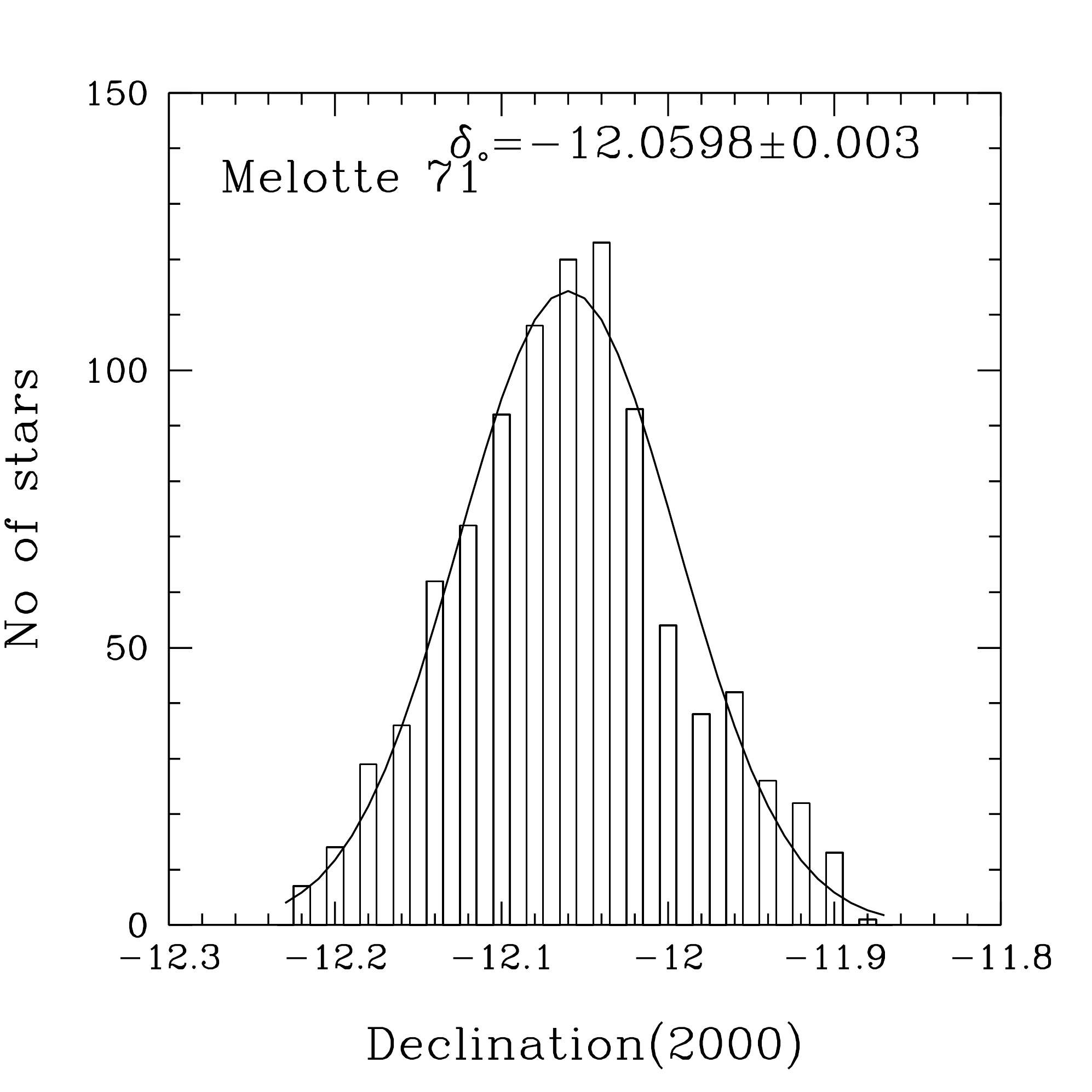}
}
\vspace{-0.1cm}\caption{Profiles of stellar counts across the region of clusters Haffner 22 and Melotte 71. The Gaussian
fits have been applied. The center of symmetry about the peaks of Right Ascension and Declination is taken to be the
position of cluster's center.}
\label{center}
\end{center}
\end{figure}

\subsection{Blue straggler stars}

The BSS are intriguing objects present in the stellar environments
like the clusters (Johnson \& Sandage 1955; Sandge 1962;
Ahumada \& Lapasset 1995.) In Brief, it  has  been recommended that BSS are the result of stellar collisions (Benz \& Hills 1987; Lombardi et al. 1996)
or because of the mass exchange in close binary systems
(McCrea 1964; Eggen \& Iben 1989; Mateo et al. 1990).
The second-generation stars (Eggen \& Iben 1988),
accretion of gas from the surrounding interstellar medium (Williams 1964), and capture of non-member stars by a star cluster are
other suggested aspects behind the formation of BSS in stellar systems.
BSS are stars lying above the main sequence (MS) turnoff region
in color-magnitude diagrams (CMDs). If the BSS had been normal single
stars, they should already have evolved away from the MS (Stryker
1993). In the 1990s, early studies gave the first hints for BSS showing this effect in GCs and some OCs
(Auriere et al. 1990; Mathieu \& Latham 1986). This paper found five and four-member BSS in clusters Haffner 22 and
Melotte 71, respectively. Those BSS members are located at a radial distance of $\sim$ 0.6-6.5 and $\sim$ 0.6-9.8
arcmin for Haffner 22 and Melotte 71, respectively. We imply that the BSSs are lying in the outer region
for these clusters, which could form
because of binarity.
Our analysis suggests that all the identified BSS are
confirmed cluster members with a membership probability
higher than 90$\%$. We have plotted all the member BSS in
Fig.~\ref{dist_age} for both clusters by blue solid dots. All the identified BSS for Haffner 22 and Melotte 71
are listed in Table \ref{bss_ha22} and Table \ref{bss_mel71} respectively. Five BSS have previously been identified by
Rain et al. (2021) in the cluster Melotte 71. We have cross-matched our BSS with Rain et al. (2021) and
found that three members (Gaia source IDs-
3033959198481332736, 3033962183479250432, and 3033962664515580672) are matched. The other two members
do not lie in
our field of view for Melotte 71. The BSS of Haffner 22 are not cataloged in the literature.

\begin{table*}
\caption{The identified BSS candidates in the cluster Haffner 22.
}
\vspace{0.5cm}
\begin{center}
\tiny
\begin{tabular}{cccccccccc}
\hline\hline
GaiaEDR3 & RA &DEC& pmRA&pmDEC&plx&Gmag&Bp-Rp & r & prob\\
& deg & deg  & mas/yr & mas/yr & mas & mag & mag & arcmin
\\
5693065398100156800 & $  123.14114760440$&$ -27.95287418172$&$ -1.562$&$ 2.814$&$ 0.3872$&$ 13.237711$&$ 0.3767$&$ 4.0767$&$ 0.99$ \\
5692972077053132288 & $  123.03294244320$&$ -27.97111049151$&$ -1.584$&$ 3.042$&$ 0.3496$&$ 13.518732$&$ 0.2865$&$ 6.4224$&$ 0.99$ \\
5693817189175756032 & $  123.08835909476$&$ -27.89216823376$&$ -1.872$&$ 2.861$&$ 0.2917$&$ 13.988064$&$ 0.4903$&$ 1.2604$&$ 0.91$ \\
5693817601492594176 & $  123.08188716773$&$ -27.85595962460$&$ -1.703$&$ 3.010$&$ 0.2733$&$ 14.345200$&$ 0.5744$&$ 2.5928$&$ 0.99$ \\
5693817257895191424 & $  123.11862294354$&$ -27.88077803456$&$ -1.569$&$ 2.892$&$ 0.3236$&$ 14.513340$&$ 0.5620$&$ 0.6554$&$ 0.99$ \\
\hline
\end{tabular}
\label{bss_ha22}
\end{center}
\end{table*}

\begin{table*}
\caption{The identified BSS candidates in the cluster Melotte 71.
}
\vspace{0.5cm}
\begin{center}
\tiny
\begin{tabular}{cccccccccc}
\hline\hline
GaiaEDR3 & RA &DEC& pmRA&pmDEC&plx&Gmag&Bp-Rp & r & prob\\
& deg & deg  & mas/yr & mas/yr & mas & mag & mag & arcmin
\\
3033959198481332736 & $ 114.40683348610$&$ -12.06451836239$&$ -2.242$&$ 4.290$&$ 0.4837$&$ 11.408119$&$ 0.1468$&$ 1.1409$&$ 0.99$ \\
3033956307964730880 & $ 114.23005068396$&$ -12.09950761098$&$ -2.600$&$ 2.077$&$ 0.3975$&$ 12.348255$&$ 0.2989$&$ 9.7571$&$ 0.98$ \\
3033962183479250432 & $ 114.39487869798$&$ -12.04600820646$&$ -2.209$&$ 4.220$&$ 0.4588$&$ 12.432313$&$ 0.2667$&$ 0.6109$&$ 0.99$ \\
3033962664515580672 & $ 114.39443602193$&$ -11.99131028286$&$ -2.392$&$ 4.281$&$ 0.4300$&$ 12.500945$&$ 0.2947$&$ 3.8302$&$ 0.90$ \\
\hline
\end{tabular}
\label{bss_mel71}
\end{center}
\end{table*}

\section{Fundamental parameters of the clusters}

\subsection{Cluster center and radial density profile}

In the earlier investigations of the open clusters, the center used to be determined just by the visual inspection (Becker \& Fenkart 1971;
Romanishim \& Angel 1980). To present the precise measurement of the fundamental parameters in clusters Haffner 22 and Melotte 71, we used
the star-count method using stars with membership probability higher than 50$\%$. Fig \ref{center} represents the histograms for
both clusters in RA and DEC directions. The Gaussian curve-fitting is performed to the star count profiles, and the estimated
center coordinates are listed in Table \ref{para}. Our evaluation is in good agreement with the values given by Dias et al. (2002).
and Cantat-Gaudin et al. (2018).

\begin{figure*}
\begin{center}
\hbox{
\includegraphics[width=7.0cm, height=7.0cm]{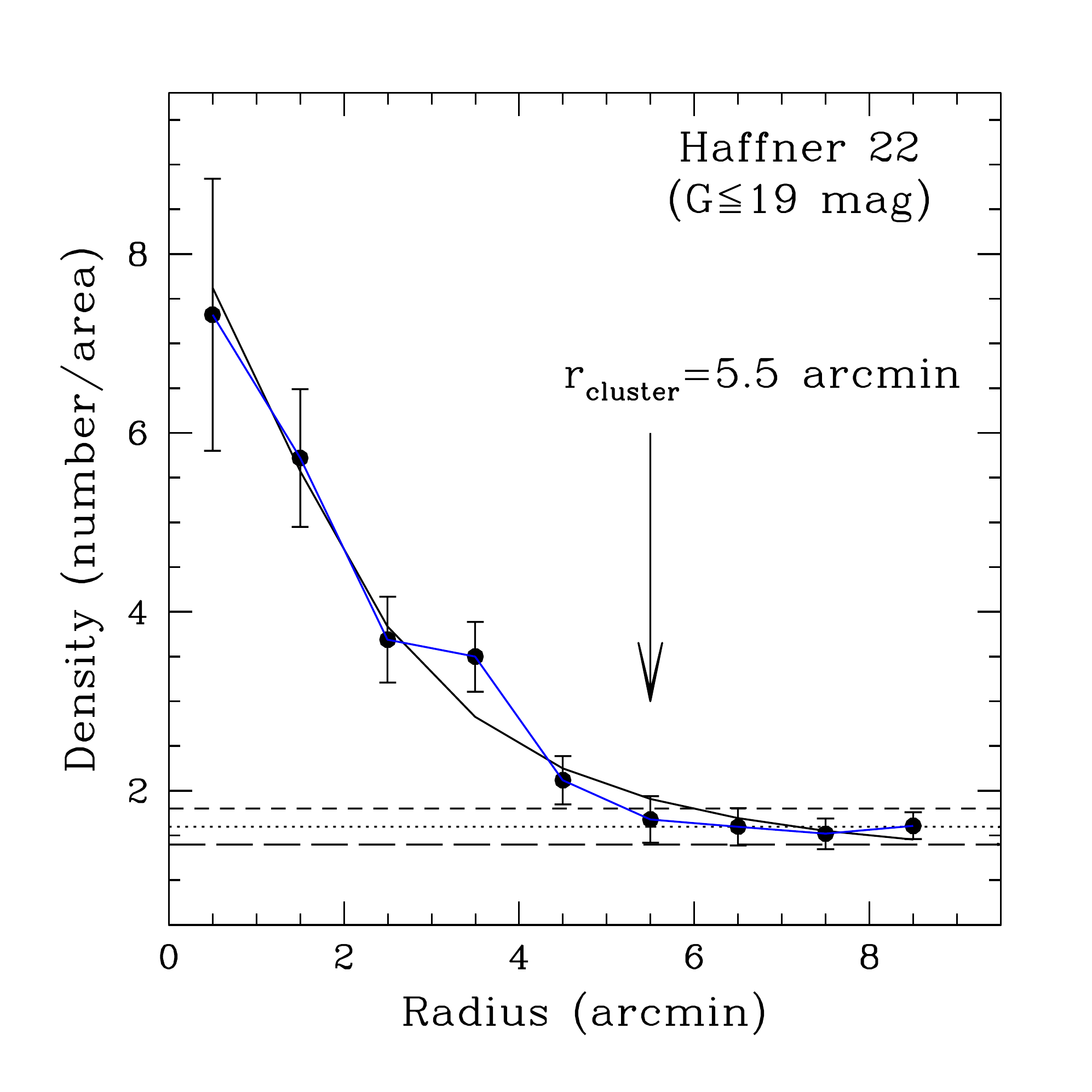}
\includegraphics[width=7.0cm, height=7.0cm]{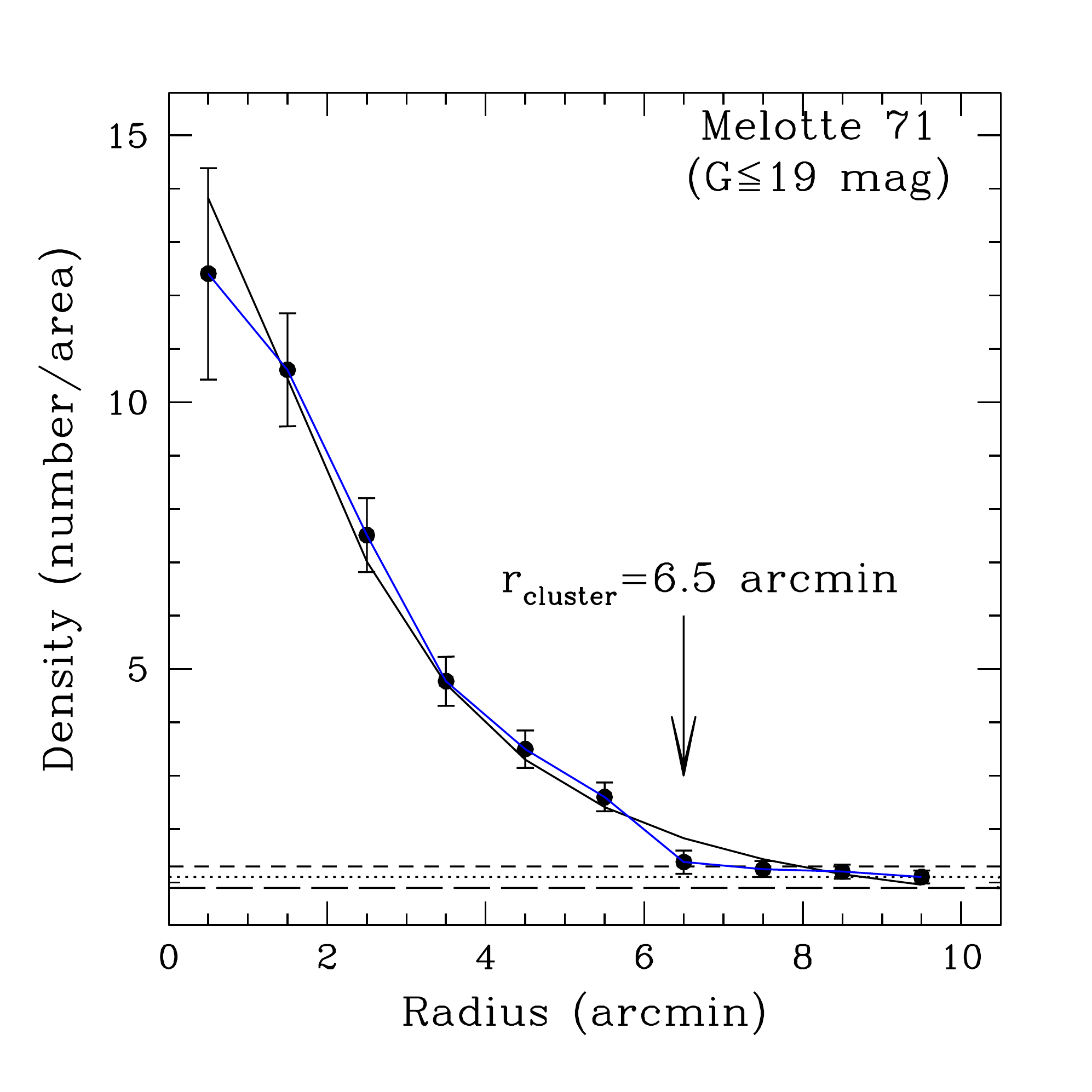}
}
\vspace{-0.1cm}\caption{Surface density distribution of the clusters Haffner 22 and Melotte 71. Errors
are determined from sampling statistics (=$\frac{1}{\sqrt{N}}$ where $N$ is the number of cluster members used in
the density estimation at that point). The smooth line represent the fitted profile of King (1962)  whereas dotted
line shows the background density level. Long and short dash lines represent the errors in background density.}
\label{dens}
\end{center}
\end{figure*}

We have plotted the radial density profile (RDP) for Haffner 22 and Melotte 71 using the above-estimated center coordinates.
The cluster area is divided into many concentric rings around the core having an equal additional radius. The number density,
$\rho_{i}$, in the $i^{th}$ zone is determined by using the formula, $\rho_{i}$ = $\frac{N_{i}}{A_{i}}$, where $N_{i}$ is the
number of cluster members, and $A_{i}$ is the area of the $i^{th}$ zone. We obtained the radii of the clusters based on the visual
inspection from RDPs. According to our criteria, the radius is the point after that cluster density merges with
the field density. The errors in background density levels are also shown by the dashed lines in Fig. \ref{dens}. We considered
5.5$^{\prime}$ and 6.5$^{\prime}$ as the cluster radius for the clusters Haffner 22 and Melotte 71. Our estimated radius value
for Haffner 22 is slightly less than the value of Carraro et al. (2016). We obtained the radius of Melotte 71 is higher than
the value cataloged by Dias et al. (2014). The enhancement of radius around 3.5 arcmin below 2$\sigma$ for cluster Haffner 22
indicates the presence of a possible corona region for this object.
The appearance of the corona region may be because of two main reasons: the mass-segregation effect in this object
(Nilakshi \& Sagar 2002), and the corona of the clusters is molded by the Galactic tidal fields (Mathieu 1985). To
estimate the spatial parameters, we fitted the King (1962) model as shown by the continuous black curve in Fig \ref{dens}.
The King (1962) profile is given by:

$f(r) = f_{b}+\frac{f_{0}}{1+(r/r_{c})^2}$\\

\begin{table*}
\caption{Structural parameters of the clusters under study. Background and central
density are in the unit of stars per arcmin$^{2}$. Core radius ($r_c$) and tidal radius ($R_t$) are
in arcmin and pc.
}
\vspace{0.5cm}
\begin{center}
\small
\begin{tabular}{ccccccccc}
\hline\hline
Name & $f_{0}$ &$f_{b}$& $r_{c}$&$r_{c}$&$\delta_{c}$&$r_{lim}$&$c$ & $R_{t}$\\
&&& arcmin & parsec  & & arcmin & & parsec
\\
Haffner 22  & $6.94$&$1.60$&$2.1$&$1.7$&$5.3$&$7.2$&$0.55$&$12.19$ \\
Melotte 71 & $14.34$&$1.10$&$2.4$&$1.6$&$14.1$&$11.6$&$0.68$&$15.13$ \\
\hline
\end{tabular}
\label{stru_para}
\end{center}
\end{table*}

where $r_{c}$ , $f_{0}$ , and $f_{b}$ are the core radius, central density, and the background density level, respectively.
By fitting the King model to RDPs, we have derived the structural parameters for both the clusters as listed in Table \ref{stru_para}. The density
contrast parameter ($\delta_{c} = 1 +\frac{f_{0}}{f_{b}}$) is calculated for both the clusters under study using the member
stars selected from proper motion data. We obtained the value of density contrast parameter ($\delta_{c}$) as 5.3 and 14.1
for Haffner 22 and Melotte 71. Our estimated value for Haffner 22 is lower than the limit ($7\le \delta_{c}\le 23$) as given
by Bonatto \& Bica (2009), which suggests that Haffner 22 is a sparse cluster. Our obtained value of $\delta_{c}$ for Melotte 71
is lying within the limit given by Bonatto \& Bica (2009), which suggests that this cluster is compact. We have also estimated the
limiting radius $(r_{lim})$ and concentration parameter (c) for both the clusters, which are listed in Table \ref{stru_para}.

It is very well known that every cluster consists of mainly two regions, core and corona (Nilakshi et al. 2008). We can describe these two regions using clusters RDP.
In Haffner 22, we can see the enhancement of radius around 3.5 arcmin. This may be because of the presence of a possible corona region in this object.

\subsection{Age and distance}

\begin{figure*}
\begin{center}
\hbox{
\includegraphics[width=7.5cm, height=7.5cm]{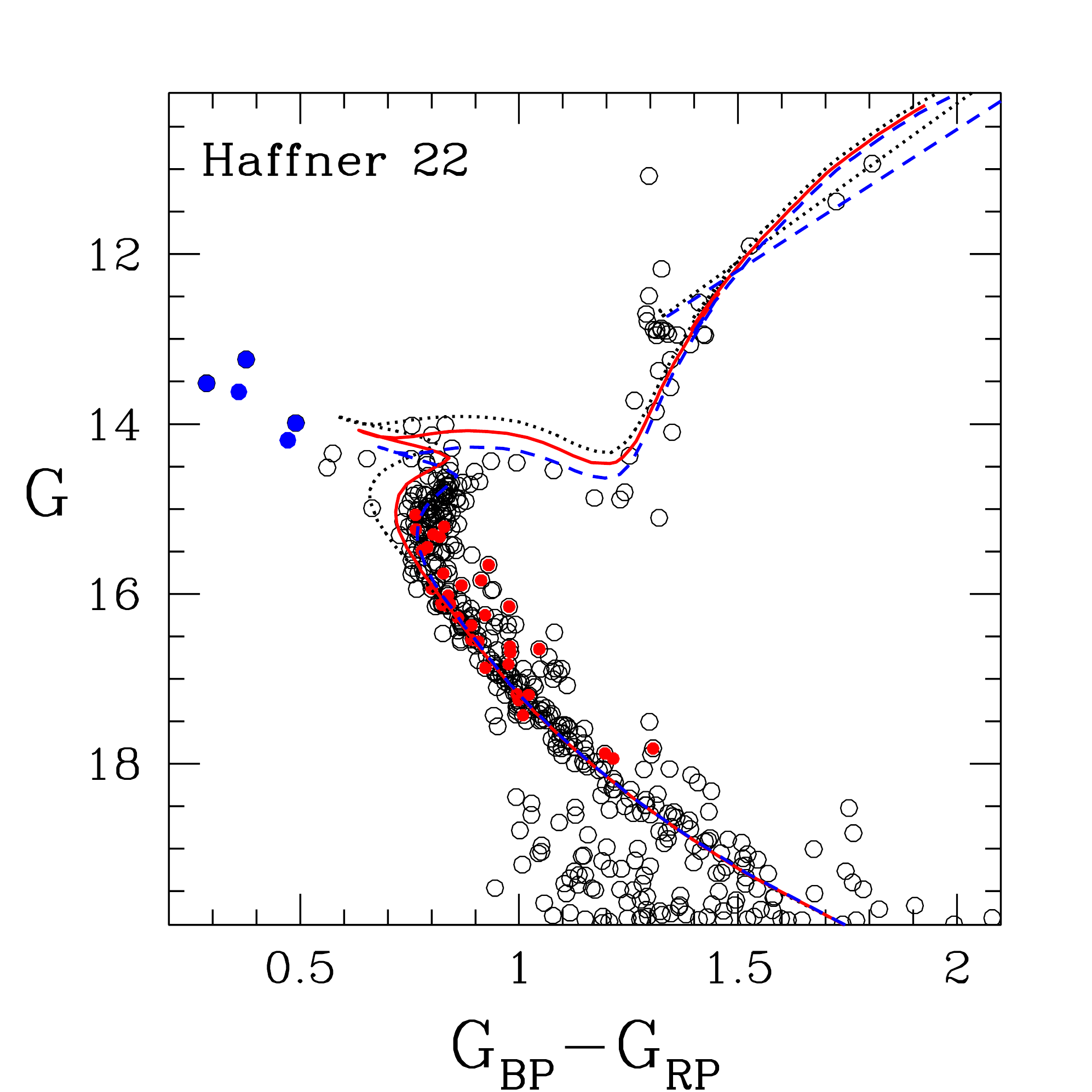}
\includegraphics[width=7.5cm, height=7.5cm]{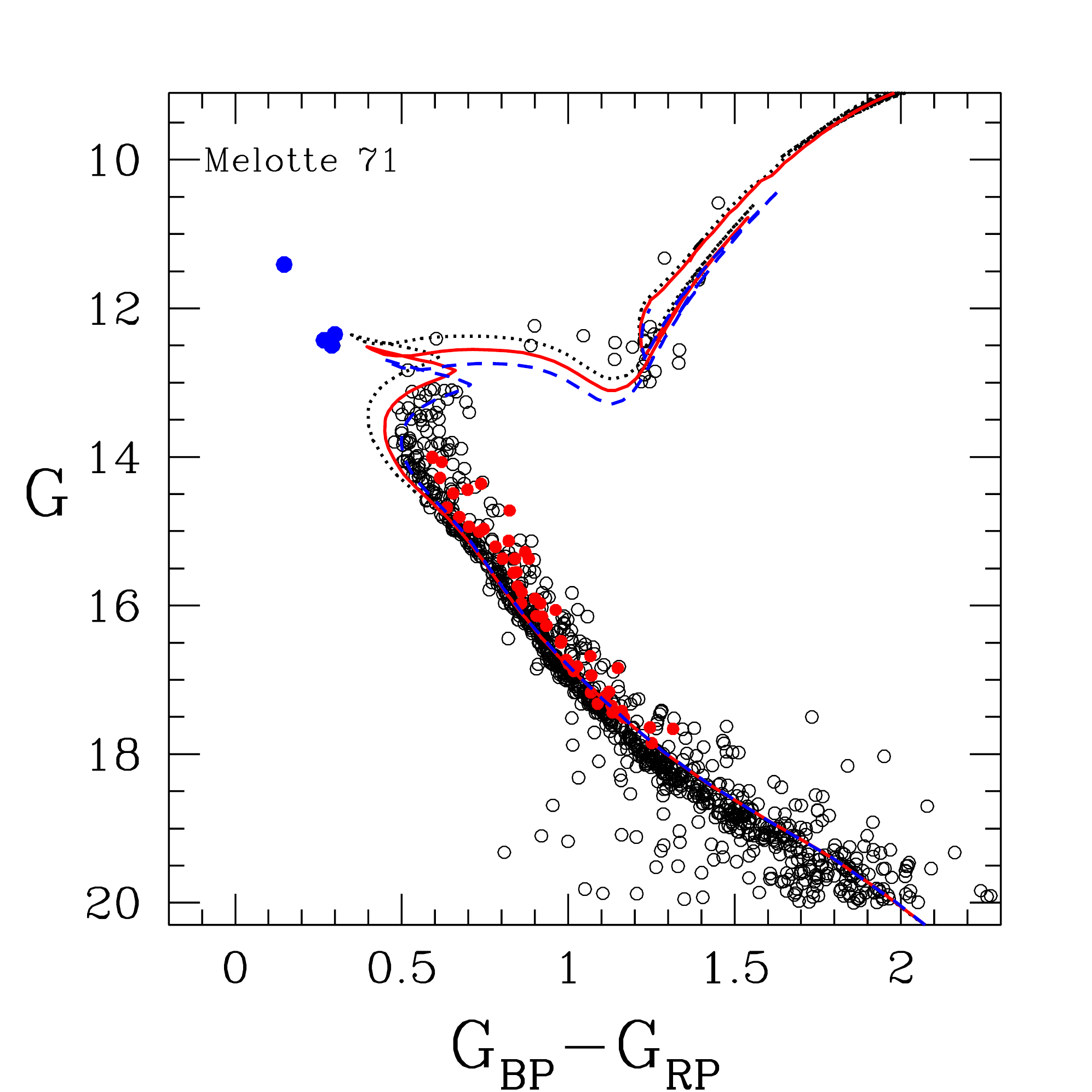}
}
\caption{ The color-magnitude diagram of the clusters under study. All stars are probable members with membership
probability higher than 50$\%$. The curves are the isochrones of (log(age) $=$  9.30 , 9.35 and  9.40))
for Haffner 22 and (log(age) $=$  9.05, 9.10 and 9.15) for Melotte 71. These ishochrones of
metallicity Z=0.005 for Haffner 22
and Z= 0.008 for Melotte 71 are taken from
Marigo et al. (2017). Blue solid dots are the possible member BSS while red dots are identified binary stars.}
\label{dist_age}
\end{center}
\end{figure*}

To trace the Galaxy's Galactic structure and chemical evolution using OCs, the distance and age of OCs play the most
crucial role (Friel \& Janes 1993). We have estimated the mean value of $A_{G}$ for the clusters Haffner 22 and Melotte 71 as
$0.63\pm0.18$ and $1.40\pm0.35$ using probable members from Gaia DR2 data. We obtained the main
fundamental parameters (age, distance, and reddening) by fitting the
isochrones with metallicity Z=0.005 for Haffner 22 and Z=0.008 for Melotte 71 of Marigo et al. (2017) to the $G, G_{BP}-G_{RP}$ CMD
as shown in Fig. \ref{dist_age}.
Our used metallicity for 
Melotte 71 is very close to the value Z=0.007 as given by Brown et al. (1996).
The above-used isochrones are derived from the stellar evolutionary tracks computed
with PARSEC (Bressan et al., 2012) and COLIBRI (Marigo et al., 2013) codes. 

The estimation of the main fundamental parameters for the clusters are given below:

{\bf Haffner 22:} We fitted the theoretical isochrones of different ages (log(age)=9.30, 9.35 and 9.40) in all the CMDs for
the cluster Haffner 22, shown in Fig. \ref{dist_age}. The best global fit is favorable for the middle isochrone with
log(age)= 9.35 to the high mass cluster members. A good fitting of isochrones provides an age of $2.25\pm0.25$ Gyr.
Our obtained value of age is close to the value cataloged by Kharchenko et al. (2013). The apparent distance modulus
($(m-M)=12.80\pm0.4$ mag) provides a distance of $2.8\pm0.50$ kpc from the Sun. Our calculated value of the distance
shows good agreement with the values obtained by Cantat-Gaudin et al. (2020) and Kharchenko et al. (2013).
\\

{\bf Melotte 71:}  The isochrones of different ages (log(age)=9.05, 9.10 and 9.15) have been overplotted on all the CMDs
for the cluster Melotte 71 as shown in Fig \ref{dist_age}. The overall fit is satisfactory for log(age)=9.10 (middle
isochrone) to the brighter stars, corresponding to $0.8\pm0.1$ Gyr. The estimated value of age is very close to the value
cataloged by Sampedro et al. (2017). The estimated distance modulus ($(m-M)=13.30\pm0.3$ mag) provides a distance from the
Sun that is $2.5\pm0.20$ kpc. Our obtained value of the distance is in fair agreement with the value given
by Kharchenko et al. (2016).

The galactocentric coordinates of the clusters
$X$ (directed towards the galactic center in the Galactic disc), $Y$ (directed towards the Galactic rotation), and distance
from the galactic plane $Z$ (directed towards the Galactic north pole) can be estimated using clusters' distances, longitude,
and latitude. The Galactocentric distance has been calculated by considering 8.3 kpc (Bajkova \& Bobylev (2016)) as the distance of Sun to the
Galactic center. The estimated Galactocentric coordinates are listed in Table \ref{para}. Our obtained values of the Galactocentric
coordinates are in fair agreement with the values obtained by Cantat-Gaudin et al. (2018).
We have compared our estimated parameters with the previously published values in the literature for both clusters.
Table \ref{comp_para} presents the comparison table for the clusters Haffner 22 and Melotte 71. All the estimated parameters
are comparable with the literature values.

\begin{table}
\centering
\tiny
\caption{A comparison of our obtained fundamental parameters for the clusters Haffner 22 and Melotte 71 with the literature values}.
\begin{tabular}{lllll}
\hline
Parameters                                       & Haffner 22 & Reference                              & Melotte 71 & Reference  \\ \hline

(Right ascension, Declination) (deg)             &  (123.1054, -27.9071)  & {\bf Present study}     &  (114.3854, -12.0598)  & {\bf Present study}\\
                                                 &  (123.108, -27.913)& Cantat-Gaudin et al. (2018)    &  (114.375, -12.069) & Liu \& Pang (2019)\\
                                                 &  (123.112, -27.90)  & Dias et al. (2014)           &   (114.383, -12.065) & Cantat-Gaudin et al. (2018)\\
                                                 &  (123.112, -27.89)  & Kharchenko et al. (2013)      &   (114.375, -12.066) & Sampedro et al. (2017)\\
                                                 &                              &                      &   (114.375, -12.06) & Dias et al. (2014)\\
                                                 &                              &                      &   (114.390, -12.055) & Kharchenko et al. (2013)\\
$(\mu_{\alpha}cos(\delta)$ (mas yr$^{-1}$) &  $-1.631\pm0.009$ & {\bf Present study}  &  $-2.398\pm0.004$ & {\bf Present study}\\
$(\mu_\delta)$ (mas yr$^{-1}$) &                $2.889\pm0.008$ & {\bf Present study}  &  $4.210\pm0.005$ & {\bf Present study}\\
                                                 &   (-1.765, 2.779) & Liu \& Pang (2019)       &   (-2.445, 4.201) & Liu \& Pang (2019)\\
                                                 &   (-1.638, 2.878) & Cantat-Gaudin et al. (2018)     &   (-2.446, 4.210) & Cantat-Gaudin et al. (2018)\\
                                                 &   (-1.95, 2.67) & Dias et al. (2014)                &   (-5.07, 5.76) & Dias et al. (2014)\\
                                                 &   (-4.52, 6.90) & Kharchenko et al. (2013)          &   (-0.94, 4.72) & Kharchenko et al. (2013)\\
Age (log)                                        &9.35 & {\bf Present study}                  &9.10 & {\bf Present study} \\
                                                 &9.39 &Cantat-Gaudin et al. (2020)                &9.06 & Liu \& Pang (2019) \\
                                                 &9.34 & Liu \& Pang (2019)                        &9.11 & Bossini et al. (2019) \\
                                                 &9.55 & Sampedro et al. (2017)                    &8.37 & Sampedro et al. (2017) \\
                                                 &9.19 & Kharchenko et al. (2013)                  &8.97 &Kharchenko et al. (2016) \\
Radius (arcmin)                                  &5.5 &{\bf Present study}                                &6.5 &{\bf Present study} \\
                                                 &6.7 & Carraro et al. (2016)                       &4.5 & Dias et al. (2014) \\
Parallax (mas)                                   &    $0.3547\pm0.006$ & {\bf Present study}                &    $0.4436\pm0.004$ & {\bf Present study}\\
                                                 &     0.325      & Liu \& Pang (2019)              &     0.44 & Liu \& Pang (2019) \\
                                                 &     0.329      & Cantat-Gaudin et al. (2018)      &    0.43 & Cantat-Gaudin et al. (2018)\\
Distance (Kpc)                                   &    $2.88\pm0.10$ & {\bf Present study}                  &    $2.28\pm0.15$ & {\bf Present study}\\
                                                 &    2.802    & Cantat-Gaudin et al. (2020)        &    2.7          & Cantat-Gaudin et al. (2018)\\
                                                 &    2.344    & Sampedro et al. (2017)             &    3.154        & Sampedro et al. (2017)\\
                                                 &    3.05     & Carraro et al. (2016)              &    2.473        & Kharchenko et al. (2016)\\
                                                 &    2.796    & Kharchenko et al. (2013)           &                 &                          \\
 \hline
\end{tabular}
\label{comp_para}
\end{table}

\subsection{Binary fraction in clusters}

\begin{figure*}
\begin{center}
\hbox{
\includegraphics[width=7.5cm, height=7.5cm]{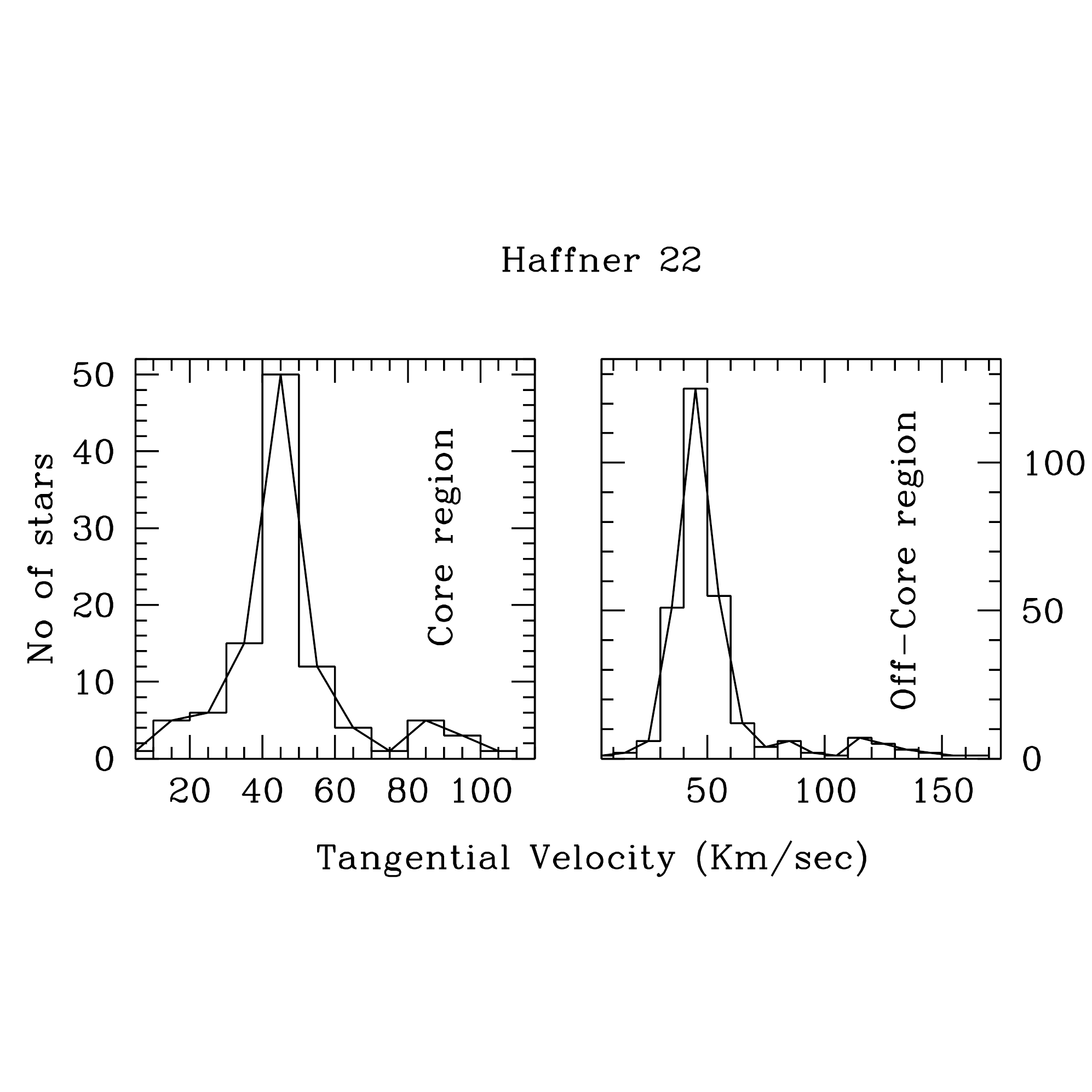}
\includegraphics[width=7.5cm, height=7.5cm]{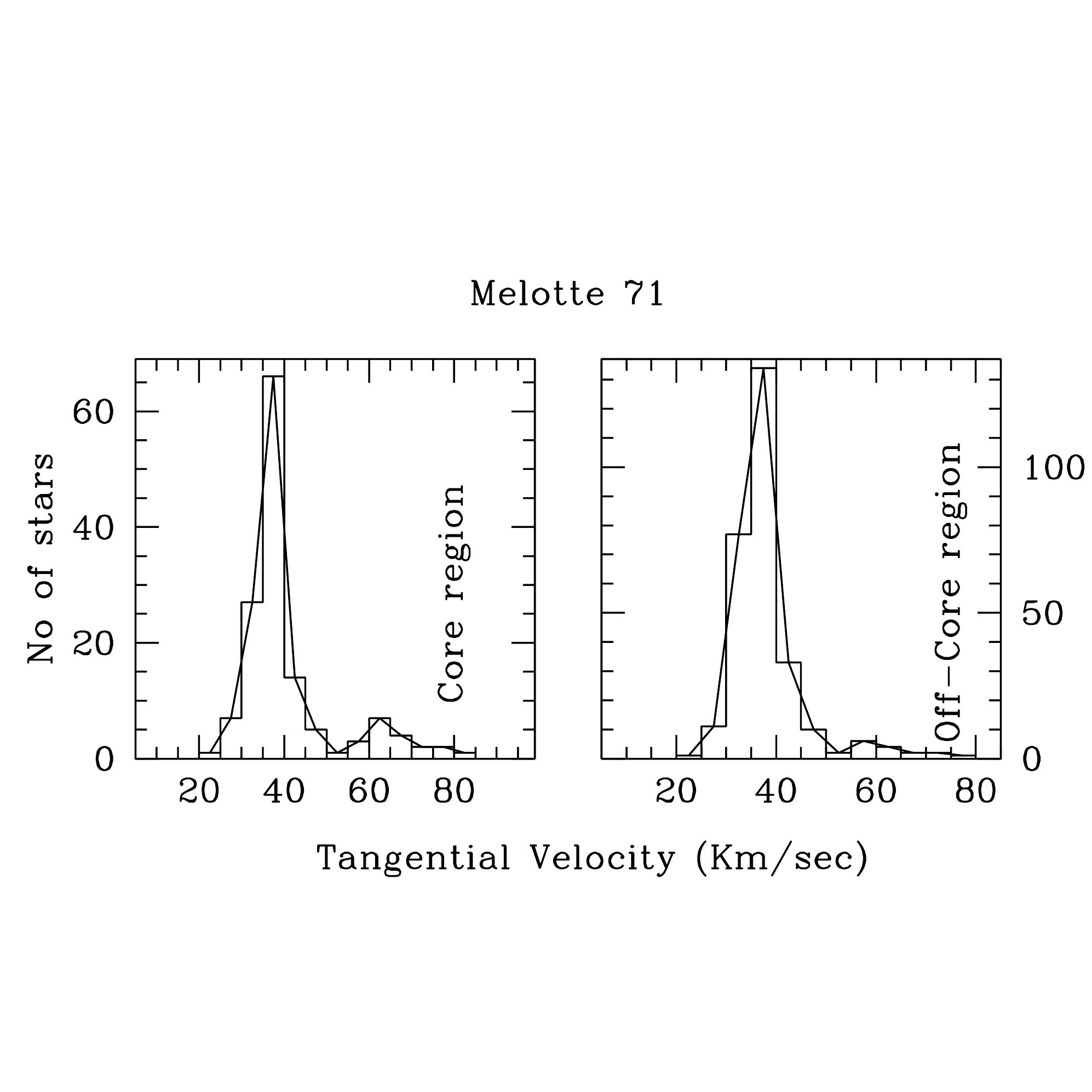}
}
\vspace{-1.5cm}
\caption{Histograms of tangential velocity in the core (left panels) and off-core (right panels) regions for clusters Haffner 22
and Melotte 71. We have used the individual distances of each star to transform proper motion into tangential velocity.
Higher and lower peak in core and off-core regions are showing single star and binary star distribution.
}
\label{binary_hist}
\end{center}
\end{figure*}

The stellar binary fraction in star clusters is a key factor in understanding the eﬀects of binary stars on the properties and
dynamical evolution of the host cluster. We have plotted histograms using the tangential velocity of member stars of the clusters
Haffner 22 and Melotte 71 as shown in Fig \ref{binary_hist}. The radius of core and off-core regions are (2.1 and 3.4) arcmin
and (2.4 and 4.1) arcmin for clusters Haffner 22 and Melotte 71, respectively. In this figure, we found two different peaks for single star and
binary star distribution (Bica \& Bonatto (2005)). In this analysis, we have used main-sequence stars
with 15$\le G \le$ 18 mag for Haffner 22 and 14$\le G \le$ 18 mag
for Melotte 71.
We have used the
weighted mean method to compute the mean value of velocity in core and off-
core regions. In the core region, there are 96 stars (1st peak) and 14 stars (2nd peak), and
in the off core region, there are 284 stars (1st peak) and 20 stars (2nd peak) for Haffner
22. In the core region, there are 140 stars (1st peak) and 13 stars (2nd peak), and in the
off core region, there are 284 stars (1st peak) and 18 stars (2nd peak) for Melotte 71.
We estimated the mean value of tangential velocity in core and off-core regions as ($45.04\pm6.25$, $44.99\pm6.16$)
km/sec and ($37.38\pm2.87$, $37.60\pm3.80$) km/sec for clusters Haffner 22 and Melotte 71, respectively. We also have obtained
the corresponding dispersion in core and off-core
region as $(5.4\pm2.2, 5.3\pm2.1)$ km/sec
and $(4.9\pm1.8, 3.7\pm1.2)$ Km/sec for both clusters.
Our estimated values of mean and dispersion for a single star
and double star distribution is approximately similar in core and off-core regions for both clusters. This analysis suggests
that some binary content is present in both objects. The binary fraction $(f_{bin})$ in a cluster can be estimated by dividing
the number of high-velocity stars by the total number of stars. Thus, in the core region, the binary fractions of Haffner 22 and
Melotte 71 are found as $f_{bin}$=$13.59\pm3$$\%$ and $14.20\pm5$$\%$. In the off-core region the binary fraction found as
$f_{bin}$=$11.10\pm4$$\%$ and $10.0\pm6$$\%$ for both objects. From here, we obtained that binary fraction is more in the core region
for both clusters. Bica and Bonatto (2005) have found a higher binary fraction in the core region for OCs NGC 2287, M 48,
NGC 6208, NGC 3680, and IC 4651. The total binary fractions are found as  $f_{bin}$=$12.34\pm3.5$$\%$ and $12.10\pm5.5$$\%$
clusters Haffner 22 and Melotte 71, respectively.
We have shown all the possible binary stars in Fig.~\ref{dist_age}

\section{Dynamical study of the clusters}

\subsection{Luminosity function and Mass function}

\begin{figure}
\begin{center}
\hbox{
\includegraphics[width=4.0cm, height=4.0cm]{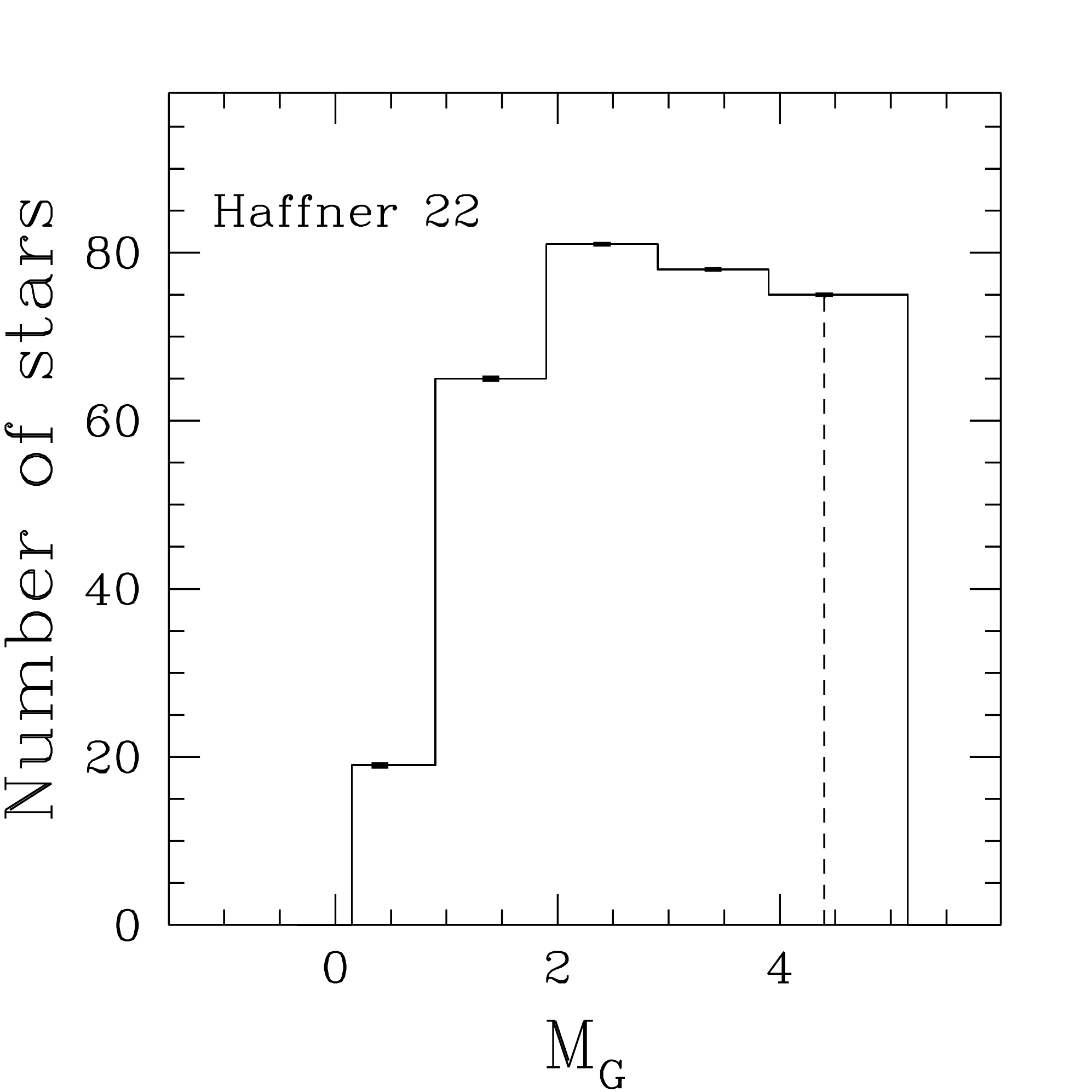}
\includegraphics[width=4.0cm, height=4.0cm]{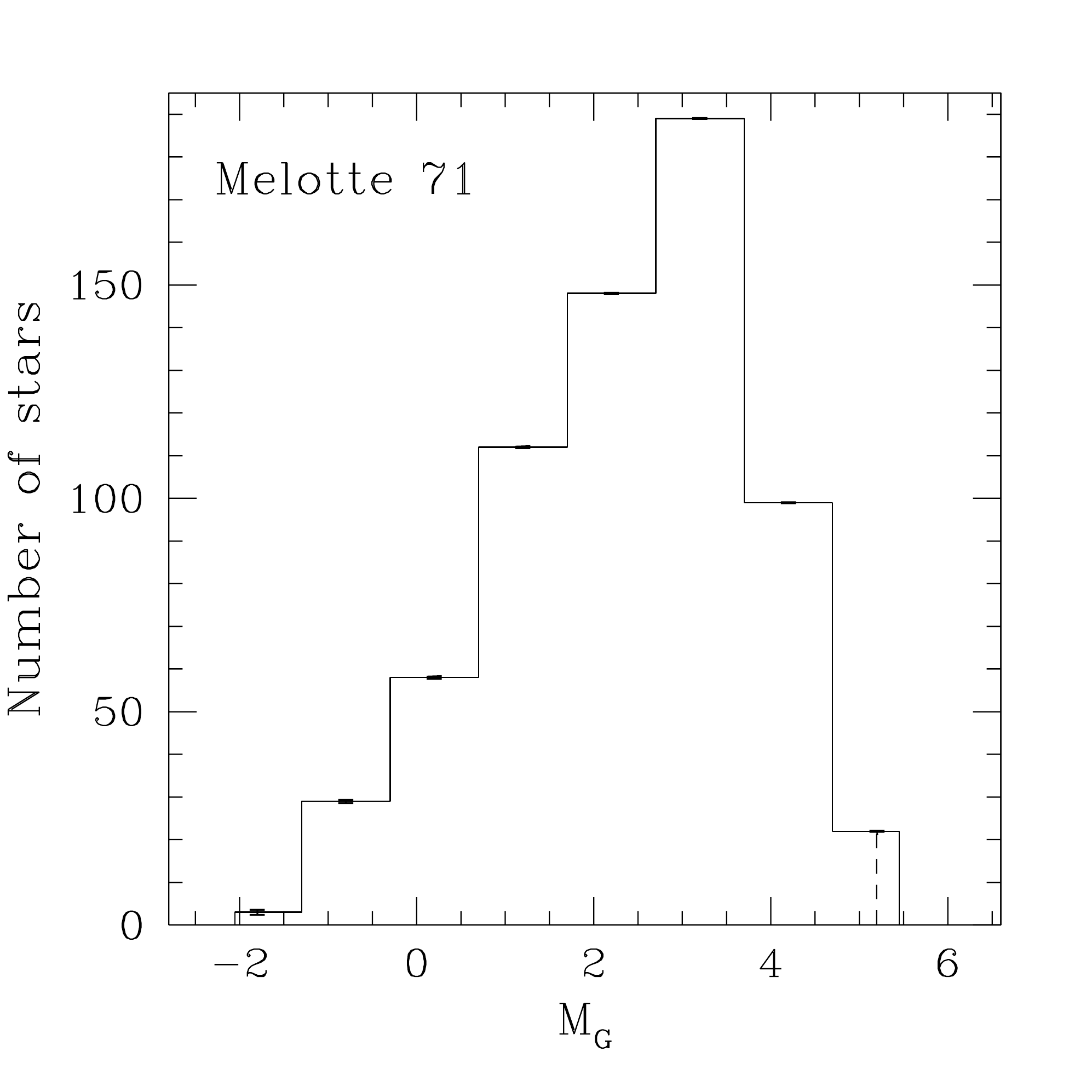}
}
\vspace{-0.1cm}
\caption{Luminosity function of stars in the region of Haffner 22 and Melotte 71. Verticle dotted line indicates our magnitude limit.}
\label{lf}
\end{center}
\end{figure}

The luminosity function (LF) and mass function (MF) depend on the number of actual cluster members, and both measures
are associated with the well-known mass-luminosity relationship. We used $G$ versus $(G_{BP}-G_{RP})$ CMDs to see the
distribution of stars with magnitude in both clusters. The distance modules can convert the $G$ magnitudes
of main-sequence stars into absolute magnitudes. We have constructed the histogram of LF with 1.0 mag intervals
as shown in Fig. \ref{lf}. This figure exhibits that the LF continues to increase up to $M_{G}\sim$ 2.5 and 3.3 mag
for the clusters Haffner 22 and Melotte 71. We have shown the used magnitude limit in Fig \ref{lf} using the verticle dotted line.

\begin{figure}
\begin{center}
\hbox{
\includegraphics[width=4.0cm, height=4.0cm]{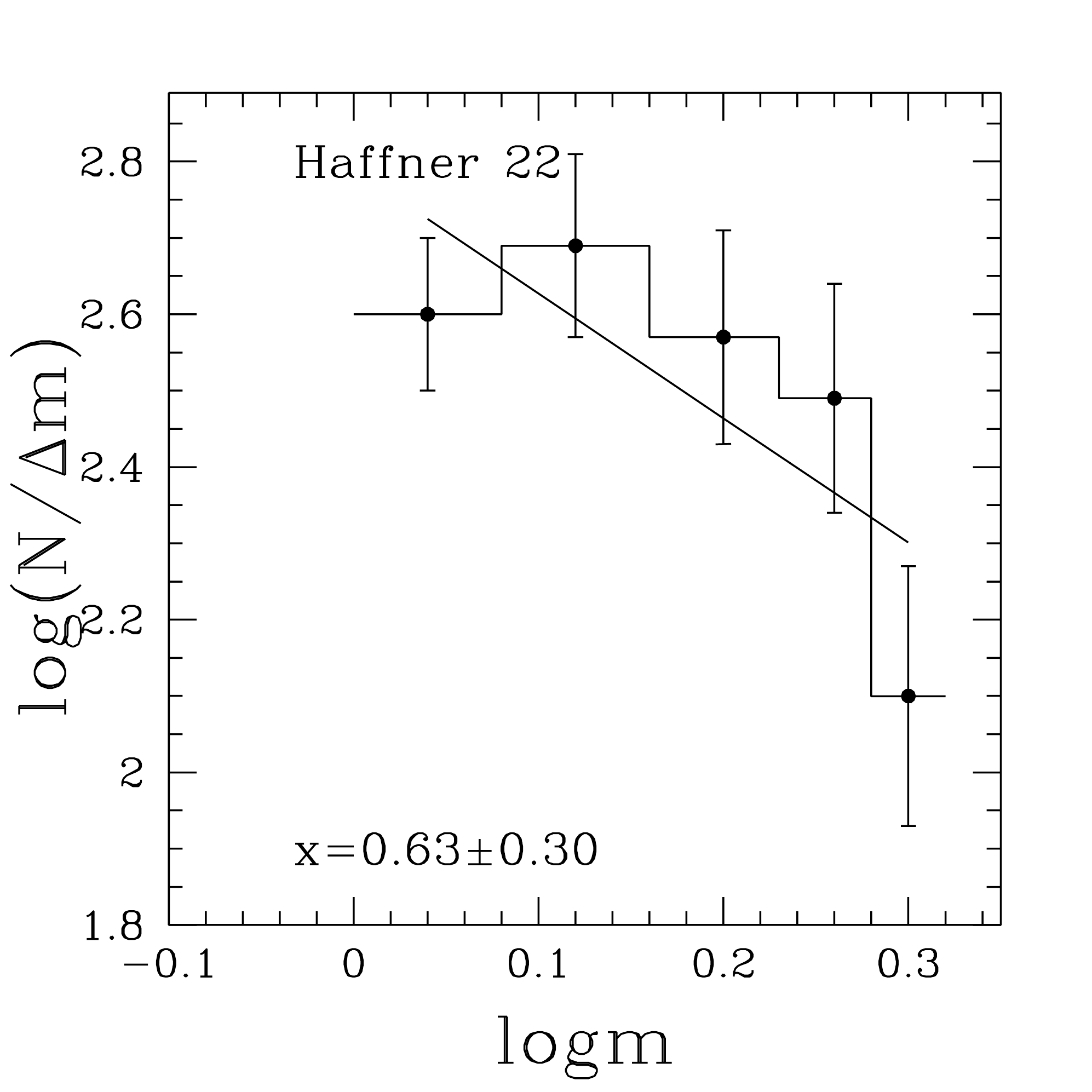}
\includegraphics[width=4.0cm, height=4.0cm]{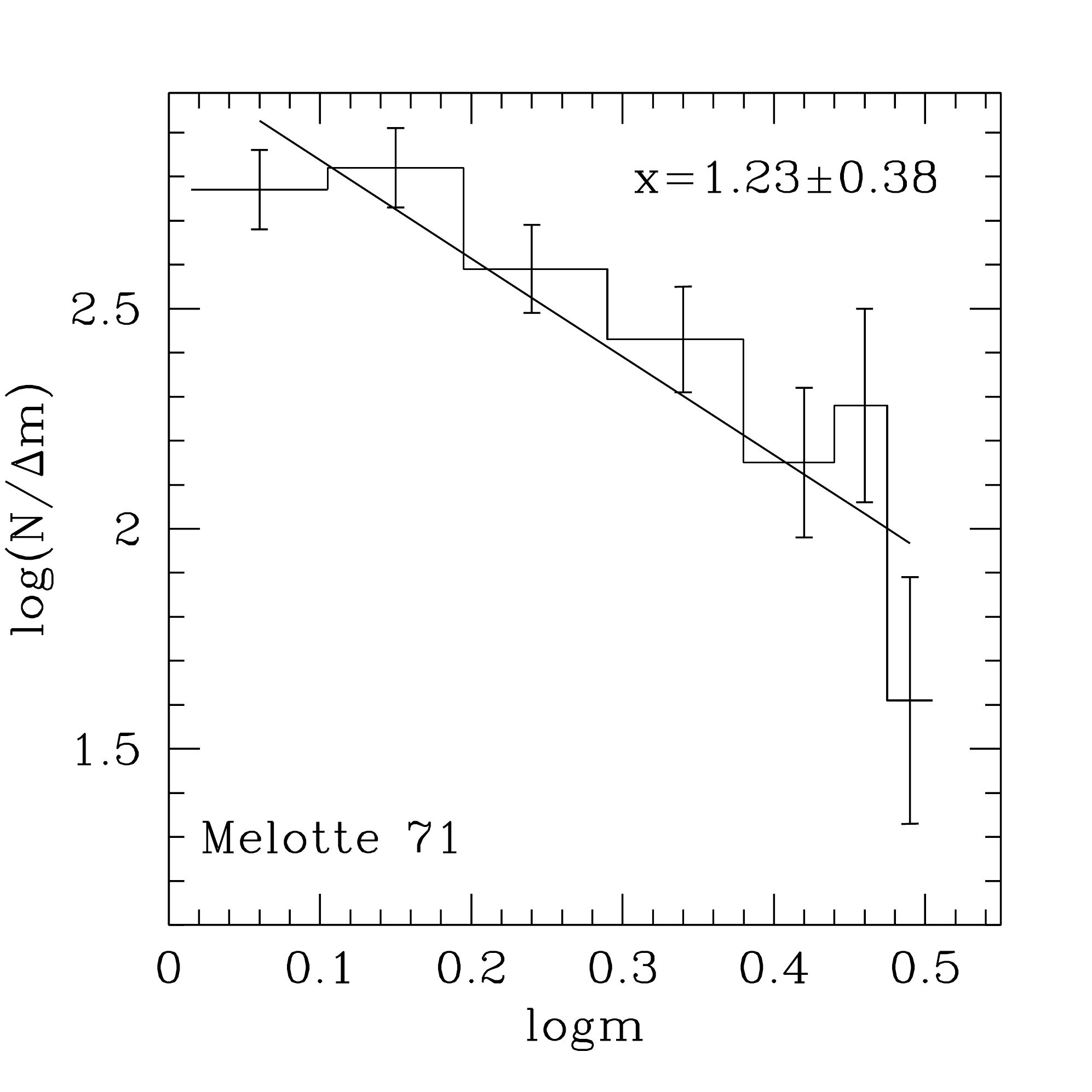}
}
\vspace{-0.1cm}
\caption{Mass function histogram derived using the most probable members, where solid line indicates the
power law given by Salpeter (1955). The error bars represent $\frac{1}{\sqrt{N}}$.}
\label{mass}
\end{center}
\end{figure}

To convert luminosity into masses, we have employed the theoretical isochrones of Marigo et al. (2017). To understand
the MF, we have transformed absolute mag bins to mass bins, and the resulting present-day mass function (PDMF) is
shown in Fig. \ref{mass}.
The first and last bin is not in the same trend because we have used the fitting
error, and the error is higher in the last bin.
The shape of the PDMF of members in Haffner 22 and Melotte 71 for masses $\ge$ 1 $M_{sol}$
can be approximated by a power law of the form

$\log\frac{dN}{dM}=-(1+x)\log(M)$+constant\\

Where $dN$ is the number of the probable cluster members in a mass bin $dM$ with central mass $M$ and $x$ is mass function slope.
Since Gaia data ($G$ mag) is not complete below G=19 mag (Arenou et al. 2018), we took only the stars brighter than this limit,
which corresponds to stars more massive than 1 $M_{\odot}$. The computed values of the MF slopes are $0.63\pm0.30$ and
$1.23\pm0.38$ for the clusters Haffner 22 and Melotte 71, respectively. The MF slope value for Haffner 22 is flatter, while
Melotte 71 is satisfactory with the Salpeter's initial mass function slope within error.
The dynamical study of Haffner 22 shows a
lack of faint stars in the inner region which leads to the mass-segregation effect. In the literature there are many dynamical studies
that suggest lack of faint stars towards the  center of the cluster (Fischer et al. 1998; Pandey et al. 1992, 2001, 2005; Kumar et al. 2008).
The complete mass has been evaluated for both
clusters using the derived mass function slope. All the MF-related parameters in this section, like mass range, mass function
slope, and the total mass measured, are listed in Table \ref{massf_tab}.

\begin{figure}
\begin{center}
\hbox{
\includegraphics[width=4.0cm, height=4.0cm]{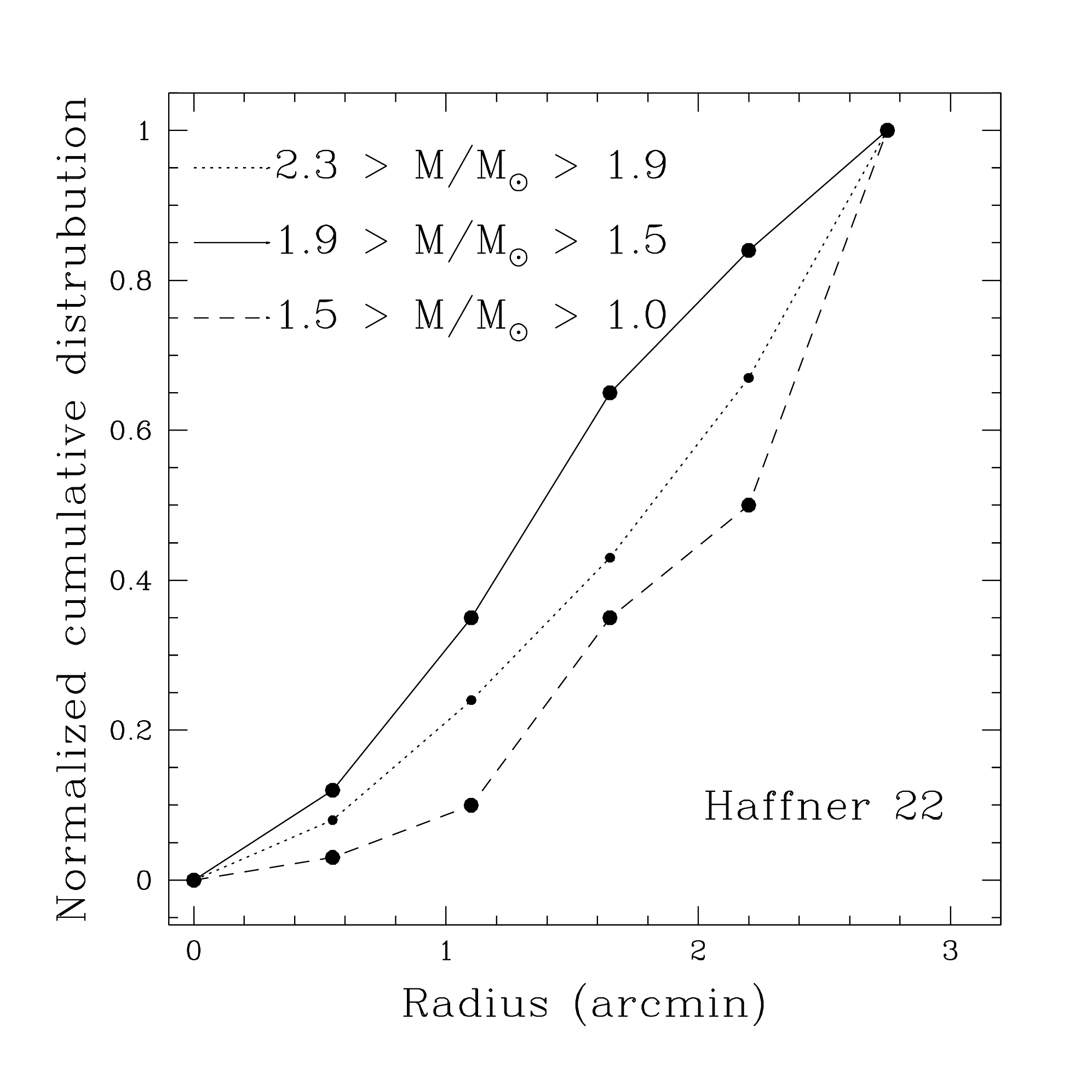}
\includegraphics[width=4.0cm, height=4.0cm]{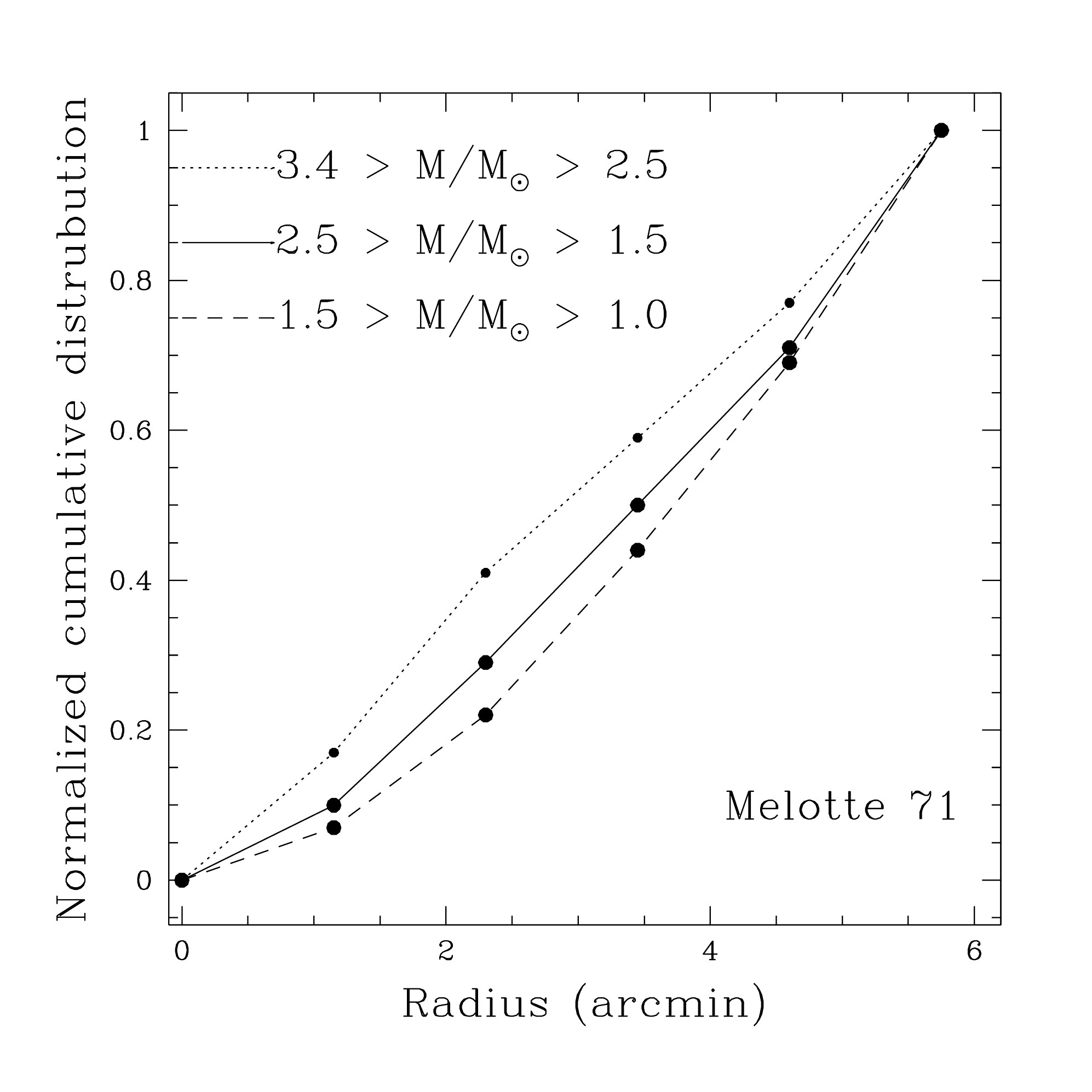}
}
\vspace{-0.1cm}
\caption{The cumulative radial distribution of stars in various mass range.}
\label{mass_seg}
\end{center}
\end{figure}

\begin{table*}
\caption{The main mass function parameters in clusters.
}
\vspace{0.5cm}
\begin{center}
\small
\begin{tabular}{lcccc}
\hline\hline
Object & Mass range & MF slope & Total mass & Mean mass \\
&      $M_{\odot}$ &  & $M_{\odot}$ & $M_{\odot}$
\\
Haffner 22  & $1.0-2.3$&$0.63\pm0.30$&$ 572$&$1.49$ \\
Melotte 71  & $1.0-3.4$&$1.23\pm0.38$&$1015$&$1.70$ \\
\hline
\end{tabular}
\label{massf_tab}
\end{center}
\end{table*}

\begin{table*}
\caption{Distribution of stars in different mass ranges along with the percentage of confidence level in mass-segregation effect for the
clusters.
}
\vspace{0.5cm}
\begin{center}
\small
\begin{tabular}{lcc}
\hline\hline
Object & Mass range & Confidence level\\
&      $M_{\odot}$ & $\%$ \\
\hline
Haffner 22  & $2.3-1.9,  1.9-1.5, 1.5-1.0$& 88 \\
Melotte 71  & $3.4-2.5,  2.5-1.5, 1.5-1.0$& 80 \\
\hline
\end{tabular}
\label{mass_seg_para}
\end{center}
\end{table*}

\subsection{Mass-segregation study}

In mass segregation, the bright stars move towards the cluster center while the low mass stars move towards
the halo region (Mathieu 1984; Kroupa 1995; de La Fuente Macross 1996). Many authors have addressed this
phenomenon in clusters (e.g., Hillenbrand \& Hartmann 1998; Meylan 2000;
Baumgardt \& Makino 2003; Dib, Schmeja \& Parker 2018; Dib \& Henning 2019; Alcock \& Parker 2019; Bisht et al. 2020a, 2021a).
We have used only the probable members to explain the mass segregation effect in our target clusters. We distributed cluster
members into three mass ranges as shown in Table \ref{mass_seg_para}. The cumulative radial stellar distribution of cluster members for the clusters
Haffner 22 and Melotte 71 is shown in Fig. \ref{mass_seg}. This diagram exhibits that the cluster members show a mass segregation
effect as bright stars seem to be more centrally concentrated than the low mass members. We found the confidence level of
mass segregation is 88 $\%$ and 80 $\%$ for the clusters Haffner 22 and Melotte 71, respectively, based on the Kolmogorov-Smirnov $(K-S)$ test.

The plausible reason for this effect may be dynamical evolution, an imprint of star formation, or both (Dib, Kim \& Shadmehri 2007;
Allison et al. 2009; Pavlik 2020). Our target objects are old-age OCs. So, the possible reasons could be both of these. The
relaxation time ($T_{R}$) is represented as the time in which the stellar velocity distribution
converts Maxwellian and denoted by the following formula (Spitzer \& Hart 1971):\\

$T_{R}=\frac{8.9\times10^5\sqrt{N}\times{R_{h}}^{3/2}}{\sqrt{\bar{m}}\times log(0.4N)}$\\

where $N$ represents the cluster members with membership probability higher than 50$\%$, $R_{h}$ is the cluster half mass
radius expressed in parsec and $\bar{m}$ is the average mass of the cluster members (Spitzer \& Hart 1971) in the solar
unit. The value of $\bar{m}$ is found as 1.49 and 1.70 $M_{\odot}$ for these objects.

We have estimated the value of $R_{h}$ based on the transformation equation as given in Larsen (2006),\\

$R_{h}=0.547\times R_{c}\times(\frac{R_{t}}{R_{c}})^{0.486}$\\

where $R_{c}$ is core radius while $R_{t}$ is tidal radius. We obtained the value of half light radius as
2.45 and 2.62 pc for the clusters Haffner 22 and Melotte 71, respectively.

The value of dynamical relaxation time $T_{R}$ is Obtained as 25 and 30 Myr for these objects. Hence, we conclude
that Haffner 22 and Melotte 71 are dynamically relaxed OCs.

\section{Orbit study of the clusters}

The study of orbits is beneficial to understand stars, clusters, and Galaxies' formation and evolution processes. We used the
Allen \& Santillan (1991) criteria for Galactic potentials to obtain Galactic orbits of Haffner 22 and Melotte 71.
Bajkova \& Bobylev (2016) and Bobylev et al. (2017) have refined Galactic potential model parameters with the
help of new observational data for the galactocentric distance R $\sim$ 0 to 200 kpc. The equations considered for
the used models are described by Rangwal et al. (2019). The main fundamental parameters
(cluster center ($\alpha$ and $\delta$), mean proper motions ($\mu_{\alpha}cos\delta$, $\mu_{\delta}$), parallax,
age and heliocentric distance ($d_{\odot}$)) have been used to determine the orbital parameters in the clusters
under study. We have used the radial velocity values as $33.23\pm0.16$ km/sec and $51.26\pm0.36$ km/sec for Haffner 22 and
Melotte 71 as taken from the catalog given by Soubiran et al. (2018).

\begin{table*}
   \caption{Position and velocity components in the Galactocentric coordinate system. Here $R$ is the galactocentric
            distance, $Z$ is the vertical distance from the Galactic disc, $U$ $V$ $W$ are the radial tangential and the vertical
            components of velocity respectively and $\phi$ is the position angle relative to the sun's direction.
}
   \begin{tabular}{ccccccccc}
   \hline\hline
   Cluster   & $R$ &  $Z$ &  $U$  & $V$  & $W$ & $\phi$   \\
   & (kpc) & (kpc) & (km/s) &  (km/s) & (km/s) & (radian)    \\
  \hline
   Haffner 22 & $9.75\pm0.278$ & $0.18\pm0.029$ & $23.29 \pm 0.25$  & $-245.75 \pm 0.23$ &  $ -12.17 \pm 0.51$ & 0.27    \\
   Melotte 71 & $10.11\pm0.213$ & $0.22\pm0.016$ & $ -17.21 \pm0.35$  & $-263.126 \pm 0.32$ &  $-10.62 \pm 0.36$ & 0.19    \\
\hline
  \end{tabular}
  \label{inp}
  \end{table*}

We have transformed equatorial space and velocity components into Galactic-space velocity components. The Galactic center is
considered at ($17^{h}45^{m}32^{s}.224, -28^{\circ}56^{\prime}10^{\prime\prime}$) and
the North-Galactic pole is considered at ($12^{h}51^{m}26^{s}.282,
27^{\circ}7^{\prime}42^{\prime\prime}.01$) (Reid \& Brunthaler, 2004). To apply a correction for Standard Solar Motion and Motion
of the Local Standard of Rest (LSR), we used position coordinates of Sun as
($8.3,0,0.02$) kpc and its velocity components
as ($11.1, 12.24, 7.25$) km/s (Schonrich et al. 2010). Transformed parameters
in Galacto-centric coordinate system are listed in Table \ref{inp}.

\begin{table*}
   \caption{Orbital parameters obtained using the Galactic potential model.
   }
   \begin{tabular}{ccccccccc}
   \hline\hline
   Cluster  & $e$  & $R_{a}$  & $R_{p}$ & $Z_{max}$ &  $E$ & $J_{z}$ & $T_{R}$ & $T_{Z}$   \\
           &    & (kpc) & (kpc) & (kpc) & $(100 km/s)^{2}$ & (100 kpc km/s) & (Myr) & (Myr) \\
   \hline\hline
   Haffner 22 & 0.001  & 10.461  & 10.439  & 0.297 & -9.988 &  -23.952  & 247 &  98 \\
   Melotte 71 & 0.004  & 11.584  & 11.491  & 0.324 & -9.335 &  -26.610  & 240 &  102 \\
 \hline
  \end{tabular}
  \label{orpara}
  \end{table*}

\begin{figure*}
\begin{center}
\hbox{
\includegraphics[width=4.2cm, height=4.2cm]{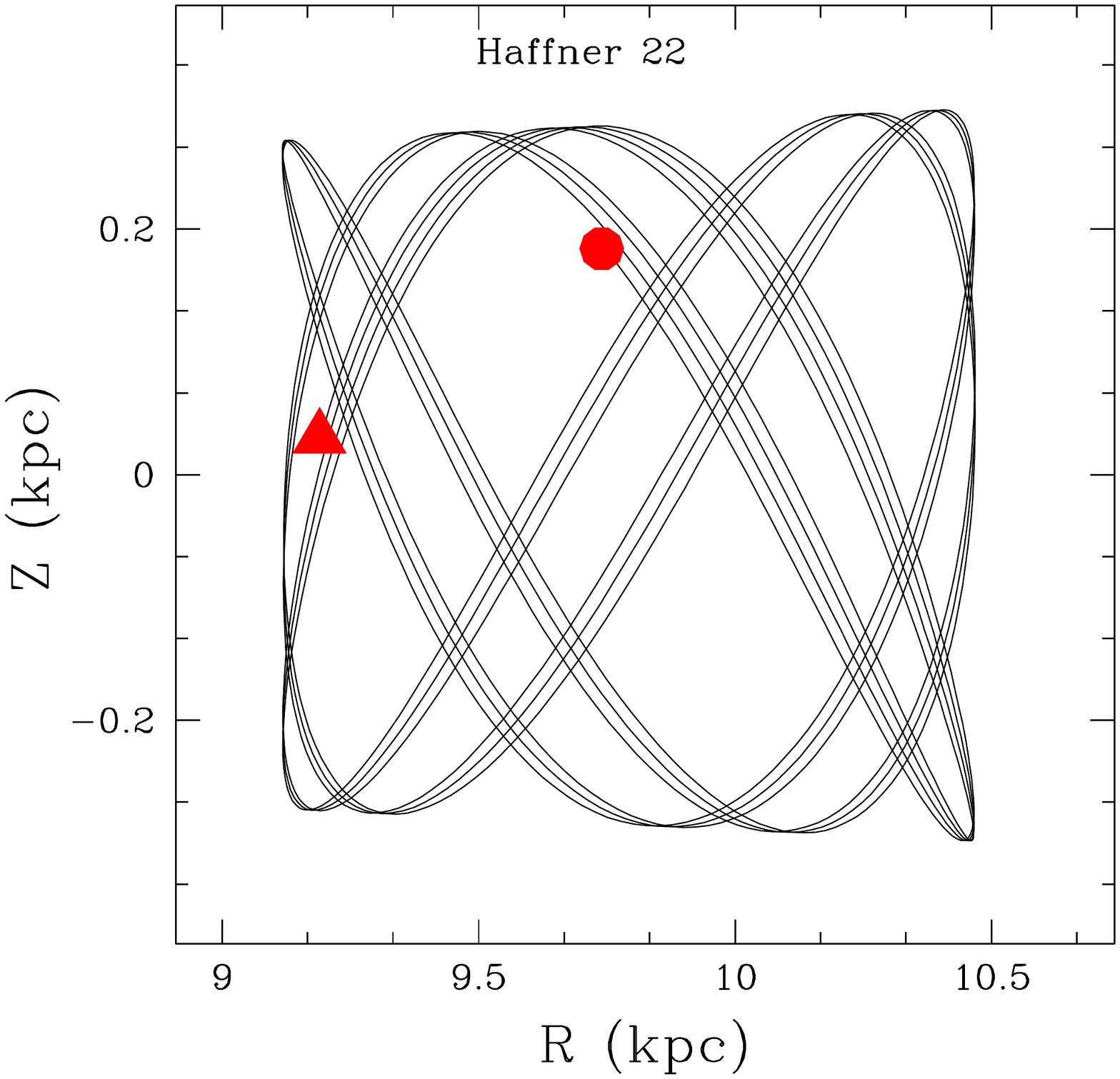}
\includegraphics[width=4.2cm, height=4.2cm]{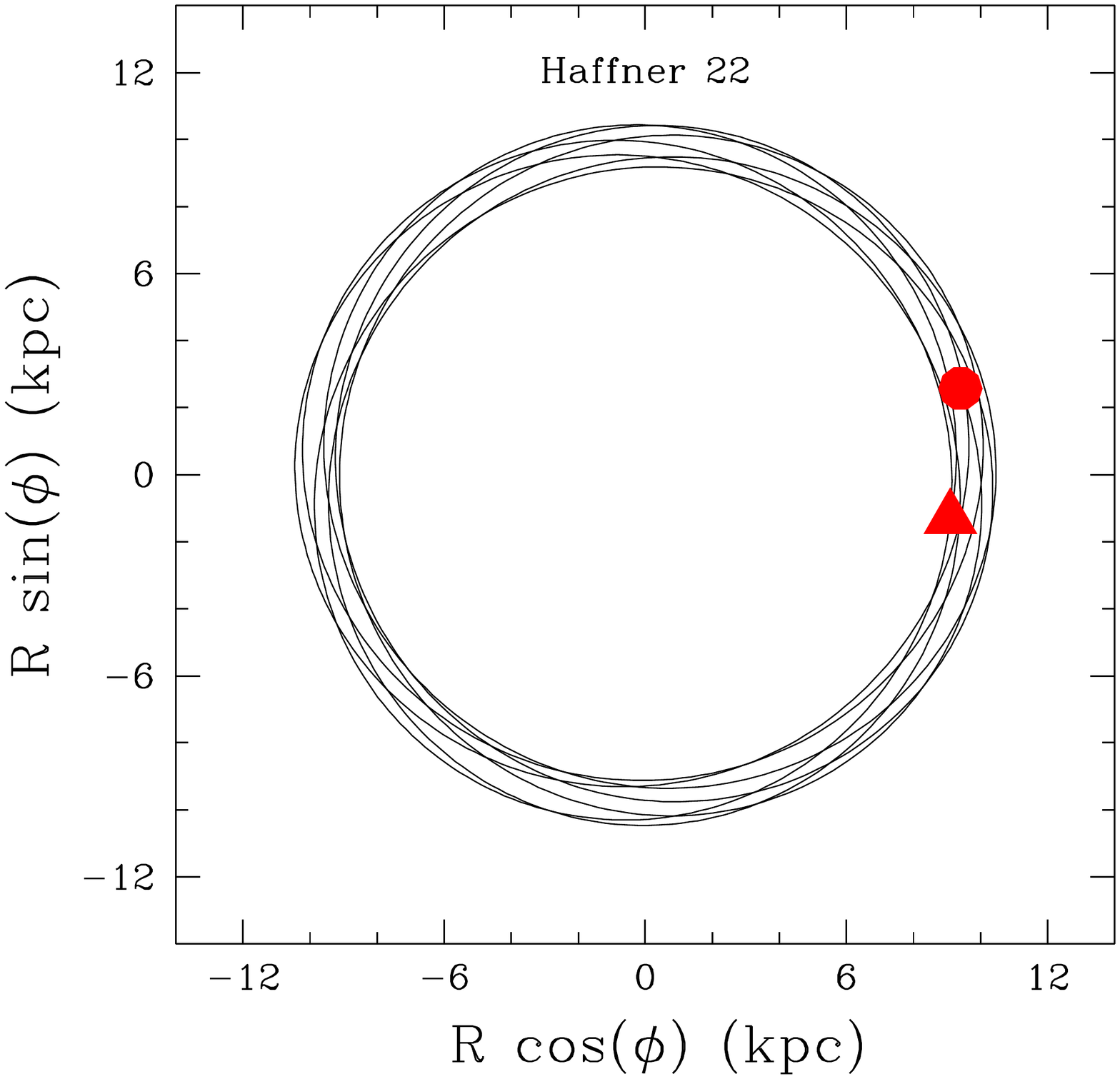}
\includegraphics[width=4.2cm, height=4.2cm]{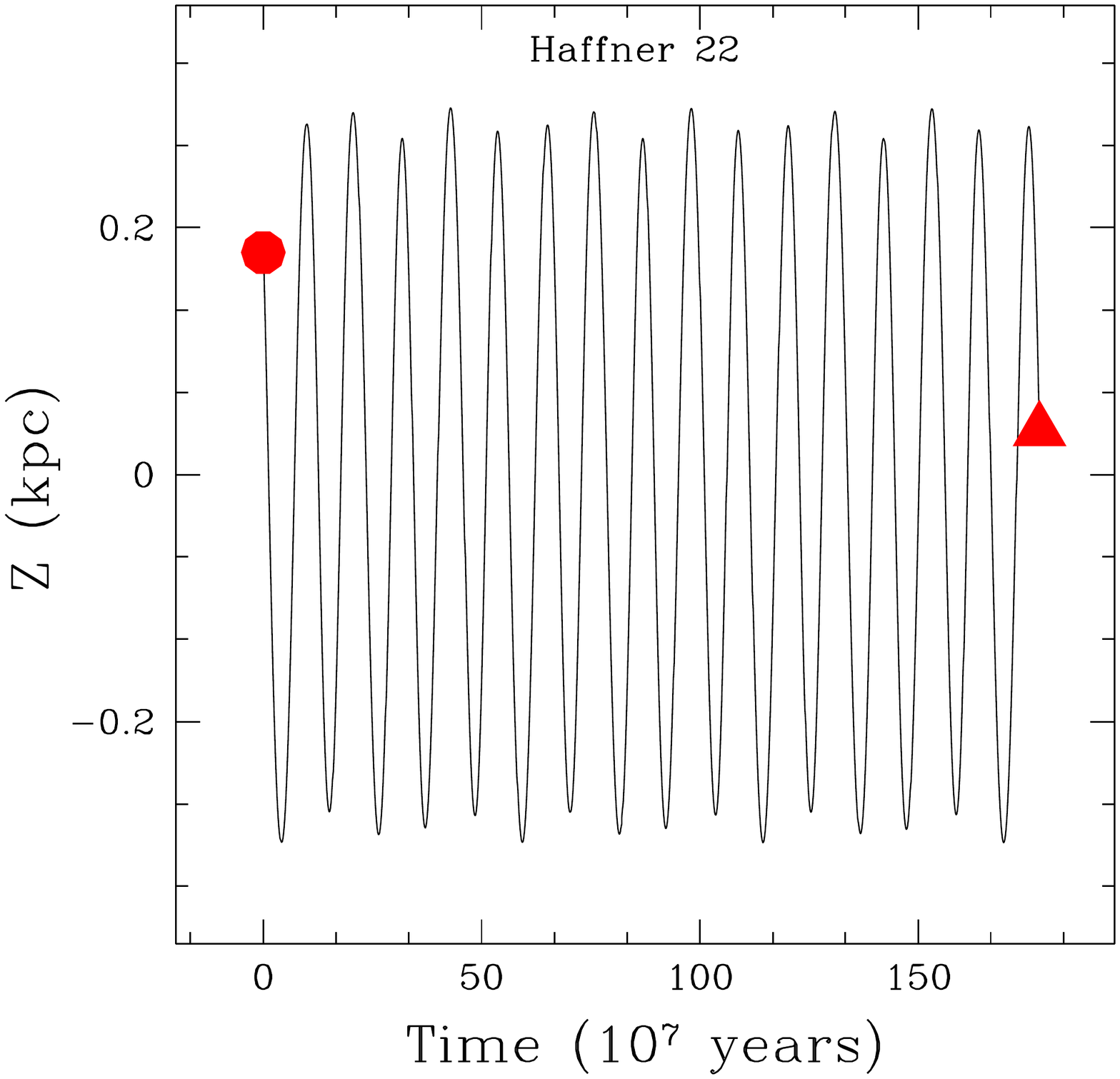}
}
\hbox{
\includegraphics[width=4.2cm, height=4.2cm]{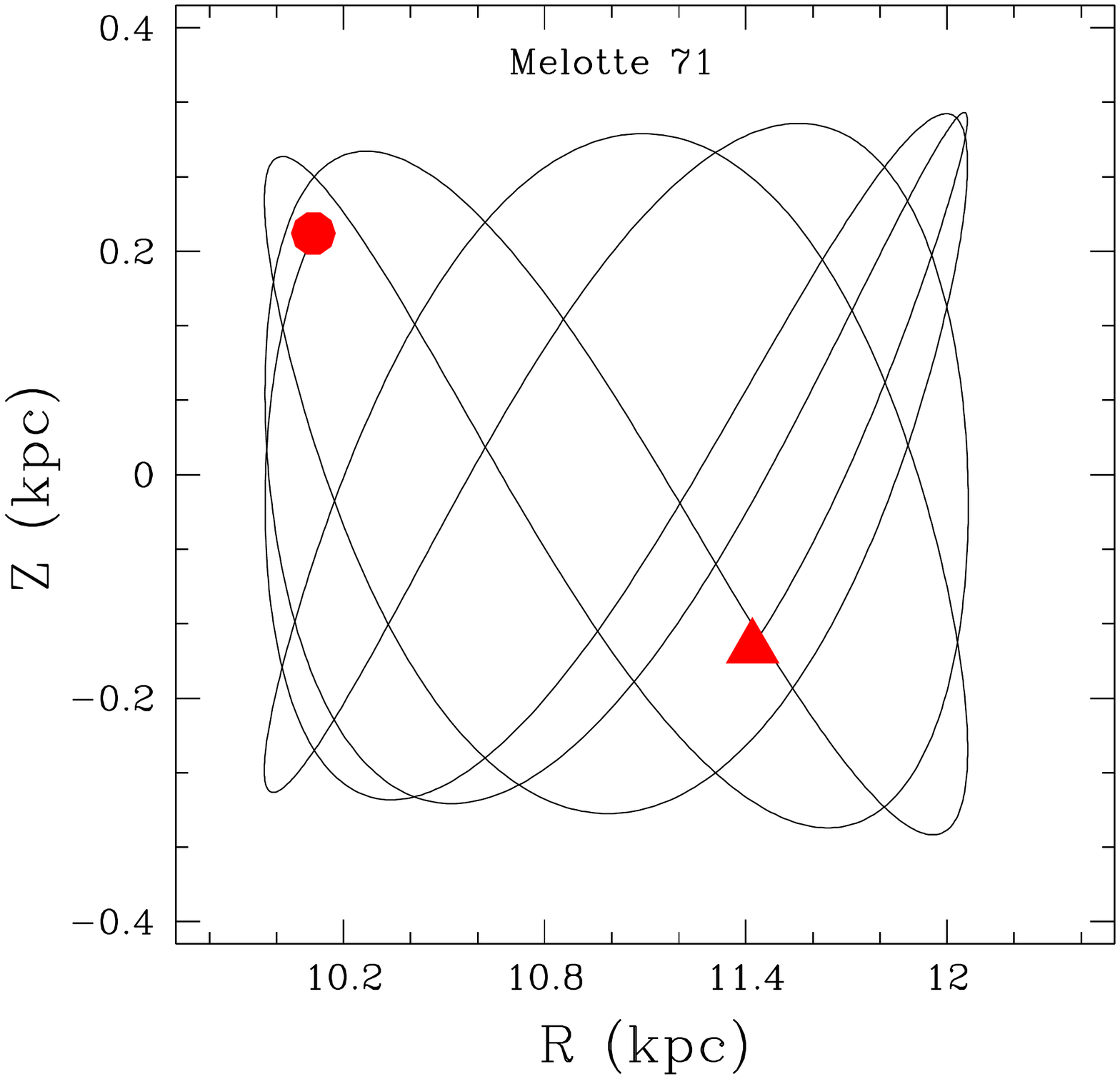}
\includegraphics[width=4.2cm, height=4.2cm]{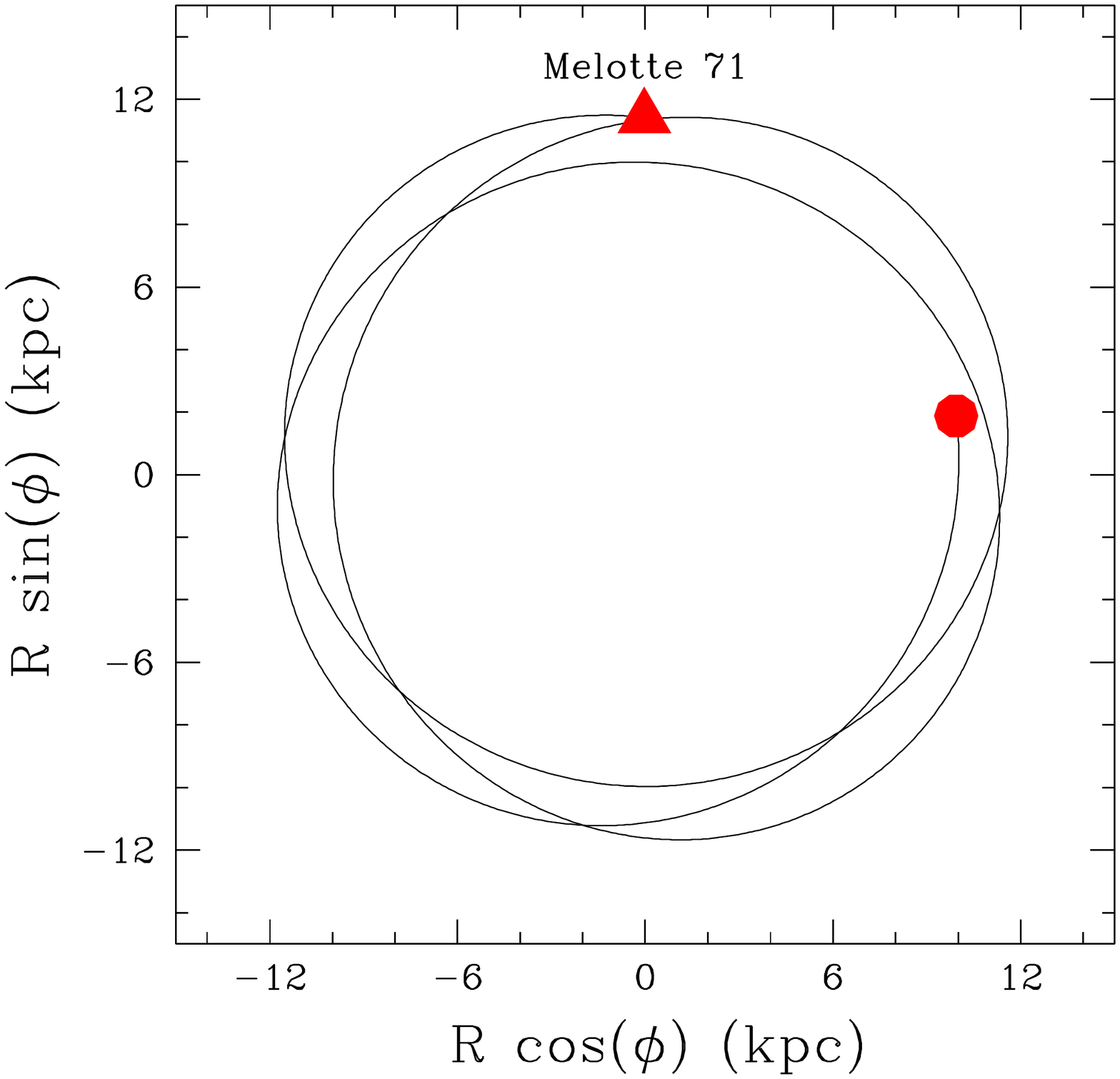}
\includegraphics[width=4.2cm, height=4.2cm]{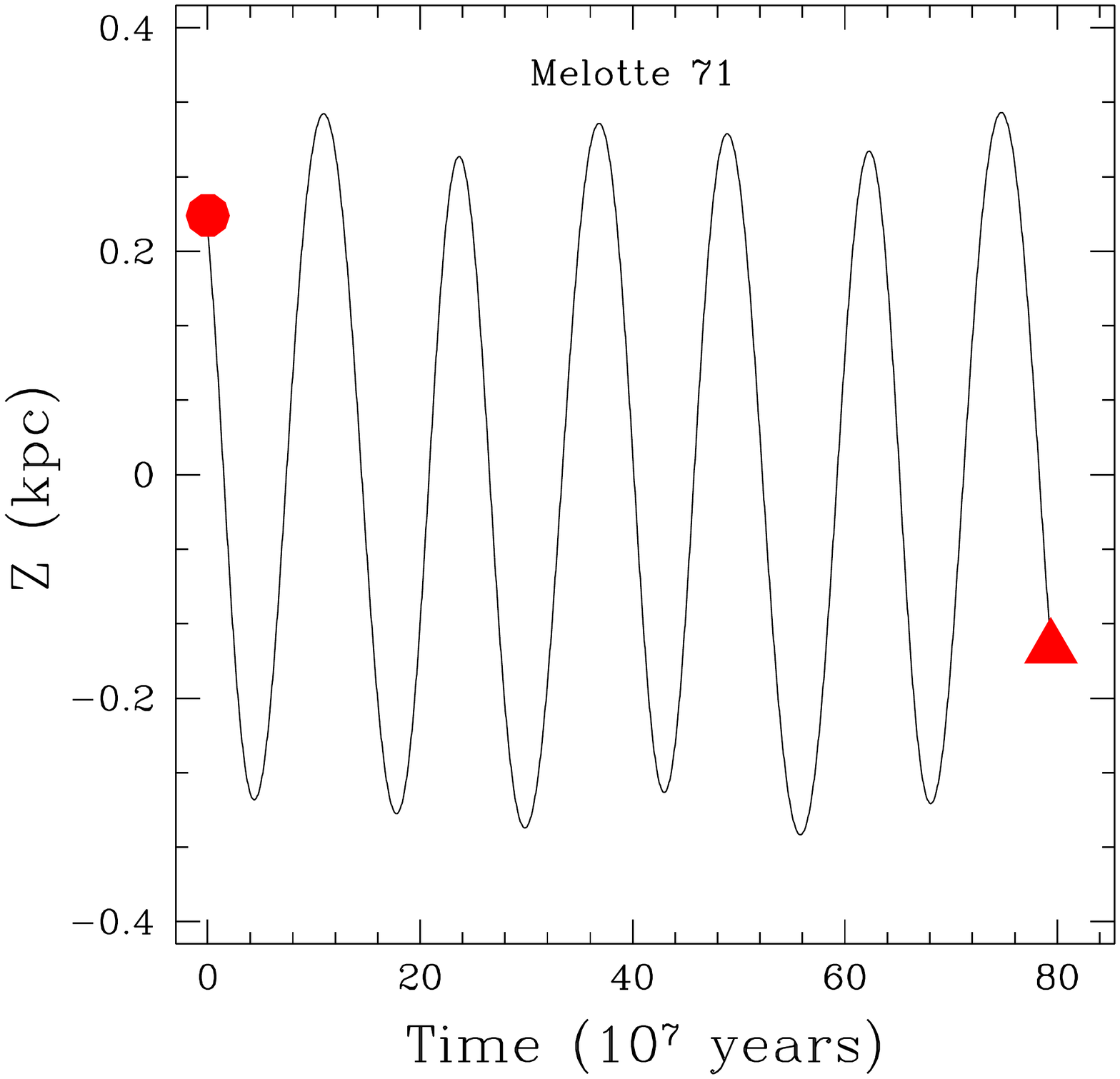}
}
\caption{Galactic orbits of the clusters Haffner 22 and Melotte 71 estimated with the Galactic potential model
described in text in the time interval of age of cluster.
The left panels show the side view and the middle panels show
the top view of the orbits.
The right panels show the motion of both the clusters in the Galactic disc with time.
The filled triangles and the circles denote the birth and the present day positions of the clusters
in the Galaxy.}
\label{orbit}
\end{center}
\end{figure*}

Fig. \ref{orbit} shows the orbits of the clusters Haffner 22 and Melotte 71. The left panel of this figure indicates the motion of
the cluster in terms of distance from the Galactic center and Galactic plane and this shows a 2-D side view of the orbits. In the middle panel,
the cluster motion is described in terms of $x$ and $y$ components of Galactocentric distance, which shows a top view of orbits.
The right panel of this figure indicates the motion of clusters under study in the Galactic disc with time. Both clusters follow a
boxy pattern according to our analysis. Our obtained values of eccentricity are nearly zero for both objects, which demonstrates that
the target clusters trace a circular path around the Galactic center. The birth and present day position in the Galaxy are
represented by filled circle and triangle as shown in Fig. \ref{orbit}.
The various orbital parameters have been obtained for these clusters, which are listed in Table \ref{orpara}. Here $e$ is
eccentricity, $R_{a}$ is the apogalactic distance, $R_{p}$ is perigalactic distance, $Z_{max}$ is the maximum distance traveled
by cluster from Galactic disc, $E$ is the average energy of orbits, $J_{z}$ is $z$ component of angular momentum, $T_{R}$ is time
period of the revolution around the Galactic center and $T_{Z}$ is the time period
of vertical motion.

In these figures, we can see that the birth positions of both the clusters are in
the thick disc of the Galaxy hence not affected by the thin disc and its
tidal forces. Also, both the clusters are orbiting outside the solar circle hence
not interacting with the inner region of the Galaxy.
The disruption of both the clusters caused by the Galactic tidal forces is slow so we expect a longer survival time for these clusters.

The tidal radius of clusters is influenced by the effects of Galactic tidal fields and internal relaxation dynamical
evolution of clusters (Allen \& Martos 1988). To find the tidal radius of Haffner 22 and Melotte 71, we have used the
formula derived by Bertin \& Varri (2008) as:\\

$r_{t}=(\frac{GM_{cl}}{\omega^{2} \nu})^{1/3}$ \\

where $\omega$ and $\nu$  are   \\

$\omega= ((d\Phi_{G}(R)/dR)_{R_{gc}}/R_{gc})^{1/2}$  \\
$\nu=4-\kappa^{2}/\omega^{2}$     \\

where $\kappa$ is \\

$\kappa=(3\omega^{2}+(d^{2}\Phi_{G}(R)/dR^{2})_{R_{gc}})^{1/2}$  \\

here $\Phi_{G}$ is Galactic potential, $M_{cl}$ mass of the cluster, $R_{gc}$ is the Galactocentric distance of
the cluster, $\omega$ is the orbital frequency, $\kappa$ is the epicyclic frequency and $\nu$ is a positive constant.
We used above discussed Galactic potentials for this calculation, the value of the Galactocentric distance is taken from
Table~\ref{para} and mass of the cluster is taken from Table \ref{massf_tab}. We obtained tidal radius as
12.19 and 15.13 pc for the clusters Haffner 22 and Melotte 71, respectively.
\section{Conclusions}
\label{con}

We have analyzed two OCs, Haffner 22 and Melotte 71, based on the Gaia~EDR3 photometric and astrometric database. We
have recognized 382 and 597 likely members for the clusters Haffner 22 and Melotte 71, respectively, with membership probabilities
higher than $50\%$. We studied the cluster structure, obtained the main fundamental parameters, described the dynamical study,
and determined the galactic orbit of these clusters. The principal outcomes of this study can be summarized as follows:

\begin{itemize}

\item  The new center coordinates, cluster radius, and proper motions are obtained for Haffner 22 and Melotte 71 and are listed in
       Table \ref{para}. The enhancement of radius around 3.5 arcmin demonstrates the presence of a possible corona region for
       Haffner 22.

\item  Our obtained distance values for both clusters from parallax are well supported by the values measured using the
isochrone fitting approach to the CMDs. Ages of $2.25\pm0.25$ and  $1.27\pm0.14$ Gyr were determined for the
clusters Haffner 22 and Melotte 71, respectively. We have compared CMDs with the theoretical isochrones of metallicity
z=0.005,  0.008 for Haffner 22 and Melotte 71, respectively, and as taken from Marigo et al. (2017).

\item  We have detected five and four-member BSS in Haffner 22 and Melotte 71, respectively, and we found those identified BSS are confirmed members of the clusters.

\item  Drawn on the relative number of high-velocity (binary) and single stars, we have determined the binary fractions for both
clusters in the range of $\sim$ $10\% \le f_{bin} \le 14\%$, for both core and oﬀ-core regions. We obtained binary content is more
in the core region for both clusters. Our investigation shows that proper motions turn out to be an essential tool for identifying
high-velocity stars as unresolved binary cluster members.

\item  The mass function slopes of $0.63\pm0.30$ and  $1.23\pm0.38$ are obtained for the clusters Haffner 22 and
Melotte 71, respectively. The MF slope for Melotte 71 is in good agreement within uncertainty with the value (1.35) given by Salpeter (1955).
We found that a flat MF slope for Haffner 22 could hint at the mass segregation in this cluster. The total mass was estimated as 572
$M_{\odot}$ and 1015 $M_{\odot}$ for both clusters.

\item The evidence of mass segregation was observed for both clusters. The $(K-S)$ test indicates $88\%$ and $80\%$ confidence
level for mass segregation in Haffner 22 and Melotte 71, respectively.
Our estimated age values for both clusters are larger than their dynamic relaxation times, indicating that both clusters are dynamically relaxed.

\item The Galactic orbits and orbital parameters were evaluated for both clusters using Galactic potential models. We found Haffner 22
and Melotte 71 are orbiting in a boxy pattern outside the solar circle, and they trace the circular path around the center of the Galaxy.
Both clusters are evolving slowly and are expected to survive for a longer lifetime.

\end{itemize}

\begin{table*}
\caption{Various fundamental parameters of the clusters Haffner 22 and Melotte 71.}
\vspace{0.5cm}
\begin{center}
\tiny
\begin{tabular}{ccc}
\hline\hline
Parameter &  Haffner 22 &  Melotte 71 \\
\hline\hline
\\
RA                                    & $123.1054\pm0.008$ deg ($8^{h} 12^{m} 25.2^{s}$)    & $114.3854\pm0.004$ deg ($7^{h} 37^{m} 32.4^{s}$)        \\
DEC                                  & $-27.9071\pm0.004$ deg ($-27^{\circ} 54^{\prime} 25.56^{\prime\prime}$)  & $-12.0598\pm0.003$ deg ($-12^{\circ} 3^{\prime} 35.28^{\prime\prime}$)          \\
Radius(arcmin)                             & 5.5                & 6.5                        \\
Radius(parsec)                             & 4.64                & 4.35                        \\
$\mu_{\alpha}cos\delta$($mas~ yr^{-1}$)    & $-1.631\pm0.009$     & $-2.398\pm0.004$            \\
$\mu_{\delta}$($mas~ yr^{-1}$)             &  $2.889\pm0.008$    &  $4.210\pm0.005$            \\
Parallax(mas)                              &  $0.3547\pm0.006$        &  $0.4436\pm0.004$          \\
Age(Gyr)                                   &  $ 2.25\pm0.25$        &  $ 1.27\pm0.14$                 \\
Metal abundance                            &  $ 0.005$            &  $ 0.008$                     \\
Distance modulus (mag)                     &  $ 12.80\pm0.40$    &  $ 12.30\pm0.30$             \\
Distance (Kpc)                             &  $ 2.88\pm0.10$     &  $ 2.28\pm0.15$             \\
$X$(Kpc)                                   &   -1.1042              &   -1.6395                        \\
$Y$(Kpc)                                   &   -2.5688             &    -1.8791                    \\
$Z$(Kpc)                                   &    0.1721              &    0.1995                       \\
$R_{GC}$(Kpc)                                   &  $ 11.1491\pm0.6$ &  $ 10.4627\pm0.7$               \\
Total Luminosity(mag)                      &   $\sim 2.5$       &   $\sim 3.3$                \\
Cluster members                            &   382              &   597                       \\
Relaxation time(Myr)                       &   25              &    30                       \\
Dynamical evolution parameter ($\tau$)     &   $ \sim  90$         &   $ \sim  42$                \\
\hline
\label{para}
\end{tabular}
\end{center}
\end{table*}

\section{ACKNOWLEDGEMENTS}

{\bf The authors thank the anonymous referee for the useful comments that improved the scientific content of the article significantly}.
This work has been financially supported by the Natural Science Foundation of China (NSFC-11590782, NSFC-11421303).
Devesh P. Sariya and Ing-Guey Jiang are supported by the grant from the Ministry of Science and Technology (MOST),
Taiwan. The grant numbers are MOST 105-2119-M-007 -029 -MY3 and MOST 106-2112-M-007 -006 -MY3. This work has made use
of data from the European Space Agency (ESA) mission Gaia (https://www.cosmos.esa.int/gaia), processed by the Gaia Data
Processing and Analysis Consortium (DPAC, https://www.cosmos.esa.int/web/gaia/dpac/consortium). Funding for the DPAC has
been provided by national institutions, particularly the institutions participating in the Gaia Multilateral Agreement.
In addition to this, It is worth mentioning that this work has been done using WEBDA.

\section{References}

\noindent Ahumada J., Lapasset E., 1995, A\&AS, 109, 375\\

\noindent Alcock, H. \& Parker R. J. 2019, MNRAS, 490, 350A\\

\noindent Allen, C. \& Santillan, A. 1991, Rev. Mexicana Astron. Astrofis., 22, 255\\ 

\noindent Allison R. J., Goodwin S. P., Parker R. J., de Grijs R., Zwart S. F. P., Kouwenhoven M. B. N., 2009, The Astrophysical Journal, 700, L99\\

\noindent Arenou, F., Luri,i X., Babusiaux, C. et al. 2018, A\&A, 616, A17\\

\noindent Auriere, M., Lauzeral, C., \& Ortolani, S. 1990, Nature, 344, 638\\

\noindent Bailer-Jones C. A. L., 2015, PASP, 127, 994\\

\noindent Bailer-Jones C. A. L., Rybizki J., Fouesneau M., Mantelet G., Andrae R., 2018, AJ, 156, 58\\

\noindent Bajkova, A. T. \& Bobylev, V. V. 2016, Astronomy Letters, 42, 9\\

\noindent Balaguer-N\'{u}\~{n}ez L., Tian, K. P., Zhao, J. L., 1998, A\&AS, 133, 387\\

\noindent Bertin G., Varri A. L., 2008, ApJ, 689, 1005\\

\noindent Baumgardt, H. \& Makino, J., 2003, MNRAS, 340, 227\\

\noindent Becker, W. \&  Fenkart, R. 1971, A\&AS, 4, 241\\

\noindent Bellini A., Piotto G., Bedin L. R., et al., 2009, A\&A, 493, 959\\

\noindent Benz, W., \& Hills, J. G. 1987, ApJ, 323, 614\\

\noindent Bica, E. \& Bonatto, C. 2005, A\&A, 431, 943\\

\noindent Bisht, D., Zhu, Q., Yadav, R. K. S. et al., 2020a, MNRAS, 482, 607\\

\noindent Bisht, D., Elsanhoury, W., Zhu, Q., et al. 2020b, AJ, 160, 119\\

\noindent Bisht, D., Zhu, Q., Yadav, R. K. S. et al., 2021a, MNRAS, 503, 5929\\

\noindent Bisht, D., Zhu, Q., Elsanhoury W., Sariya, D. P. et al., 2021b, PASJ, 73, 677B\\ 

\noindent Bobylev, V. V., Bajkova, A. T. \& Gromov, A. O. 2017, Astronomy Letters, 43, 4\\

\noindent Bonatto, C., Bica, E. 2009, MNRAS, 397, 1915\\

\noindent Brown, J. A., Wallerstein, G., Geisler, D. \& Oke, J. B. 1996, AJ, 112, 1551B\\

\noindent Bressan A., Marigo P., Girardi L., et al., 2012, MNRAS, 427, 127\\

\noindent Bukowiecki, L. et al. 2011, AcA, 61, 231\\

\noindent Cantat-Gaudin, T. \& Anders, F. 2020, A\&A, 633, A99\\

\noindent Cantat-Gaudin, T., et al. 2020, A\&A, 640, A1\\

\noindent Cantat-Gaudin, T., Jordi, C., Vallenari, A., et al. 2018, A\&A, 618A, 93C\\

\noindent Cantat-Gaudin, T., Anders F., 2020, A\&A, 633, A99\\

\noindent Carraro, G., Selezine, A. F., Baume, G. \& Turner, D. G. 2016, MNRAS, 455, 4031C\\

\noindent de La Fuente Marcos, R. 1996, A\&A, 314, 453\\

\noindent Dias, W. S., Alessi, B. S., Moitinho, A., Lepine, J. R. D., 2002, A\&A, 389, 871\\

\noindent Dias, W. S., Monteiro H., Caetano T. C. et al. 2014, A\&A, 564A, 79D\\

\noindent Dib, S. \& Henning, T. 2019, A\&A, 629, 135\\

\noindent Dib, S., Schmeja, S., \& Parker, R. J. 2018, MNRAS, 473, 849\\

\noindent Dib, S., Kim, J., \& Shadmehri, M. 2007, MNRAS, 381, L40\\

\noindent Eggen, O. J., \& Iben, I. J. 1988, AJ, 96, 635\\

\noindent Eggen, O. J., \& Iben, I. J. 1989, AJ, 97, 431\\

\noindent Ferreira, F. A., Corradi, W. J. B., Maia, F. F. S., Angelo, M. S., Santos, J. F. C. J. 2020, MNRAS, 496, 2021\\

\noindent Fischer, P., Pryor, C., Murray, S., Mateo, M., \& Richtler, T. 1998, AJ, 115, 592\\

\noindent Friel E. D., Janes K. A., 1993, A\&A, 267, 75\\

\noindent Gaia Collaboration, Brown, A. G. A., et al. 2020, arXiv:2012.01533\\

\noindent Geller A. M., Mathieu R. D., 2012, AJ, 144, 54\\

\noindent Genzel R., Townes C. H., 1987, ARA\&A, 25, 377\\

\noindent Girard, T. M., Grundy, W. M., Lopez, C. E., \& van Altena, W. F. 1989, AJ, 98, 227\\

\noindent Griffin R. F., Suchkov A. A., 2003, ApJS, 147, 103\\

\noindent Goodwin S. P., Kroupa P., 2005, A\&A, 439, 565\\

\noindent Hillenbrand L. A., Hartmann L. W., ApJ, 492, 540\\

\noindent Johnson H. L., Sandage A. R., 1955, ApJ, 121, 616\\

\noindent King, I., 1962, AJ 67, 471\\

\noindent Kharchenko N. V., Piskunov A. E., Schilbach, S., Roeser, S. and Scholz R. D. 2013, A\&A, 558A, 53K\\

\noindent Kharchenko N. V., Piskunov A. E., Schilbach, S., Roeser, S. and Scholz R. D. 2016, A\&A, 585A, 101K\\

\noindent Kroupa P., 1995a, MNRAS, 277, 1491\\

\noindent Kroupa P., 1995b, MNRAS, 277, 1507\\

\noindent Kroupa, P. 1995, MNRAS, 277, 1522\\

\noindent Kouwenhoven M. B. N., Brown A. G. A., Zinnecker H., Kaper L., Portegies
Zwart S. F., 2005, A\&A, 430, 137\\

\noindent Kumar, B., Sagar, R., \& Melnick, J. 2008, MNRAS, 386, 1380\\

\noindent Larsen S. S., 2006, An ISHAPE Users Guide. p. 14\\

\noindent Lindegren L., et al. 2020, arXiv e-prints, arXiv:2012.01742\\

\noindent Liu, L. \& Pang, X. 2019, ApJS, 245, 32L\\

\noindent Lombardi, J. C., Rasio, F. A., \& Shapiro, S. L. 1996, ApJ, 468, 797\\

\noindent Maciejewski, G. \& Niedzielski, A., 2007, A\&A, 467, 1065\\

\noindent Marigo P., Bressan A., Nanni A., Girardi L., Pumo M. L., 2013, MNRAS, 434, 488\\

\noindent Mathieu, R. D. 1984, ApJ, 284, 643\\

\noindent Mathieu, R. D. 1985, IAU Symp., 113, 427\\

\noindent Marigo, P. et al. 2017, ApJ, 835, 77\\

\noindent Mateo, M., Harris, H. C., Nemec, J., \& Olszewski, E. W. 1990, AJ, 100, 469\\

\noindent Mathieu R. D., Latham D. W., 1986, AJ, 92, 1364\\

\noindent McCrea, W. H. 1964, MNRAS, 128, 147\\

\noindent Mermilliod, J. C., Claria, J. J., Andersen J. \& Mayor, M. 1997, A\&A, 324, 91M\\

\noindent Meylan, G., 2000, Massive stellar clusters, conference held in strasbourg,
france. In:Lanon, A., Boily, C. (Eds.), A Astronomical Society of the pacific Conference
Series, p.215\\

\noindent Milone A. P. et al., 2012, A\&A, 540, A16\\

\noindent Monteiro, H., Dias, W. S., Moitinho, A., et al. 2020, MNRAS, 499, 1874\\

\noindent Nilakshi, \& Sagar, R. 2002, A\&A, 381, 65\\

\noindent Pandey, A. K., Mahra, H. S., \& Sagar, R. 1992, BASI, 20, 287\\

\noindent Pandey, A. K., Nilakshi, Ogura, K., Sagar, R., \& Tarusawa, K. 2001, A\&A, 374, 504\\

\noindent Pandey, A. K., Upadhyay, K., Ogura, K., Sagar, R., Mohan, V., Mito, H., Bhatt,
H. C., \& Bhatt, B. C. 2005, MNRAS, 358, 1290\\

\noindent Pavlik, Vaclav. 2020, A\&A, 638, A155\\

\noindent Rain, M. J., Ahumada, J. A. \& Carraro, G. 2021, A\&A, 650A, 67R\\

\noindent Reid M J., Brunthaler A., 2004, ApJ, 616, 872\\

\noindent Rangwal, G., Yadav, R. K. S., Durgapal, A., Bisht, D. \& Nardiello, D. 2019, MNRAS, 490, 1383\\

\noindent Rastegaev D. A., 2010, AJ, 140, 2013\\

\noindent Romanishim W., Angel J. R. P., 1980, ApJ, 235, 992\\

\noindent Sampedro, L., Dias, W. S., Alfaro, E. J., Monteiro, H. and Molino, A. 2017, MNRAS, 470, 3937S\\

\noindent Schonrich, Ralph., Binney, James., Dehnen, Walter. 2010, MNRAS, 403, 1829S\\

\noindent Salpeter, E. E., 1955, ApJ, 121, 161\\

\noindent Sampedro, L., Dias, W. S., Alfaro, E. J., Monteiro, H. and Molino, A. 2017, MNRAS, 470, 3937S\\

\noindent Sandage A. R., 1962, ApJ, 135, 333\\

\noindent Sariya, D. P., Jiang, Ing-Guey., Sizova, M. D., et al. 2021a, AJ, 161, 101\\

\noindent Sariya, D. P., Jiang, Ing-Guey., Bisht, D., et al. 2021b, AJ, 161, 102\\

\noindent Shao Z., Zhao J., 1996, Acta Astronomica Sinica, 37, 377\\

\noindent Sim, G., Lee, S. H., Ann, H. B., Kim, S. 2019, JKAS, 52, 145\\

\noindent Spitzer, L., \& Hart, M. H., 1971, ApJ, 164,399\\

\noindent Soubiran ,C., Cantat-Gaudin, T., et al. 2018, A\&A, 619\\

\noindent Sollima A., 2008, MNRAS, 388, 307\\

\noindent Stryker, L. L. 1993, PASP, 105, 1081\\

\noindent Twarog, B. A., Corder, S. \& Anthony-Twarog, B. J. 2006, AJ, 132, 299T\\

\noindent Williams, I. P. 1964, Annales d’Astrophysique, 27, 198\\

\noindent Wu, Z. Y., Zhou, X., Ma, J. \& Du, C. H. 2009, MNRAS, 399, 2146\\

\noindent Yang, X., Mo, H. J., van den Bosch, F. C., et al. 2005, MNRAS, 356, 1293\\

\end{document}